\documentclass[aps,reprint, prb, superscriptaddress]{revtex4-2}

\usepackage{dcolumn}% Align table columns on decimal point
\usepackage{graphicx}
\usepackage{bm}
\usepackage{amssymb}
\usepackage{amsmath}
\usepackage{amsthm}
\usepackage{hyperref}
\usepackage{comment}
\usepackage{mathtools}
\usepackage{xcolor}
\usepackage{float}
\usepackage{xcolor}
\usepackage{mathrsfs}
\usepackage{placeins}
\usepackage{multirow}
\usepackage{enumitem}
\hypersetup{colorlinks=true, citecolor=blue,linkcolor=blue,urlcolor=blue}

\newcommand\Eq[1]{Eq.~(\ref{#1})}
\newcommand\Fig[1]{Fig.~\ref{#1}}

\newcommand\ket[1]{|{#1}\rangle}
\newcommand\bra[1]{\langle{#1}|}
\newcommand\s{\bm{S}}
\newcommand\sx{{S}^x}
\newcommand\sy{{S}^y}
\newcommand\sz{{S}^z}
\newcommand\KT{U_{\textnormal{KT}}}
\newcommand{\ZZ}{\mathbb{Z}_2^y \times \mathbb{Z}_2^z}
\newcommand{\ZfZ}{\mathbb{Z}_4^y \rtimes \mathbb{Z}_2^z}
\newcommand{\HSD}{H_{\textnormal{SD}}}
\newcommand{\XXZ}{H_{\text{XXZ}}}
\newcommand{\XYZ}{H_{\text{XYZ}}}
\newcommand\sigx{{\sigma}^x}
\newcommand\sigy{{\sigma}^y}
\newcommand\sigz{{\sigma}^z}
\newcommand{\Hqbit}{\mathscr{H}_{1/2}}
\newcommand{\Hone}{\mathscr{H}_{1}}
\newcommand{\z}{\mathbb{Z}_2}

\newtheoremstyle{prop*}% name
{3pt}% Space above
{3pt}% Space below
{\itshape}% Body font
{}% Indent amount
{\bfseries}% Theorem head font
{:}% Punctuation after theorem head
{.5em}% Space after theorem head
{}% Theorem head spec (can be left empty, meaning ‘normal’ )
\theoremstyle{prop*}

\newtheorem*{prop*}{Proposition}

\newtheorem*{conj*}{Conjecture}

\begin{document}

% Use the \preprint command to place your local institutional report number in the upper righthand corner of the title page in preprint mode.
% Use the 'preprintnumbers' class option to override journal defaults to display numbers if necessary
%\preprint{}

\title{Duality, Criticality, Anomaly, and Topology in Quantum Spin-1 Chains
}


%\author{Hong Yang}
%\thanks{}
%\affiliation{Department of Physics, Beijing Normal University, Beijing 100875, China}

\author{Hong Yang}
\affiliation{
Department of Physics, Graduate School of Science, The University of Tokyo, 7-3-1 Hongo, Bunkyo-ku,  Tokyo 113-0033, Japan
}

\author{Linhao Li}
\affiliation{
Institute for Solid State Physics, The University of Tokyo, Kashiwa, Chiba 277-8581, Japan
}

\author{Kouichi Okunishi}
\affiliation{
Department of Physics, Faculty of Science, Niigata University, Niigata 950-2181, Japan
}

\author{Hosho Katsura}
\affiliation{
Department of Physics, Graduate School of Science, The University of Tokyo, 7-3-1 Hongo, Bunkyo-ku, Tokyo 113-0033, Japan
}
\affiliation{
Institute for Physics of Intelligence, The University of Tokyo, 7-3-1 Hongo, Bunkyo-ku, Tokyo 113-0033, Japan
}
\affiliation{
Trans-scale Quantum Science Institute, The University of Tokyo, 7-3-1 Hongo, Bunkyo-ku, Tokyo 113-0033, Japan
}


\date{\today}

\begin{abstract}
In quantum spin-1 chains, 
there is a nonlocal unitary transformation
known as 
the Kennedy-Tasaki transformation $\KT$,
which 
defines a duality between
the Haldane phase and the 
$\mathbb{Z}_2 \times \mathbb{Z}_2 $ symmetry-breaking phase.
In this paper,
we find that $\KT$ also defines a duality between
a topological Ising critical phase and a trivial Ising critical phase,
which provides a ``hidden symmetry breaking" interpretation of 
the topological criticality.
Moreover, since the duality relates different phases of matter,
we argue that a model with self-duality
(i.e., invariant under $\KT$)
is natural to 
be at a critical or multicritical point.
We study concrete examples to demonstrate this argument.
In particular, when $H$ is 
the Hamiltonian of
the spin-1 antiferromagnetic
Heisenberg chain, 
we prove that 
the self-dual model
$H + \KT H \KT$ is exactly equivalent to a 
gapless spin-$1/2$ XY chain,
which 
also
implies 
%not only 
an emergent quantum anomaly. 
%but also
%%immediately implies 
%a Gaussian criticality.
On the other hand, we show that the topological and trivial
Ising 
%criticalities
critical phases
that are dual to each other meet at a 
multicritical point which is indeed self-dual.
%Our discussions can be generalized
%%to other symmetries 
%beyond spin-1 chains.
\end{abstract}

\maketitle

\section{Introduction}

\mbox{Symmetry-protected topological (SPT)} phases are 
distinct from trivially gapped phases,
provided that certain symmetry is imposed.
A paradigm of SPT phases is the Haldane phase
in the spin-1 antiferromagnetic (AFM)
Heisenberg model in (1+1)D~\cite{PTBO_2010,PTBO_2012,PhysRevB.80.155131}.
Protected by the $\mathbb{Z}_2 \times \mathbb{Z}_2$ spin rotation symmetry,
%(comprising $\pi$ rotations of spins around $x,y$ and $z$ axes),
the Haldane phase is characterized by
a unique gapped ground state (GS) in the bulk,
nonlocal string order, 
and gapless edge states~\cite{PTBO_2010,PTBO_2012}.
%, and degenerate entanglement spectrum
These properties generally hold for SPT phases of
(1+1)D quantum systems protected by an on-site, unitary, 
and linear representation of
an arbitrary symmetry group $G$, and
the most general understanding of the (1+1)D SPT phases protected by $G$
is based on the projective representations classified by the
second cohomology group 
$H^2[G, U(1)]$~\cite{Chen1604,PhysRevB.83.035107,PhysRevB.87.155114,zeng2019quantum}.
Nevertheless, a broad class of (1+1)D SPT phases including the Haldane phase
can also be understood from 
a different perspective: 
\textit{hidden symmetry breaking}~\cite{PhysRevB.88.085114,Kennedy_Tasaki,kennedy1992hidden,Oshikawa_1992,PTBO_2012}.
For example, for any short-range interacting odd-integer-spin chains respecting the $\mathbb{Z}_2 \times \mathbb{Z}_2$ symmetry, a nonlocal unitary transformation,
known as the Kennedy-Tasaki (KT)
 transformation $\KT$,
defines a duality between the Haldane phase and 
the $\mathbb{Z}_2 \times \mathbb{Z}_2$ spontaneous symmetry breaking (SSB) phase~\cite{Kennedy_Tasaki,kennedy1992hidden,Oshikawa_1992,PTBO_2012}.
%In contrast, a $\mathbb{Z}_2 \times \mathbb{Z}_2$ 
%trivial phase remains trivial under $\KT$. (save???)
The SPT order of the Haldane phase is thus interpreted as
\textit{hidden $\mathbb{Z}_2 \times \mathbb{Z}_2$ symmetry breaking}.

While it has been a well-known fact that gapped phases can be further classified 
with additional symmetries imposed, 
it was recently realized that for critical systems,
a universality class can also split into distinct subclasses when
additional symmetries are
imposed, yielding the concept of
symmetry-protected 
(or symmetry-enriched) quantum criticality~\cite{PhysRevX.11.041059,PhysRevX.7.041048,PhysRevLett.120.057001,PhysRevB.97.165114,jones2019asymptotic,PhysRevB.103.L100207,PhysRevB.104.075132}. 
In particular, when two subclasses can be distinguished by 
symmetry properties of certain nonlocal operators,
an SPT/trivial classification of quantum criticalities becomes possible~\cite{PhysRevX.11.041059}.
In this work, we find that the KT transformation also defines a duality 
between an SPT Ising criticality and a trivial Ising criticality.
%((The SPT IC is protected by
%$\mathbb{Z}_2 \times \mathbb{Z}_2$ symmetry
%and hosts two-fold degenerate edge modes,
%while the trivial IC describes the transition between 
%the $\mathbb{Z}_2 \times \mathbb{Z}_2$ fully broken phase
%and a $\mathbb{Z}_2$ broken phase.))
We thus argue that the ``topological" nature of the SPT Ising criticality 
can also be interpreted as
\textit{hidden symmetry breaking}.

When a duality becomes a symmetry
(i.e., the system is \textit{self-dual}),
the self-duality must force the system to
stay on the phase boundary between
the two duality-related phases,
often leading to criticality or multicriticality~\cite{PhysRevB.104.205142, 10.21468/SciPostPhys.9.6.088, tantivasadakarn2021pivot, aasen2020topological, PhysRevB.96.125104}.
A prominent example is the quantum transverse field Ising chain
$H_{\text{Ising}} = -\sum_{j} (\sigma^z_j \sigma^z_{j+1} +h \sigma^x_j )$,
in which the Kramers-Wannier duality~\cite{PhysRev.60.252, Aasen_2016} 
exchanges the symmetric phase
and the $\mathbb{Z}_2$ SSB phase.
At the self-dual point $h=1$, $H_{\text{Ising}}$ is at a
critical point described by the Ising conformal field theory (CFT).
%i.e., the CFT with central charge $c=1/2$.

In this paper, we focus on the KT duality
and study a Hamiltonian of the form
$H(\lambda) = (1-\lambda) H_{\text{Hal}} + (1+\lambda) \KT H_{\text{Hal}}\KT $,
where $H_{\text{Hal}}$ is 
an
SO(3) symmetric and short-range interacting spin-1 chain 
in the SPT Haldane phase (for example, the AFM Heisenberg model).
In other words, $H(\lambda)$ with $-1 \leqslant \lambda \leqslant 1$
interpolates between the Haldane phase and its KT-dual phase ($\mathbb{Z}_2 \times \mathbb{Z}_2$ SSB phase).
Note that $\KT H(\lambda) \KT = H(-\lambda)$.
Surprisingly, we find that the self-dual model $H(0)$
is \textit{exactly equivalent} to 
a $(1+1)$D spin-1/2 XXZ model
doped by immobile holes,
and the holes are completely absent from the low-energy theory.
This means that the self-dual point is indeed
a critical point described by 
a Gaussian CFT (with central charge $c=1$).
Furthermore, we find that the effective model for $H(|\lambda|\ll 1)$
is given by the famous $(1+1)$D spin-1/2 XYZ model,
which implies that there is an \textit{emergent quantum anomaly}
around the self-dual point $\lambda=0$.
To our knowledge, the idea of
emergent anomaly 
can be found in Refs.~\cite{PhysRevB.96.195105, PhysRevB.104.075132,PhysRevB.98.085140,PhysRevX.8.031048,PhysRevB.98.125108}.

\begin{figure}[hbt]
  \includegraphics[width=0.5\textwidth]{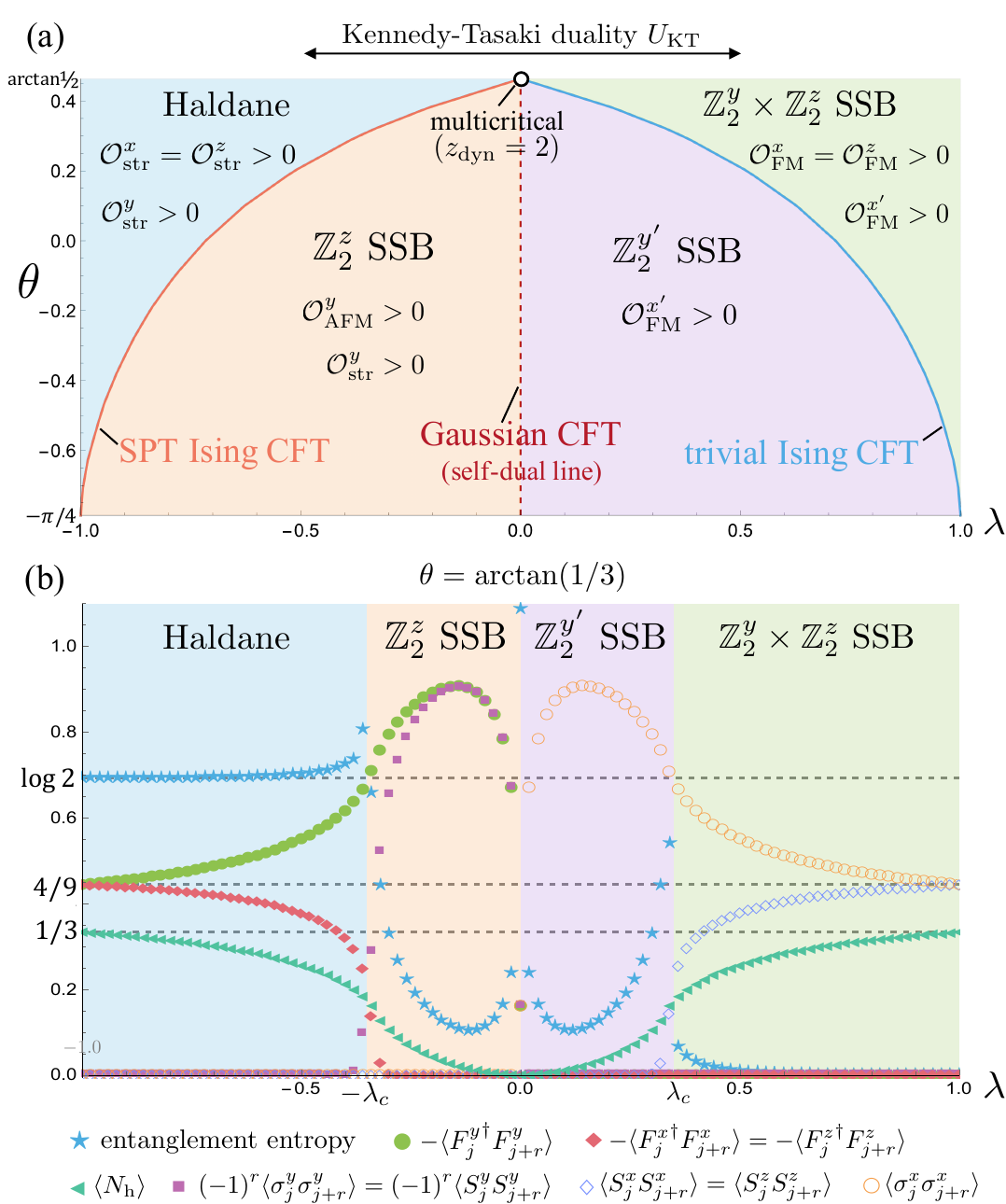}
  \caption{(a) Schematic phase diagram of $H(\lambda,\theta)$ in the region
  $(\lambda,\theta)\in [-1,1] \times (-\pi/4, \arctan\frac{1}{2}]$.
  There is an emergent Lieb--Schultz--Mattis (LSM) anomaly when $|\lambda| \ll 1$.
  Note that $\mathbb{Z}^{y}_2$ is a normal subgroup of $\mathbb{Z}^y_4$,
  while $\mathbb{Z}^{y'}_2 = \mathbb{Z}^{y}_4 / \mathbb{Z}^{y}_2$.
  Due to the global $\mathbb{Z}^y_4$ symmetry, 
  $\mathcal{O}_{\text{str}}^x = \mathcal{O}_{\text{str}}^z$ and 
  $\mathcal{O}_{\text{FM}}^x = \mathcal{O}_{\text{FM}}^z$.
  The $\mathbb{Z}^{y'}_2$ SSB phase can also be viewed as a 
  $\mathbb{Z}^z_2$ SSB phase, while
  the $\ZZ$ SSB phase is also a fully $\ZfZ$ breaking phase.
  (b) DMRG results at $\theta=\arctan(1/3)$. 
  The total hole number $\langle N_{\rm h} \rangle$ is obtained by infinite DMRG,
  while 
  the correlation functions
%  other quantities 
  are calculated on an open chain with $L=64$ and $r=40$. 
  Due to \Eq{FyFy_sigysigy}, $ \mathcal{O}_{\text{str}}^y \approx \mathcal{O}_{\text{AFM}}^y  $ when $-\lambda_c \ll \lambda <0$.
  It is also clear that $\mathcal{O}^y_{\text{str}}>0$ at the SPT Ising critical point $-\lambda_c $.
  The half-chain entanglement entropy ($L=64$) shows a sudden change at the critical points.}
  \label{phase_diagram}
\end{figure}

In fact, $H(\lambda)$ hosts more phases other than
the Haldane and the $\mathbb{Z}_2 \times \mathbb{Z}_2$ SSB phases:
there are two $\mathbb{Z}_2$ SSB phases in the region
$-\lambda_c< \lambda <0$ and $0< \lambda < \lambda_c$ (with $\lambda_c <1$),
where $\pm \lambda_c$ 
are two Ising critical points;
the former one is a $\mathbb{Z}_2 \times \mathbb{Z}_2$ trivial Ising criticality
while the latter is a $\mathbb{Z}_2 \times \mathbb{Z}_2$
SPT Ising criticality.
This means that the KT transformation also defines a
duality between the SPT and trivial Ising criticalities.
If we introduce an additional parameter $\theta$ to the model,
then the two critical lines $\pm \lambda_c(\theta)$
meet at $(\lambda, \theta)=(0,\arctan\frac{1}{2})$,
%which is a multicritical point
on which the model $H(\lambda,\theta)$ is exactly equivalent to
a \mbox{spin-1/2} ferromagnetic (FM) Heisenberg chain 
doped by immobile holes.
This means that the two KT duality-related Ising critical lines
meet at a self-dual point which is indeed multicritical.
See~\Fig{phase_diagram}(a) for the phase diagram of $H(\lambda,\theta)$.

%\FloatBarrier

\smallskip

\section{Kennedy-Tasaki (KT) transformation}

For a quantum spin-$S$ chain
where $S$ is a nonzero \mbox{\textit{integer}},
%with open boundary condition (OBC),
let $\s_j = (\sx_j,\sy_j,\sz_j)$ be the \mbox{spin-$S$} operator on 
the lattice site $j\in\{1,2,...,L\}$. 
The on-site spin rotation operators can be written as
\begin{equation}
	\begin{split}
		Y_{\theta} &= \prod\nolimits_{j} \exp(-i\theta \sy_j),\\
		Z_{\theta} &= \prod\nolimits_{j} \exp(-i\theta \sz_j),\\
		X_{\pi} &= Y_{\pi}Z_{\pi} =\prod\nolimits_{j} \exp(-i\pi \sx_j).
	\end{split}
\end{equation}
%$Y_{\theta}= \prod_{j} \exp(-i\theta \sy_j)$,
%$Z_{\theta}= \prod_{j} \exp(-i\theta \sz_j)$,
%$X_{\pi}= Y_{\pi}Z_{\pi} =\prod_{j} \exp(-i\pi \sx_j)$,
We define several rotation groups as~\footnote{The 
direct product group $\ZZ$ is isomorphic to the dihedral group $D_2$
(the dihedral group of order $4$),
while the
semidirect product group $\ZfZ$ is isomorphic to $D_4$.}
\begin{equation}
	\begin{split}
		\mathbb{Z}_4^y &= \{1, Y_{\pi/2}, Y_{\pi}, Y_{3\pi/2} \},\\
		\mathbb{Z}_2^y &= \{1, Y_{\pi} \},\\
		\mathbb{Z}_2^z &=\{ 1, Z_\pi\},\\
		\ZZ &= \{1,X_{\pi},Y_{\pi}, Z_{\pi} \},\\
		\mathbb{Z}_4^y \rtimes \mathbb{Z}_2^z &= \{1,X_{\pi},Y_{\pi}, Z_{\pi}, Y_{\pi/2}, Y_{3\pi/2}, Z_{\pi} Y_{\pi/2} , Y_{\pi/2} Z_{\pi} \}.
	\end{split}
\end{equation}
The ``symmetry flux" of $\ZZ$ is a nonlocal operator defined as
\begin{equation}
	F_j^{\alpha} =  \exp(-i\pi \sum_{k<j} S^{\alpha}_{k}) S^{\alpha}_j, \quad \alpha = x,y,z.
\end{equation}
The correlation of two symmetry fluxes gives the
nonlocal \textit{string order parameter}~\cite{StringOrder}.
%$\mathcal{O}^\alpha_{\text{str}}=-\lim_{r\uparrow\infty} \langle {F_j^{\alpha}}^\dagger F_{j+r}^{\alpha} \rangle = -\lim_{r\uparrow\infty} \langle S^\alpha_j \exp(i\pi \sum_{k=j+1}^{j+r-1} S^\alpha_k ) S^\alpha_{j+r} \rangle$
\begin{equation}
	\mathcal{O}^\alpha_{\text{str}} =-\lim_{r\rightarrow\infty} \langle {F_j^{\alpha}}  F_{j+r}^{\alpha} \rangle, \quad \alpha = x,y,z.
\end{equation}
It is known that $\mathcal{O}^\alpha_{\text{str}} > 0$
serves as an order parameter
for the Haldane phase protected by $\ZZ$,
while $\mathcal{O}^\alpha_{\text{str}} = 0$ for 
the trivial phase~\cite{Kennedy_Tasaki, kennedy1992hidden, Oshikawa_1992,TopoIndex_PhysRevB.86.125441,PhysRevLett.100.167202}.

The KT transformation is defined on a spin-$S$ chain  
with open boundary condition (OBC)
as~\cite{Kennedy_Tasaki, kennedy1992hidden, Oshikawa_1992, PhysRevB.83.104411}
\begin{equation}
	\KT = \prod_{1\leqslant u<v \leqslant L} \exp(i\pi \sz_u \sx_v),
	\label{U_KT}
\end{equation}
which satisfies $\KT=\KT^\dagger$ and $\KT^2=1$. 
%$\mathbb{Z}_4^y = \{1, Y_{\pi/2}, Y_{\pi}, Y_{3\pi/2} \}$,
%$\mathbb{Z}_2^y = \{1, Y_{\pi} \}$,
%and
%$\mathbb{Z}_2^z=\{ 1, Z_\pi\}$.
The operator $\KT$ obviously has the on-site $\ZZ$ symmetry,
which guarantees a nice property of $\KT$:
If a $(1+1)$D Hamiltonian $H$ has the on-site
$\ZZ$ symmetry, then the dual Hamiltonian $\widetilde{H} = \KT \ H \ \KT$ must also have the same on-site $\ZZ$ symmetry.
Spin operators transform under $\KT$ as~\cite{Tasaki2020,kennedy1992hidden}
\begin{equation} \label{KT_transf_integer_spin}
	\begin{split}
		\sx_j &\xleftrightarrow{\quad \KT \quad} \KT \sx_j \KT = \sx_j \ e^{ i \pi \sum\limits_{k=j+1}^L \sx_k },\\
		\sy_j &\xleftrightarrow{\quad \KT \quad } \KT \sy_j \KT = e^{ i \pi \sum\limits_{k=1}^{j-1} \sz_k } \ \sy_j \  e^{ i \pi \sum\limits_{k=j+1}^L \sx_k },\\
		\sz_j &\xleftrightarrow{\quad \KT \quad} \KT \sz_j \KT = e^{ i \pi \sum\limits_{k=1}^{j-1}  \sz_k } \ \sz_j.
	\end{split}
\end{equation}
We can thus see that in $x$ and $z$ directions,
the following duality holds:
\begin{equation}
	-{F_j^{\alpha}} F_{j+r}^{\alpha} = -
	S^\alpha_j \ e^{i\pi \sum\limits_{k=j+1}^{j+r-1} S^\alpha_k } S^\alpha_{j+r} 
	\xleftrightarrow[\alpha=x,z]{\ \KT \ }
	S^\alpha_j S^\alpha_{j+r}. \label{KT_order_parameter}
\end{equation}
[As for the $y$ direction, see \Eq{KT_FyFy}.]
It is thus clear that the Haldane phase with $\mathcal{O}^\alpha_{\text{str}} > 0$
is KT-dual to a $\ZZ$ SSB phase with an FM order
\mbox{$\mathcal{O}^\alpha_{\text{FM}}=  \lim_{r\rightarrow\infty} \langle S^\alpha_j S^\alpha_{j+r} \rangle >0 $} ($\alpha=x,z$)~\cite{Kennedy_Tasaki, kennedy1992hidden, Oshikawa_1992}.
As an example, 
it can be easily seen that for the $S=1$ case,
an AFM Heisenberg
interaction $\s_j \cdot \s_{j+1}$ is KT-dual to 
an FM interaction in both the $x$ and $z$ directions~\footnote{\label{footnote0}When $S=1$, we have
$\exp(i\pi \sx_{j+1}) \sx_{j+1} = -  \sx_{j+1}$ and $\exp(i\pi \sz_{j}) \sz_j   = -\sz_j$.}:
%%which can be seen from the identity
%more precisely,
\begin{equation}
	\s_j \cdot \s_{j+1} \xleftrightarrow[S=1]{\KT} -\sx_j \sx_{j+1} + \sy_j e^{i\pi(\sz_j + \sx_{j+1})} \sy_{j+1} - \sz_j \sz_{j+1}. \label{KT_int}
\end{equation}

On the contrary, \Eq{KT_order_parameter}
suggests that the KT dual
of a trivial phase with $\mathcal{O}^\alpha_{\text{str}} = 0$
is again a trivial phase with
$\mathcal{O}^\alpha_{\text{FM}} =0$.
This means that the trivial phase is distinct from the 
SPT phase in that the former has \textit{no} hidden symmetry breaking.
As a simple example, it can be seen that
the trivial model
\begin{equation}
	H_{\text{triv}}=\sum_j (S_j^z)^2
\end{equation}
is invariant under the KT transformation.

\smallskip

\section{Model}
From now on let us focus on the case with $S=1$.
For a spin-1 chain with only nearest-neighbor interaction 
and SO(3) spin rotation symmetry, the most general Hamiltonian is 
the bilinear-biquadratic (BLBQ) model~\cite{PhysRevA.68.063602,BLBQ_Lauchli,PhysRevResearch.3.023210}
\begin{equation}
	H_{\text{BLBQ}}(\theta)
	= \sum_{j=1}^{L-1} \big[ \cos\theta (\s_j \cdot \s_{j+1} ) 
	+ \sin\theta  (\s_j \cdot \s_{j+1} )^2 \big]. \label{BLBQ}
\end{equation}
%The Heisenberg model corresponds to $\theta=0$.
In particular, $\theta=0$ and $\arctan(1/3)$
correspond to the Heisenberg model and
the Affleck-Kennedy-Lieb-Tasaki (AKLT) model~\cite{AKLT1987,AKLT1988},
respectively.
In fact, the GS of $H_{\text{BLBQ}}(\theta)$
is in the SPT Haldane phase protected by 
%$\ZZ \subset \mathbb{Z}_4^y \rtimes \mathbb{Z}_2^z \subset \text{SO(3)} $
$\ZZ $
as long as $-\pi/4<\theta<\pi/4$~\cite{BLBQ_Lauchli},
and in that case the dual Hamiltonian $\widetilde{H}_{\text{BLBQ}}(\theta) =  \KT H_{\text{BLBQ}}(\theta) \KT$
is in the $\ZZ$ SSB phase.
%[An explicit expression of $\widetilde{H}_{\text{BLBQ}}$ can be obtained by 
%substituting \Eq{KT_int} into \Eq{BLBQ}.]
In the following, we study a one-parameter interpolation 
between the two duality-related models as
\begin{equation}
	H(\lambda,\theta) = (1-\lambda) H_{\text{BLBQ}}(\theta) + (1+ \lambda) \widetilde{H}_{\text{BLBQ}}(\theta)
	\label{model_Hamiltonian}
\end{equation}
where we have assumed the model is defined on a chain of
length $L$ with OBC 
and $-1\leqslant \lambda \leqslant 1$. 
The Hamiltonian actually has the on-site $\ZfZ$ symmetry
due to the fact that 
the right hand side of \Eq{KT_int}
%The operator $\KT$ obviously 
respects the on-site 
$\mathbb{Z}_4^y \rtimes \mathbb{Z}_2^z$ symmetry~\footnote{\label{footnote2}
The Hamiltonian $\sum_j \s_j \cdot \s_{j+1}$ has an on-site $\text{SO(3)}$ symmetry whose group element $g$
looks like $g = \prod_j \exp(-i \theta_x \sx_j - i \theta_y \sy_j - i \theta_z \sz_j)$. 
The dual Hamiltonian $ \KT  (\sum_j \s_j \cdot \s_{j+1})  \KT$ also has an $\text{SO(3)}$ symmetry,
but elements in this $\text{SO(3)}$ group take the form $\tilde{g}= \KT \ g  \ \KT$ which are \textit{not} on-site in general. 
We are only interested in on-site symmetries. 
The on-site $\ZZ$ symmetry of $\KT$ guarantees the on-site $\ZZ$ symmetry of $ \KT \ (\sum_j \s_j \cdot \s_{j+1}) \ \KT$.
However, the on-site $\ZfZ$ symmetry of $ \KT  (\sum_j \s_j \cdot \s_{j+1})  \KT$
is rather a coincidence for spin-1 chains due to the identities in footnote~[36].
For an integer spin $S>1$, $ \KT \ (\sum_j \s_j \cdot \s_{j+1}) \ \KT$
in general has no on-site $\ZfZ$ symmetry.}.
In the thermodynamic limit $L \to \infty$, 
the model has translation symmetry
(denote the group as $\mathbb{Z}^{\text{trn}}$),
and thus the whole symmetry group $G$ of $H(\lambda,\theta)$ is 
\begin{equation}
	G= \mathbb{Z}_4^y \rtimes \mathbb{Z}_2^z \times \mathbb{Z}^{\text{trn}}.
\end{equation}
%$\mathbb{Z}_4^y \rtimes \mathbb{Z}_2^z \times \mathbb{Z}^{\text{trn}}$.
%for a general $(\lambda,\theta)$.
%For a general $(\lambda,\theta)$,
%the largest on-site symmetry group of $H(\lambda,\theta)$
%is \mbox{$\mathbb{Z}_4^y \rtimes \mathbb{Z}_2^z$},
%where $\mathbb{Z}_4^y = \{1, Y_{\pi/2}, Y_{\pi}, Y_{3\pi/2} \}$.
%Obviously, $ \ZZ \subset \mathbb{Z}_4^y \rtimes \mathbb{Z}_2^z$.
%In the thermodynamic limit $L \to \infty$, 
%$H(\lambda,\theta)$ also has translation symmetry
%(denote the group as $\mathbb{Z}^{\text{trn}}$),
%thus the whole symmetry group is 
%$\mathbb{Z}_4^y \rtimes \mathbb{Z}_2^z \times \mathbb{Z}^{\text{trn}}$.
%{\color{orange}The model $H(\lambda,\theta)$ might look obscure at first glance,
%but we will show that it actually provides many intuitive pictures
%for quantum criticalities and emergent quantum anomaly. 
%In particular, $H(\lambda,\theta)$ reduces to
%a \textit{spin-1/2 XYZ model} when $|\lambda| \ll 1$.}
A phase diagram for $H(\lambda, \theta)$
is presented in \Fig{phase_diagram}(a).
Note that the $\ZZ$ SSB phase can alternatively be regarded as a fully $\ZfZ$ breaking phase
since both $\ZZ$ and $\ZfZ$ have four different 1D representations which give
rise to four (quasi)degenerate GSs.

\smallskip

\section{Self-duality} 

Since $\KT H(\lambda, \theta) \KT = H(-\lambda,\theta)$,
the model $\HSD(\theta) = H(0,\theta)$ is self-dual at $\lambda=0$.
%has self-duality at $\lambda=0$ in the sense that
%$[\HSD(\theta),\KT]=0$ 
%where $\HSD(\theta)=H(0,\theta)$.
Let $\{\ket{+}_j, \ket{0}_j, \ket{-}_j \}$ be a basis of local Hilbert space satisfying
$\sz_j \ket{\pm}_j=\pm \ket{\pm}_j$ and $\sz_j \ket{0}=0$. We define a ``$p$-wave basis" 
$\{ \ket{\uparrow}_j, \ket{\downarrow}_j, \ket{\text{h}}_j \}$
as~\cite{Kennedy_1994, PhysRevB.89.134422}
\begin{equation}
	\begin{split}
		\ket{\uparrow}_j &= \frac{1}{\sqrt{2}}(\ket{+}_j - \ket{-}_j),\\
		\ket{\downarrow}_j &= \ket{0}_j,\\
		\ket{\text{h}}_j &= \frac{1}{\sqrt{2}}(\ket{+}_j + \ket{-}_j).
	\end{split}
\end{equation}
In the following,
we will often abbreviate $\ket{\cdot}_j$ 
 to $\ket{\cdot}$ for simplicity.
Let us treat $\ket{\uparrow}/\ket{\downarrow}$ as the up/down spin 
of a spin-1/2 particle (qubit) and $\ket{\text{h}}$ as a hole.
Define Pauli operators as
\begin{equation}
	\begin{split}
		\sigma^x_{j} &= \ket{\uparrow}\bra{\downarrow} + \ket{\downarrow}\bra{\uparrow},\\
		\sigma^y_{j} &= -i\ket{\uparrow}\bra{\downarrow} + i\ket{\downarrow}\bra{\uparrow} =\sy_j,\\
		\sigma^z_{j} &= \ket{\uparrow}\bra{\uparrow} - \ket{\downarrow}\bra{\downarrow}.
	\end{split}
\end{equation}
Also define two number operators
\begin{equation}
	\begin{split}
		n_{j} &= \ket{\uparrow}\bra{\uparrow} + \ket{\downarrow}\bra{\downarrow},\\
		{h}_{j} &= \ket{\text{h}}\bra{\text{h}}
	\end{split}
\end{equation}
satisfying $n_j + h_j =1$.
The self-dual Hamiltonian can then be \textit{exactly} written as
%\begin{widetext}
\begin{equation}
	\HSD(\theta)
%	=H(0,\theta) 
%	= (\cos\theta-\sin\theta) \sum_{j=1}^{L-1} \Big[-\sigx_j \sigx_{j+1} + \sigy_{j}\sigy_{j+1} + \delta(\theta) \sigz_{j}\sigz_{j+1}  \Big] 
%	+ \sin\theta\sum_{j=1}^{L-1} \left( 
%	 2 h_j h_{j+1} + n_j n_{j+1} +2 \right),
=  \XXZ 
	+ \sin\theta\sum_{j=1}^{L-1} \left( 
	 2 h_j h_{j+1} + n_j n_{j+1} +2 \right),
	 \label{HSD}
\end{equation}
%\end{widetext}
where 
\begin{equation}
\begin{split}
	\XXZ =& (\cos\theta-\sin\theta) \sum_{j=1}^{L-1} (-\sigx_j \sigx_{j+1} + \sigy_{j}\sigy_{j+1}) \\
	 &+ \sin\theta  \sum_{j=1}^{L-1} \sigz_{j}\sigz_{j+1}
\end{split}
\label{XXZXXZXXZ}
\end{equation}
is the \textit{spin-$1/2$ XXZ model}.
The minus sign in front of $\sigx_j \sigx_{j+1}$ 
can be eliminated by a 
unitary transformation
\begin{equation}
	V=\prod_{k=\text{odd}} \sigy_k. \label{V_odd}
\end{equation}
%and
%$\delta(\theta) =  \sin\theta/(\cos\theta-\sin\theta)$.
%The minus sign in front of $\sigx_j \sigx_{j+1}$ in \Eq{HSD}
%can be eliminated by a 
%With the
%unitary transformation
%$V=\prod_{k=\text{odd}} \sigy_k$,
%the Hamiltonian
%$\mathcal{H}_{\text{SD}} = V \HSD V$
%is thus the standard \textit{spin-$1/2$ XXZ model} 
%(denoted as $\XXZ$)
%doped with holes, where
%$\XXZ = (\cos\theta-\sin\theta) \sum_{j} [\sigx_j \sigx_{j+1} + \sigy_{j}\sigy_{j+1} - \delta(\theta) \sigz_{j}\sigz_{j+1} ] $.
%Note that both the holes and spins are immobile,
%and there is no double occupancy.
The Hilbert space of a spin-1 chain is given by
\begin{equation}
	\mathscr{H}_{1}=\bigotimes_{j=1}^L \text{span}(\ket{+}_j, \ket{0}_j, \ket{-}_j),
\end{equation}
while we define the Hilbert space of a spin-1/2 chain by
\begin{equation}
	\mathscr{H}_{1/2}=\bigotimes_{j=1}^L \text{span}(\ket{\uparrow}_j, \ket{\downarrow}_j).
\end{equation}
For $\HSD$, the holes are completely 
decoupled from the qubits, 
making $\Hqbit$ an invariant subspace.
We emphasize that a system is specified by a pair consisting of
the Hamiltonian
and its underlying Hilbert space.
The pair $(\HSD, \mathscr{H}_{1})$ completely 
specifies the self-dual model.
On the other hand, 
%$(\mathcal{H}_{\text{SD}}, \mathscr{H}_{1/2})$ 
$({H}_{\text{SD}}, \mathscr{H}_{1/2})$ 
is equal to
$(\XXZ, \mathscr{H}_{1/2})$ up to a constant,
meaning that
%$P\mathcal{H}_{\text{SD}}P = P\XXZ P + 3(L-1)\sin\theta P$,
\begin{equation}
	P{H}_{\text{SD}}P = P\XXZ P + 3(L-1)\sin\theta P, \label{embed}
\end{equation}
where
\begin{equation}
	P=\bigotimes_{j=1}^L n_j = \bigotimes_{j=1}^L \Big( \ket{\uparrow}\bra{\uparrow} + \ket{\downarrow}\bra{\downarrow} \Big)
\end{equation}
%$P=\prod_j n_j$ 
is the projection operator onto $\mathscr{H}_{1/2}$.
Luckily, $(\XXZ, \mathscr{H}_{1/2})$
is exactly solvable by the
Bethe ansatz~\cite{PhysRev.150.321, PhysRev.150.327}:
Let 
\begin{equation}
	\Delta(\theta)= \frac{\sin\theta}{|\cos\theta-\sin\theta|}.
\end{equation}
%$\Delta(\theta)=\sin\theta/|\cos\theta-\sin\theta|$.
The GS of $(\XXZ, \mathscr{H}_{1/2})$ is 
\begin{enumerate}[label=(\roman*)]
	\item a Gaussian CFT ($c=1$) when $-1\leqslant \Delta <1$;
	\item gapped and degenerate when $|\Delta|>1$;
	\item gapless, degenerate, and has a dynamical critical exponent $z_{\text{dyn}}=2$ when $\Delta=1$. (In this case, the model is unitarily equivalent to the FM Heisenberg chain.)
\end{enumerate}
%(i) a $c=1$ CFT
%when $-1\leqslant \Delta <1$;
%%while the GS is 
%(ii) gapped degenerate when $|\Delta|>1$;
%and the point $\Delta=1$ represents a spin-1/2 FM Heisenberg model
%whose GS is (iii) gapless, degenerate, 
%and has a dynamical critical exponent $z_{\text{dyn}}=2$.
%which 
% $(\XXZ, \mathscr{H}_{1/2})$
%is  $(\XXZ, \mathscr{H}_{1})$.
For the low-energy theory of $(\HSD, \mathscr{H}_{1})$
in the thermodynamic limit,
we have the following
% \mbox{{Proposition}}:
%\smallskip
\begin{prop*} 
%Proposition  
When $-\infty< \Delta <1$,
holes do not appear in the low-energy eigenstates of $(\HSD, \mathscr{H}_{1})$,
meaning that states with holes are ``gapped degrees of freedom (DOF)".
On the other hand, 
when $\Delta \geqslant 1$ or $\Delta \to -\infty$,
let $W^{}_{1}$ and $W^{}_{1/2}$ be the GS
eigenspace of $(\HSD, \mathscr{H}_{1})$ and $(\HSD, \mathscr{H}_{1/2})$,
respectively. 
Then 
\mbox{$W^{}_{1} \supsetneq W^{}_{1/2}$},
implying that holes appear in $W^{}_{1}$.
\end{prop*}
%\smallskip
An intuitive picture of the case $-\infty< \Delta <1$ is shown in~\Fig{prop_schematic}.
A ``general proof" of the {Proposition} is presented in 
Appendix~\ref{app_holes}, 
%the Supplemental Material~\footnote{\label{footnote1}See Supplemental Material at \href{http://}{http://...}, which also cites Refs.~\cite{PhysRev.150.321, PhysRev.150.327,hamer1987conformal,Kapustin_1996,PhysRevX.6.041068,PhysRevLett.123.180201,PhysRevLett.114.031601,PhysRevX.8.011040, PhysRevX.11.031043,PhysRevB.87.155114, PhysRevB.83.035107,PhysRevB.104.075132,10.21468/SciPostPhys.8.1.015,francesco2012conformal,PhysRevB.12.3908,Oshikawa_1992,PhysRevB.96.125104,tantivasadakarn2021pivot,zeng2019quantum}.},
but let us take a look at two special points as
intuitive examples.
At $\theta=0$, $( V {H}_{\text{SD}} V,\mathscr{H}_{1})$ simply becomes
a \mbox{spin-1/2} XX model
doped by immobile holes.
%which is exactly equivalent to
A \mbox{spin-1/2} XX chain in $\mathscr{H}_{1/2}$ can be
exactly mapped to 
a free fermion chain.
When $L\to\infty$, 
the GS 
energy density of the fermion chain is given by 
$e_{\infty} = -4/\pi$~\cite{PhysRev.150.321, PhysRev.150.327}.
In this case, if we cut the fermion chain somewhere,
two edges will be created,
which will raise the GS energy by 
$f=2-4/\pi$~\cite{hamer1987conformal}.
Now at certain site $j$, if we replace a qubit by a hole $\ket{\text{h}}_j$,
the total length of the \mbox{spin-1/2} chain will be shortened by one,
and in the mean time
two edges will be created on both sides of $j$.
%Replacing a qubit (in the XX chain)
%by an immobile hole 
%is equivalent to a cut 
%and in the mean time shorten the 
%total length of the fermion chains by one.
This in total changes the GS energy
by $f-e_{\infty} = 2 >0$.
Therefore, holes are energetically unfavorable
when $\theta=0$.
On the other hand, 
at $\theta=\pi/4$,
$(\HSD, \mathscr{H}_{1})$
%$\HSD(\theta=\pi/4)$
is precisely the classical AFM \mbox{three-state} Potts model
when representing in the \mbox{$p$-wave} basis [see~\Eq{eq_3statepotts}],
implying that $\{\ket{\text{h}}_j\}$ are involved in
the GS eigenspace.

\begin{figure}[hbt]
  \includegraphics[width=0.5\textwidth]{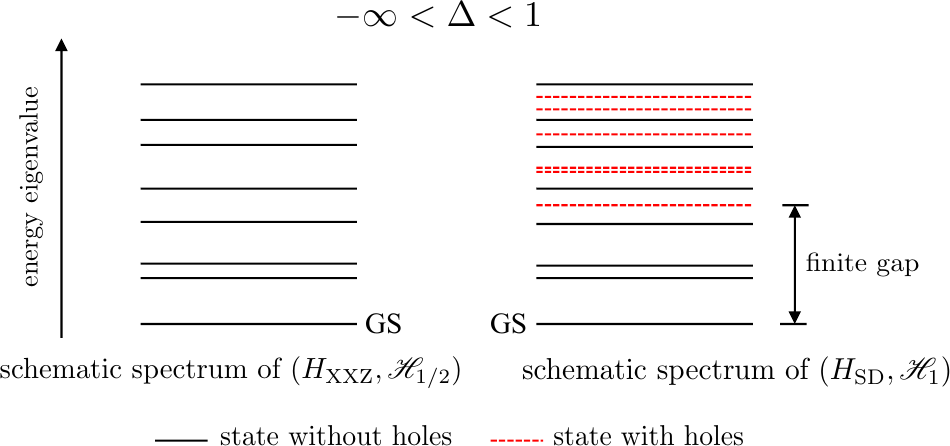}
  \caption{An intuitive picture of the case $-\infty< \Delta <1$ described in the Proposition. According to~\Eq{embed}, the spectrum of $(\XXZ, \mathscr{H}_{1/2})$ is completely embedded in that of $({H}_{\text{SD}}, \mathscr{H}_{1})$. When $-\infty< \Delta <1$, in the thermodynamic limit, eigenstates of $({H}_{\text{SD}}, \mathscr{H}_{1})$ with holes are separated from the GS by a finite energy gap, making the low-energy physics of $(\XXZ, \mathscr{H}_{1/2})$ and $({H}_{\text{SD}}, \mathscr{H}_{1})$ identical. In this figure, we have ignored the constant energy shift $3(L-1)\sin\theta$.}
  \label{prop_schematic}
\end{figure}

A direct corollary of the {Proposition} is that
following the same $\Delta$ dependence
of the spin-1/2 XXZ model,
the GS of $(\HSD, \mathscr{H}_{1})$ can only be
in any of the three cases
(i), 
(ii), and 
(iii)
as $(\XXZ, \Hqbit)$.
%In particular, 
Let us now focus on the region where
$-\infty < \Delta(\theta) <1$;
in such a case, the low-energy theories of
%$(\mathcal{H}_{\text{SD}},\mathscr{H}_{1})$
$({H}_{\text{SD}},\mathscr{H}_{1})$
and $(H_{\text{XXZ}},\mathscr{H}_{1/2})$ are identical.
Let 
%$X'_{\pi}= \prod_{j} \exp(-i\pi \sigx_j /2)$, 
\begin{equation}
\begin{split}
	Y'_{\pi} &= Y_{\pi/2} = \prod_{j} \exp(-i\pi \sigy_j /2),\\
	Z'_{\pi} &=\prod_{j} \exp(-i\pi \sigz_j /2).
\end{split}
\end{equation}
%$P_{1/2}=\prod_j n_j$,
%$P_{h}=\prod_j h_j$,
%and $\mathbb{Z}_4^{y}=\{1 , Y'_{\pi}, (Y'_{\pi})^2, (Y'_{\pi})^3 \}$.
%Let us call states with nonzero holes as ``gapped DOF";
%Note that 
Since 
\begin{equation}
	\exp(-i\pi \sigy_j ) = \exp(-i\pi \sigz_j ) = -n_j + h_j, 
\end{equation}
%and $P (Y'_{\pi})^2 P = P (Z'_{\pi})^2 P = (-1)^L P$,
we see that $Y_{\pi}=(Y'_{\pi})^2$ and $(Z'_{\pi})^2$ can
only act nontrivially on 
the gapped DOF.
We thus call $\mathbb{Z}_2^y$ a ``gapped symmetry".
In the low-energy theory (which lies in $\Hqbit$), 
$\mathbb{Z}_4^{y}$ reduces to the quotient group
\begin{equation}
	\mathbb{Z}_2^{y'} = \mathbb{Z}_4^{y}/\mathbb{Z}_2^y.
\end{equation}
%$\mathbb{Z}_2^{y'} = \mathbb{Z}_4^{y}/\mathbb{Z}_2^y$.
Similarly, one can also define 
\begin{equation}
	\mathbb{Z}_2^{z'} = \{1, Z'_{\pi}, (Z'_{\pi})^2, (Z'_{\pi})^3 \}/\{ 1, (Z'_{\pi})^2 \}.
\end{equation}

The fact that the GS of $(H_{\text{XXZ}},\mathscr{H}_{1/2})$ 
always belongs to the cases
(i), 
(ii), and 
(iii)
is nowadays understood as
a \textit{Lieb--Schultz--Mattis (LSM) anomaly}.
The anomaly essentially states that 
a spin-1/2 system with certain symmetries 
can never have 
a unique gapped GS~\cite{LSM1961,LSM1986,PhysRevLett.84.1535,PhysRevB.93.104425, ogata2019lieb, ogata2021general,yao2020twisted}.
The LSM anomaly of $H_{\text{XXZ}}$ (and also $\HSD$) is a result of
the
%$\mathbb{Z}_2^{y'} \times \mathbb{Z}_2^{z'} = \{  1,  X'_{\pi},  Y'_{\pi},  Z'_{\pi} \} $
$\mathbb{Z}_2^{y'} \times \mathbb{Z}_2^{z'} \times \mathbb{Z}^{\text{trn}}$ symmetry 
in $\mathscr{H}_{1/2}$~\cite{PhysRevB.93.104425, ogata2019lieb, ogata2021general,yao2020twisted}.
Since a self-duality in various cases implies a quantum phase transition,
one may wonder if the anomaly of $\HSD$ can also be
regarded as a result of the KT self-duality 
$\mathbb{Z}_2^{\text{KT}} = \{1, \KT \}$.
Let us consider a trivial model $H_{\text{triv}}=\sum_j (S_j^z)^2$
satisfying $[H_{\text{triv}}, \KT]=0$.
Clearly, $(H_{\text{triv}}, \mathscr{H}_{1/2})$ has a unique gapped GS,
because $PH_{\text{triv}}P=\sum_j P(\sigz_j/2 +1/2)P$.
This tells us that neither $\mathbb{Z}_2^{\text{KT}}$
nor $\mathbb{Z}_2^{\text{KT}} \times \mathbb{Z}^{\text{trn}}$
in $\mathscr{H}_{1/2}$ has an anomaly.
%However, we find that $\mathbb{Z}_2^{\text{KT}} \rtimes \mathbb{Z}_2^{y'} \times \mathbb{Z}^{\text{trn}}$ in $\mathscr{H}_{1/2}$
%indeed has an anomaly.
However, a direct calculation shows that
$P Y'_{\pi} \KT Y'_{\pi} \KT P = (-i)^L P Z'_{\pi} P$
when $L$ is even,
which means that 
within $\Hqbit$,
a system with both
$\mathbb{Z}_2^{\text{KT}} $ and $ \mathbb{Z}_2^{y'}$ symmetries
must also have
$\mathbb{Z}_2^{y'} \times \mathbb{Z}_2^{z'}$ symmetry.
% (isomorphic???).
%Let us further note that $\sigy_j = \sy_j$, so
%$\mathbb{Z}_4^{y'} = \mathbb{Z}_4^{y}$ and
%$\mathbb{Z}_2^{y'} = \mathbb{Z}_2^{y} = \{ 1, Y_{\pi}\}$. 
Therefore, we claim that the anomaly
of $\HSD$ is protected by
$\mathbb{Z}_2^{\text{KT}} $, $ \mathbb{Z}_2^{y'} $, and $ \mathbb{Z}^{\text{trn}}$
symmetries together in $\mathscr{H}_{1/2}$.
This is actually an emergent anomaly,
details will be discussed in Sec.~\ref{sec_em_ano}.

The remainder of this paper will
focus on the $\theta$ where
$H(-1,\theta)$ is in the Haldane phase
while $\HSD(\theta)$ is critical,
namely
\begin{equation}
	\theta \in \mathcal{R}=(-\pi/4, \arctan\frac{1}{2}),
\end{equation}
which is included in the region $-\infty<\Delta(\theta)<1$.

Within $\mathcal{R}$, 
it follows from
$(V H_{\text{XXZ}} V,\mathscr{H}_{1/2})$ 
that the low-energy theory of 
$(V H_{\text{SD}} V,\mathscr{H}_{1})$ 
can be exactly mapped
to a spinless fermion chain with U(1) symmetry.
See Appendix~\ref{app_KT_lowE} for
the details of $\KT$ in the spinless fermion language.

\smallskip

\section{Perturbation theory}

Our model can be written as
\begin{equation}
	H(\lambda,\theta) = \HSD(\theta) + \lambda H_{\text{pert}}(\theta),
\end{equation}
where in the $p$-wave basis, $\HSD$ is given by \Eq{HSD},
and the second term reads
\begin{widetext}
	\begin{equation} \label{H_pert}
\begin{split}
	H_{\text{pert}}(\theta) &=  \KT {H}_{\text{BLBQ}}(\theta) \KT  - H_{\text{BLBQ}}(\theta) \\
	&=- \cos\theta \sum_{j} \Big( \sigma^x_j \sigma^x_{j+1} + \sigma^y_j \sigma^y_{j+1} \Big)\\
	&\quad - 2\cos\theta \sum_{j} \Big( \ket{\uparrow \text{h}}\bra{\text{h}\uparrow} + \ket{\downarrow\mathrm{h}}\bra{\text{h} \downarrow} + \ket{\text{h}\uparrow}\bra{\uparrow \text{h}} + \ket{\text{h} \downarrow}\bra{\downarrow\text{h}}  \Big)\\
	&\quad -2(\cos\theta -\sin\theta)\sum_{j} \Big( \ket{\uparrow \uparrow}\bra{\text{hh}} + \ket{\downarrow \downarrow}\bra{\text{hh}} + \ket{\text{hh}}\bra{\uparrow \uparrow} + \ket{\text{hh}}\bra{\downarrow \downarrow}  \Big), \\
\end{split}
\end{equation}
\end{widetext}
where $\ket{\cdot \cdot}$ stands for a two-site state $\ket{\cdot\cdot}_{j,j+1}$.
Around the self-dual point, we can treat $\lambda H_{\text{pert}}$
as a perturbation to $\HSD$. 
Thanks to the {Proposition} and \Eq{H_pert},
we know that 
holes are absent from the low-energy states
of $ \big( H(\lambda,\theta ), \Hone \big)$
when $|\lambda| \ll 1 $ and $\theta \in \mathcal{R}$.
%see Supplemental Material for details.
Let $N_{\rm h} = \sum_j h_j$. Holes being absent means that
$\lim_{\lambda \rightarrow 0} \langle N_{\rm h} \rangle_{\text{GS}} =0$,
which is also verified by our numerical calculations;
see \Fig{phase_diagram}(b). 
Up to first order in $\lambda$,
we find that the effective theory for $ \big( H(\lambda,\theta ), \Hone \big)$
is given by $( \XYZ , \Hqbit )$, where
\begin{widetext}
\begin{equation}
	\XYZ(\lambda,\theta) = 
	-\big[(1+\lambda)\cos\theta -\sin\theta \big] 
	\sum_{j=1}^{L-1} \sigx_j \sigx_{j+1} 
	+ \big[(1-\lambda)\cos\theta -\sin\theta \big]
	\sum_{j=1}^{L-1} \sigy_j \sigy_{j+1}
	+\sin\theta \sum_{j=1}^{L-1} (\sigz_j \sigz_{j+1}+3).
	\label{XYZ_eff}
\end{equation}	
\end{widetext}
The \textit{spin-1/2 XYZ model} $\XYZ$ obviously has
$\mathbb{Z}_2^{y'} \times \mathbb{Z}_2^{z'} \times \mathbb{Z}^{\text{trn}}$ 
symmetry in $\Hqbit$.
Let us note that 
%$\sigy_j = \sy_j$
%and 
\begin{equation}
	P Z_{\pi} P = i^L P Z'_\pi P, \label{PzP=PzP}
\end{equation}
%$P Z_{\pi} P = i^L P Z'_\pi P$,
%The former means that 
%$\mathbb{Z}_4^{y'} = \mathbb{Z}_4^{y}$ 
%while the latter 
which means
that we can identify $Z_{\pi}$ with $Z'_\pi$ 
in the low-energy theory.
The exact solutions by 
Bethe ansatz~\cite{LUKYANOV2003323, Baxter_1974, PhysRevLett.26.832, PhysRevLett.26.834}
tell us that
when $\lambda<0$,
 $( \XYZ , \Hqbit )$
is in a phase with 
\begin{equation}
	\mathcal{O}_{\text{AFM}}^y = \lim_{r\to \infty} (-1)^{r} \langle \sigy_j \sigy_{j+r}  \rangle = \lim_{r\to \infty} (-1)^{r} \langle \sy_j \sy_{j+r}  \rangle > 0,
\end{equation}
implying the breaking of $\mathbb{Z}_2^{z'}$ in $\Hqbit$
(or equivalently, $\mathbb{Z}_2^{z}$ SSB in $\Hone$).
On the other hand, when $\lambda >0$,
the XYZ model is in the
$\mathbb{Z}_2^{y'}$ SSB phase
with 
\begin{equation} \label{FMx'>0SSB}
	\mathcal{O}_{\text{FM}}^{x'} = \lim_{r\to \infty} \langle \sigx_j \sigx_{j+r}  \rangle > 0.
\end{equation}
In fact, from
the following duality
\begin{equation}
	-{F_j^{y}} F_{j+r}^{y} \ \xleftrightarrow{\ \ \KT \ \  } \ \sigx_j \sigx_{j+r}, \label{KT_FyFy}
\end{equation}
it is clear that the two $\mathbb{Z}_2$ SSB phases are dual to each other,
because 
\begin{equation}
	-P{F_j^{y}} F_{j+r}^{y}P = (-1)^{r} P\sigy_j \sigy_{j+r}P.
	\label{FyFy_sigysigy}
\end{equation}

The whole phase diagram for $(\lambda, \theta) \in [-1,1]\times \mathcal{R}$
is determined by density matrix renormalization group (DMRG) calculations,
and the results are presented in \Fig{phase_diagram}.
%It turns out that as $\lambda$ increases from $-1$ to $1$,
%the GS successively experiences 
%the Haldane phase, the $\mathbb{Z}_2^{z}$ SSB phase,
%the $\mathbb{Z}_2^{y}$ SSB phase,
%and the $\mathbb{Z}_2^{y} \times \mathbb{Z}_2^{z}$ SSB phase.
The DMRG results suggest that 
a direct transition between
the $\mathbb{Z}_2^{y}$ SSB phase
and the $\mathbb{Z}_2^{y} \times \mathbb{Z}_2^{z}$ SSB phase
happens at $\lambda_c (\theta) >0$.
Due to the KT duality, there is also a direct transition between
the Haldane phase and the $\mathbb{Z}_2^{z}$ SSB phase
at $-\lambda_c (\theta) <0$.

On the other hand, when $\theta \in \mathcal{S} = [\arctan\frac{1}{2}, \pi/2 ]$,
although the GS eigenspace of $\HSD$ contains holes,
\Eq{E_m_holes} tells us that two holes are never adjacent.
Therefore, the effective Hamiltonian around the self-dual point 
is given by $(H_{tJ}, \Hone )$, where
\begin{widetext}
	\begin{equation}
\begin{split}
	H_{tJ} ( \lambda, \theta \in \mathcal{S} ) =
	&-\big[(1+\lambda)\cos\theta -\sin\theta \big] 
	\sum_{j} \sigx_j \sigx_{j+1} 
	+ \big[(1-\lambda)\cos\theta -\sin\theta \big] 
	\sum_{j} \sigy_j \sigy_{j+1} \\
	&+\sin\theta \sum_{j} (\sigz_j \sigz_{j+1}+ n_j n_{j+1} +2) \\
	&+ 2\lambda \cos \theta \sum_{j} \Big( \ket{\uparrow \text{h}}\bra{\text{h}\uparrow} + \ket{\downarrow\mathrm{h}}\bra{\text{h} \downarrow} + \ket{\text{h}\uparrow}\bra{\uparrow \text{h}} + \ket{\text{h} \downarrow}\bra{\downarrow\text{h}}  \Big).
\end{split}
\end{equation}
\end{widetext}
We can see that $H_{tJ}$ is like a $t$-$J$ model without double occupancy.
A detailed study of $H_{tJ}$ will be a future direction.
In the following, we will keep focusing on the region
$\mathcal{R}=(-\pi/4, \arctan\frac{1}{2})$.

\section{Emergent anomaly} \label{sec_em_ano}

The $G=\mathbb{Z}_4^{y} \rtimes \mathbb{Z}_2^{z} \times \mathbb{Z}^{\text{trn}}$ 
symmetry of the complete theory reduces to 
\begin{equation}
	G' = \mathbb{Z}_2^{y'} \times \mathbb{Z}_2^{z'} \times \mathbb{Z}^{\text{trn}}
\end{equation}
in the low-energy theory.
In other words, in the low-energy theory,
$G=\mathbb{Z}_4^{y} \rtimes \mathbb{Z}_2^{z} \times \mathbb{Z}^{\text{trn}}$
leads to an LSM anomaly,
which accounts for 
the absence of a unique gapped GS for $( \XYZ , \Hqbit )$~\cite{PhysRevB.93.104425, ogata2019lieb, ogata2021general,yao2020twisted}.
However, 
$G$ 
in $\Hone$ has no anomaly, which can be seen from 
the toy model
%$H_{\text{toy}} = -\sum_j [ (\sx_j)^2 + (\sz_j)^2 ] $
$H_{\text{toy}} = \sum_j  (\sy_j)^2  $
whose GS is trivially gapped.
In other words, $H(|\lambda| \ll \lambda_c, \theta \in \mathcal{R})$
has an \textit{emergent} anomaly.
In order to recover the complete anomaly-free theory in $\Hone$,
the emergent anomaly has to be canceled by some mechanism.
Note that for the gapped symmetry $\mathbb{Z}_2^y$,
$\exp(-i\pi \sy_j ) $ is identical to $-1$
in the low-energy theory.
This indicates that the GS
%, despite gapless,
is ``stacked" on a gapped (weak) SPT phase protected by
$\mathbb{Z}_2^y \times \mathbb{Z}^{\text{trn}}$.
It is this SPT phase that cancels the emergent anomaly,
because $Y'_{\pi} Z'_{\pi} = Y_{\pi} Z'_{\pi} Y'_{\pi}$.
Below we will explain how this works in detail.

The model
$H(|\lambda| \ll 1,\theta \in \mathcal{R})$
is effectively described by $(\XYZ, \Hqbit)$
and hence has an emergent LSM anomaly protected by 
$G'$ 
in the low-energy eigenspace.
Let ${M}_{d}$ be a $d$D manifold in
the real space and $S^1$ be a circle standing for
the imaginary time with 
periodic boundary condition (PBC).
Now let us put the model on a circle $M_1 = S^1$
[i.e., a chain with PBC. In the low-energy theory, 
the KT duality also holds for PBC;
see \Eq{PBC_qubit}.],
and consider the anomaly as the boundary 
of an SPT phase 
defined on $M_2 \times S^1$ with $\partial M_2 = M_1 $.
Due to the bulk-boundary correspondence~\cite{PhysRevX.6.041068,PhysRevLett.123.180201}, 
the SPT phase
is also protected by 
$G' = \mathbb{Z}_2^{y'} \times \mathbb{Z}_2^{z'} \times \mathbb{Z}^{\text{trn}}$.
The partition function on $M_2 \times S^1$
coupled to the $G'$-gauge field should be~\cite{PhysRevX.8.011040, PhysRevX.11.031043}~\footnote{In general, the partition function
of an SPT phase on $M_{d} \times S^1$ and the partition function of its corresponding anomalous theory on $\partial M_{d} \times S^1$ differ by a gauge invariant term, which is not important.}:
\begin{equation}
\begin{split}
	&Z[A^{y'}, A^{z'}, A^{\text{trn}}]\\
	=&  Z[0,0,0] \exp \bigg(i\pi \int_{M_2 \times S^1} A^{y'} \wedge A^{z'} \wedge A^{\text{trn}} \bigg),
\end{split}
		\label{Z_tentative}
\end{equation}
where $Z[0,0,0]$ is the partition function in the absence of the $G'$-gauge field.
$A^{y'}$, $A^{z'}$, and $A^{\text{trn}}$
are gauge fields associated with 
$\mathbb{Z}_2^{y'}$, $\mathbb{Z}_2^{z'}$, 
and $ \mathbb{Z}^{\text{trn}} $, respectively~\footnote{\label{footnote_cochain}Strictly speaking, the gauge field of a discrete group is a cochain rather than a differential form. Therefore, the product of such gauge fields should be
%a more precise expression of \Eq{Z_tentative} is to use 
the cup product $\smile$ instead of the wedge product $\wedge$. 
See Appendix~B of Ref.~\cite{kapustin2014coupling}, Appendix~J.4.e of Ref.~\cite{PhysRevB.87.155114}, and Refs.~\cite{steenrod1947products, PhysRevB.95.205142} for details.
Nevertheless, the calculus on a differential manifold and the calculus on a
simplicial complex are almost parallel. In this article, we sacrifice the mathematical rigor and
use differential forms and wedge products for simplicity.
}. 
In general, a $\z$-gauge field should satisfy the following restriction (taking $A^{y'}$ as an example)~\cite{PhysRevLett.114.031601}:
\index{Z2@$\z$-gauge field}
\index{gauge field!$\z$}
\begin{equation}  
\begin{split}
	 \int A^{y'}_{\mu} \ \mathrm{d}x^\mu 
		 &=  \ 0, 1  \mod 2, \quad  \forall \mu,  \\
  \mathrm{d}A^{y'}  \ &=  \ 0 \quad \text{ (almost everywhere)},\\
  \mathrm{d}A^{y'} \ &\neq  \ 0 \quad \text{ (at monodromy defects)},\\
   \int_{N_2}  \ \mathrm{d}A^{y'}  &=  \ 0 \mod 2 , \quad \forall N_2 \subset M_2 \times S^1 ,
\end{split}
\end{equation}
where $N_2$ is any 2D closed submanifold and
the gauge field $A^{y'}$ is a 1-form~\footnote{See footnote [64].}:
\begin{equation}
	A^{y'} = \sum_{\mu=1}^3  A^{y'}_\mu (x^1, x^2, x^3) \ \mathrm{d}x^\mu.
\end{equation}
%\begin{equation}
%	\int A_{\mu} \ \mathrm{d}\mu \mod 2 \
%		 =  \ 0, 1, \quad \forall \mu
%\end{equation}
%\begin{equation}
%	\int A_{\mu} \ \mathrm{d}\nu \mod 2 \
%		 =  \ 0, 1, \quad \forall \mu, \nu
%\end{equation}
On the other hand,
the $\mathbb{Z}$-gauge field $A^{\text{trn}}$ satisfies the restriction
\index{Z@$\mathbb{Z}$-gauge field}
\index{gauge field!$\mathbb{Z}$}
\begin{equation}
\begin{split}
	\int A^{\text{trn}}_{\mu} \ \mathrm{d}x^\mu  \
		 &=  \ 0, 1, 2,3,..., \quad \forall \mu,\\
		 \mathrm{d}A^{\text{trn}} &=0.
\end{split}
\end{equation}
%Similar to the example in \Eq{Z_Haldane},
Equation~(\ref{Z_tentative})
is not gauge invariant due to the 
(emergent) anomaly on $\partial M_2$.
Be aware that introducing the $G'$-gauge field
is sort of ``illegal" because $G'$ is not
really the symmetry of the complete theory
which is anomaly free.
The true symmetry of the complete theory is
$G = \mathbb{Z}^y_4 \rtimes \mathbb{Z}^z_2 \times \mathbb{Z}^{\text{trn}}$
in $\Hone$.
Nevertheless, $G$ reduces to $G'$ in the low-energy theory.
Therefore, for consistency,
the partition function coupled to the $G$-gauge field should
take the form
\begin{equation}
	Z[A^{\bar{y}}, A^{z}, A^{\text{trn}} ]
	= Z[ A^{y'}, A^{z'}, A^{\text{trn}} ] \ Z_{\text{others}},
	\label{Z[A]_init}
\end{equation}
where $A^{\bar{y}}$ is a $\mathbb{Z}^y_4$-gauge field, and
\index{Z4y@$\mathbb{Z}^y_4$}
\index{gauge field!$\mathbb{Z}^y_4$}
$Z_{\text{others}}$ is some other phase factor
that should be able to cancel the emergent anomaly on $\partial M_2 \times S^1$
and thus guarantees the gauge invariance of $Z[A^{\bar{y}}, A^{z}, A^{\text{trn}}]$.

Recall that $\mathbb{Z}_2^y$ is a symmetry associated with gapped DOF.
An important observation is that 
$\exp(-i\pi \sy_j)$ is identical to $ -1 $ in the low-energy theory,
which means that each lattice site is in a (0+1)D ``gapped" SPT phase 
protected by $\mathbb{Z}_2^y$.
Together with the translation symmetry, 
the GS of our model can be regarded as
``stacking" on
 a (1+1)D gapped SPT phase protected by
the $\mathbb{Z}_2^y \times \mathbb{Z}^{\text{trn}}$ symmetry~\cite{PhysRevB.87.155114, PhysRevB.83.035107}.
[This is a weak SPT phase
because it is essentially equivalent to a translational copy of (0+1)D SPT phases.]
Under the $G$-gauge field,
this (1+1)D weak SPT phase manifests itself 
via the following contribution to $Z_{\text{others}}$:
\begin{equation}
	\exp \bigg(i\pi \int_{\partial M_2 \times S^1} A^y \wedge A^{\text{trn}}  \bigg),
	\label{Z_weakSPT}
\end{equation}
where $A^{y}$ is a $\mathbb{Z}^y_2$-gauge field.
We now show that \Eq{Z_weakSPT} cancels the emergent anomaly in \Eq{Z_tentative}.
%To evaluate \Eq{Z_tentative},
Note that there is an identity
\begin{equation}
	Y'_{\pi} Z'_{\pi} = Y_{\pi} Z'_{\pi} Y'_{\pi},
	\label{YZYZY}
\end{equation}
which implies that~\cite{10.21468/SciPostPhys.8.1.015}
\begin{equation}
	 \int_{N_2} \mathrm{d} A^{y} -
	  \int_{N_2} A^{y'} \wedge A^{z'}   \ = \ 0 \mod 2,
	  \label{gauge_field_additional_constraint}
\end{equation}
where $N_2$ refers to any 2D closed submanifold of ${M_2 \times S^1}$.
Using \Eq{gauge_field_additional_constraint}
and the Stokes theorem,
\Eq{Z_tentative} becomes
\begin{equation}
	Z[ A^{y'}, A^{z'}, A^{\text{trn}} ]
	= Z[0,0,0] \ \exp\bigg(i\pi \int_{\partial M_2 \times S^1} A^y \wedge A^{\text{trn}} \bigg). \label{eqS34}
\end{equation}
%where we have assumed $\mathrm{d} A^{\text{trn}} = 0$.
Since the two phases in \Eq{Z_weakSPT} and \Eq{eqS34} combine into
\begin{widetext}
	\begin{equation}
	\exp \left(2\pi i \int_{\partial M_2 \times S^1} A^y \wedge A^{\text{trn}} \right) = \exp\left(2\pi i \int_{\partial M_2} A^y_1 \mathrm{d}x^1 \int_{S^1} A^{\text{trn}}_3 \mathrm{d}x^3 - 2\pi i\int_{S^1} A^y_3 \mathrm{d}x^3 \int_{\partial M_2} A^{\text{trn}}_1 \mathrm{d}x^1 \right) =1 ,
\end{equation}
\end{widetext}
it is now clear that \Eq{Z[A]_init}
is indeed gauge invariant and anomaly-free.

In fact, \Eq{YZYZY} can also be written as
$Y'_{\pi} Z_{\pi} = Y_{\pi} Z_{\pi} Y'_{\pi}$,
which
implies the following
\textit{short exact sequence}: \index{short exact sequence}
\begin{equation}
	1 \to  \mathbb{Z}_2^y
	\to \mathbb{Z}^y_4 \rtimes \mathbb{Z}^z_2 
	\to \mathbb{Z}_2^{y'} \times \mathbb{Z}_2^{z'}  \to 1.
\end{equation}
We say that $\mathbb{Z}^y_4 \rtimes \mathbb{Z}^z_2 $ is the \textit{extension} of
 $\mathbb{Z}_2^{y'} \times \mathbb{Z}_2^{z'}$.
We also notice that
the idea of the anomaly cancellation by symmetry extension  
\index{symmetry extension}
can be found in
in Refs.~\cite{PhysRevB.96.195105, PhysRevB.104.075132,PhysRevB.98.085140,PhysRevX.8.031048,PhysRevB.98.125108}.

At the end of this section, we note that 
\textit{higher-dimensional emergent anomaly}
might be realized by simply defining 
our model $H(\lambda,\theta)$ 
on higher-dimensional lattices,
where $\widetilde{H}_{\text{BLBQ}}(\theta)$
is defined by the right-hand side of \Eq{KT_int}.
If one can show that holes are absent
from the low-energy theory (though that might not be an easy task),
then the effective spin-1/2 Hamiltonian $H_{\text{XYZ}}$ in \Eq{XYZ_eff} 
holds regardless of dimensions.
In that case,
we have emergent LSM anomaly protected
by $\mathbb{Z}_2^{y'} \times \mathbb{Z}_2^{z'} \times (\text{crystalline symmetry})$
in higher dimensions.

\smallskip

\section{Duality of SPT/trivial Ising criticality}
\label{sec_duality_Ising}

From Eq.~(\ref{FMx'>0SSB}), one can see that
the $\mathbb{Z}_2^{y'}$ SSB phase
also breaks the $\mathbb{Z}_2^{z'}$ symmetry.
Furthermore, since the two symmetries  
$\mathbb{Z}_2^{z'}$ and $\mathbb{Z}_2^{z}$ are the same 
in the low-energy theory [see~\Eq{PzP=PzP}],
the $\mathbb{Z}_2^{y'}$ SSB phase is also a 
$\mathbb{Z}_2^{z}$ SSB phase.
Thus the $\mathbb{Z}_2^{y}$ symmetry is broken or restored
every time we cross the
critical line $\lambda_c (\theta)>0$,
indicating 
%that the criticality should belong to 
the Ising universality class.
Similarly, $-\lambda_c(\theta)<0$ is also an Ising critical line.
See Appendix~\ref{app_num_Ising} for numerical evidence.

%Note that the $\mathbb{Z}_2^{y'}$ SSB phase
%can alternatively be viewed as a $\mathbb{Z}_2^{z}$ SSB phase,
%so the $\mathbb{Z}_2^{y}$ symmetry is broken or restored
%every time we cross the
%critical line $\lambda_c (\theta)$,
%indicating 
%%that the criticality should belong to 
%the Ising universality class; see also 
%Supplemental Material for numerical evidence.
%Every time we cross the
%critical lines at $\pm \lambda_c (\theta)$, one $\mathbb{Z}_2$ symmetry 
%is broken or restored, indicating that the criticalities should belong to 
%the Ising universality class; see also 
%Supplemental Material for numerical evidence.
The transition at $\lambda_c>0$ is a \textit{trivial Ising criticality} 
(protected by $\ZZ$)
in the sense that the phases on both sides have SSB.
At $\lambda_c$, $\mathbb{Z}_2^y$ is the ``critical symmetry", and thus~\cite{francesco2012conformal}
\begin{equation}
	\langle \sx_j \sx_{j+r} \rangle = \langle \sz_j \sz_{j+r} \rangle \sim r^{-1/4}.
\end{equation}
Due to the duality in \Eq{KT_FyFy},
we know that $\mathcal{O}_{\text{FM}}^{x'} >0$ for the $\ZZ$ SSB phase.
Since $\mathcal{O}_{\text{FM}}^{x'}$ is also nonzero for the 
$\mathbb{Z}_2^{y'}$ SSB phase, it is easy to believe that 
at the Ising critical point between the two SSB phases ($\lambda_c>0$),
\begin{equation}
	\mathcal{O}_{\text{FM}}^{x'} = \lim_{r\rightarrow\infty} \langle \sigx_j \sigx_{j+r}  \rangle > 0.
\end{equation}
%On the other hand, $\mathcal{O}_{\text{FM}}^{x'} >0$ even at $\lambda_c$,
%because $\mathbb{Z}_4^y $ is
%always broken (or partially broken) when $0<\lambda \leqslant 1$.
It then directly follows from \Eq{KT_order_parameter} and \Eq{KT_FyFy} that
at $-\lambda_c<0$,
\begin{subequations}
	\begin{equation}
		\langle F_j^x F_{j+r}^x \rangle = \langle F_j^z F_{j+r}^z \rangle \sim r^{-1/4},
	\end{equation}
	\begin{equation}
		\mathcal{O}^y_{\text{str}}=-\lim_{r\rightarrow\infty} \langle {F_j^{y}} F_{j+r}^{y} \rangle >0.
	\end{equation}
\end{subequations} 
Since the nonlocal symmetry fluxes $F_j^x$ and $F_j^z$ 
carry nontrivial charges under $\ZZ$
(for example, $Z_\pi F_j^x Z_\pi = - F^x_j$)~\footnote{\label{footnote2}
Instead of $\ZfZ$, it is sufficient to consider its subgroup $\ZZ$,
because their second cohomology groups are the same:
$H^2[\ZfZ,U(1)] = H^2[\ZZ,U(1)] = \mathbb{Z}_2$~\cite{PhysRevB.87.155114, PhysRevX.11.041059}.},
we claim that the transition at $-\lambda_c$ represents a
\textit{$\ZZ$ SPT Ising criticality}~\cite{PhysRevX.11.041059}.
Since $\mathbb{Z}_2^z$ is broken as long as $0<\lambda \leqslant 1$,
it is obvious that the GS has two-fold
degeneracy at $\lambda_c$; so is that at $-\lambda_c$ 
due to the KT duality
(remember that we always assume OBC).
The two-fold degeneracy at $-\lambda_c$ is actually 
associated with topological edge states~\cite{PhysRevX.11.041059};
this 
can be seen by noting that
%follows from the clustering property
$0 \neq \langle F_1^{y} F_L^y \rangle
= \langle S^y_1 e^{i\pi S^y_1} Y_\pi e^{i\pi S^y_L} S^y_L \rangle
= \pm \langle S^y_1 e^{i\pi S^y_1} e^{i\pi S^y_L} S^y_L \rangle 
 $
implies edge magnetization 
$\langle S^y_1 e^{i\pi S^y_1} \rangle = - \langle S^y_1 \rangle  \neq 0$ 
and $\langle S^y_L \rangle \neq 0 $,
where we have used the clustering property
$\langle S^y_1 e^{i\pi S^y_1} e^{i\pi S^y_L} S^y_L \rangle  \approx  \langle S^y_1 e^{i\pi S^y_1} \rangle \langle e^{i\pi S^y_L} S^y_L \rangle$~\cite{PhysRevB.104.075132}.
Moreover,
$\mathcal{O}^y_{\text{str}}>0$ at $-\lambda_c$ indicates that 
the topological criticality is partially protected by 
the gapped symmetry $\mathbb{Z}^y_2$,
which further implies that the
twofold (quasi)degenerate GS
has an energy splitting proportional to $e^{-L/\xi}$~\cite{PhysRevX.11.041059}. 
%For a general theory of symmetry-enriched Ising criticalities, 
%see Ref.~\cite{PhysRevX.11.041059}.
From the above discussions, we can see that the KT duality
also provides a \textit{hidden $\ZZ$ symmetry breaking} 
picture for the SPT criticality:
The algebraic decay or the long-range order
of $\langle F_j^{\alpha } F_{j+r}^\alpha \rangle$ at $-\lambda_c$
can be easily understood from the classical Landau transition at $\lambda_c$.
For more details about the interpretation of hidden symmetry breaking,
see Appendix~\ref{hidd_symm_brk}.

%Furthermore, it is clear that the GS has two-fold
%degeneracy at $\lambda_c$???
%(hidden symmetry-breaking picture)???
%Moreover,
%$\mathcal{O}^y_{\text{str}}>0$ indicates that
%the SPT criticality is partially protected by 
%the gapped symmetry $\mathbb{Z}^y_2$,
%which further implies that 
%the model with OBC at $-\lambda_c$
%has a two-fold (quasi-)degenerate GS
%whose energy splitting is proportional to $e^{-L}$
%for a large $L$~\cite{PhysRevX.11.041059}. 
%The two-fold degeneracy at $-\lambda_c$ is actually 
%associated with topological edge states~\cite{PhysRevX.11.041059};
%this can be seen by noting that
%$0 \neq \langle F_1^{y\dagger} F_L^y \rangle
%= \langle S^y_1 e^{i\pi S^y_1} Y_\pi e^{i\pi S^y_L} S^y_L \rangle
%= \pm \langle S^y_1 e^{i\pi S^y_1} e^{i\pi S^y_L} S^y_L \rangle$
%implies $\langle S^y_1 e^{i\pi S^y_1} \rangle = - \langle S^y_1 \rangle  \neq 0$ 
%and $\langle S^y_L \rangle \neq 0 $~\cite{PhysRevB.104.075132}.
%For a general theory of symmetry-enriched Ising criticalities, 
%see Ref.~\cite{PhysRevX.11.041059}.

In \Fig{phase_diagram}(a), one can see that the trivial and topological Ising criticalities
related by the KT duality meet at 
the self-dual point $(\lambda,\theta)=(0,\arctan\frac{1}{2})$,
forcing the model $\HSD(\arctan\frac{1}{2})$ to be
at a multicritical point.
Indeed, 
$\theta=\arctan\frac{1}{2}$ corresponds to $\Delta=1$, in which case
%as we have seen before,
$\big( \HSD, \Hone \big)$ is equivalent to a
spin-1/2 FM Heisenberg model doped by immobile holes,
which has $z_{\text{dyn}}=2$.
On the other direction, 
the Ising critical lines terminate at 
$H(\pm 1, -\pi/4)$, whose low-energy physics is the CFT
with $c=3/2$~\cite{TAKHTAJAN1982479, BABUJIAN1982479, PhysRevB.59.11358}.

\section{Discussion}

We have been focusing on the spin-1 chains,
but in fact, the KT transformation
directly applies to \textit{any integer} spin quantum number $S$,
as long as we take $\sz_u$ and $\sx_v$ in 
\Eq{U_KT} to be the \mbox{spin-$S$} operators~\cite{Oshikawa_1992}.
Let $H_S = \sum_j \bm{S}_j \cdot \bm{S}_{j+1} $ be the
spin-$S$ AFM Heisenberg chain; it is believed that the GS of
$H_S$ is in the $\ZZ$ SPT phase when $S$ is odd, while the GS is
trivial when $S$ is even~\cite{Oshikawa_1992,Tasaki2020}.
Note that the KT dual of 
the trivial phase is still trivial.
Therefore, we propose the following 
%\textit{Conjecture}:
%\smallskip
\begin{conj*}
	Let $H_{\textnormal{SD}}^{(S)} = H_S + \KT H_S \KT$;
the GS of $H_{\textnormal{SD}}^{(S)}$ is gapless when $S$ is odd,
while it is trivially gapped when $S$ is even.
\end{conj*}
%\smallskip
Moreover, the KT transformation can be generalized to 
(1+1)D systems with a broad class of symmetries beyond $\mathbb{Z}_2 \times \mathbb{Z}_2$, 
such as $\mathbb{Z}_n \times \mathbb{Z}_n$ and SO($2n-1$)~\cite{PhysRevB.88.085114, PhysRevB.78.094404, PhysRevB.88.125115, PhysRevB.89.125112},
thus providing the hidden symmetry breaking picture for 
the SPT phases in such systems.
Exploring the relationship between the
KT duality, criticality, anomaly, and topology
in such systems will also be interesting.

\smallskip

\begin{acknowledgments}
We acknowledge stimulating discussions with \mbox{Yunqin Zheng.}
H.~Y. was supported by the Grant-in-Aid for
JSPS Research Fellowship for Young Scientists (DC1), Grant No.~20J20715.
%K.~O. was supported by Grant-in-Aid for Scientific Research No.~21H05191 
%and No.~21H05182.
%H.~K. was supported in part by JSPS Grant-in-Aid for Scientific Research on Innovative Areas No.~JP20H04630, JSPS KAKENHI Grant No.~JP18K03445, 
%Grant-in-Aid for Transformative Research Areas (A) “Extreme Universe” No.~JP21H05191[D02], and the Inamori Foundation.
K.~O. and H.~K. were supported by funding including
JSPS Grants-in-Aid for Scientific Research 
on Innovative Areas (Grant No.~JP20H04630) 
and for Transformative Research Areas (A) 
``Extreme Universe” (Grant No.~JP21H05191 and No.~21H05182).
H.~K. was also supported in part by JSPS KAKENHI Grant No.~JP18K03445 
and the Inamori Foundation. 
\end{acknowledgments}

\FloatBarrier 
\appendix

\section{Existence or absence of holes}
\label{app_holes}

\begin{figure}
\centering
\includegraphics[width=0.5\textwidth]{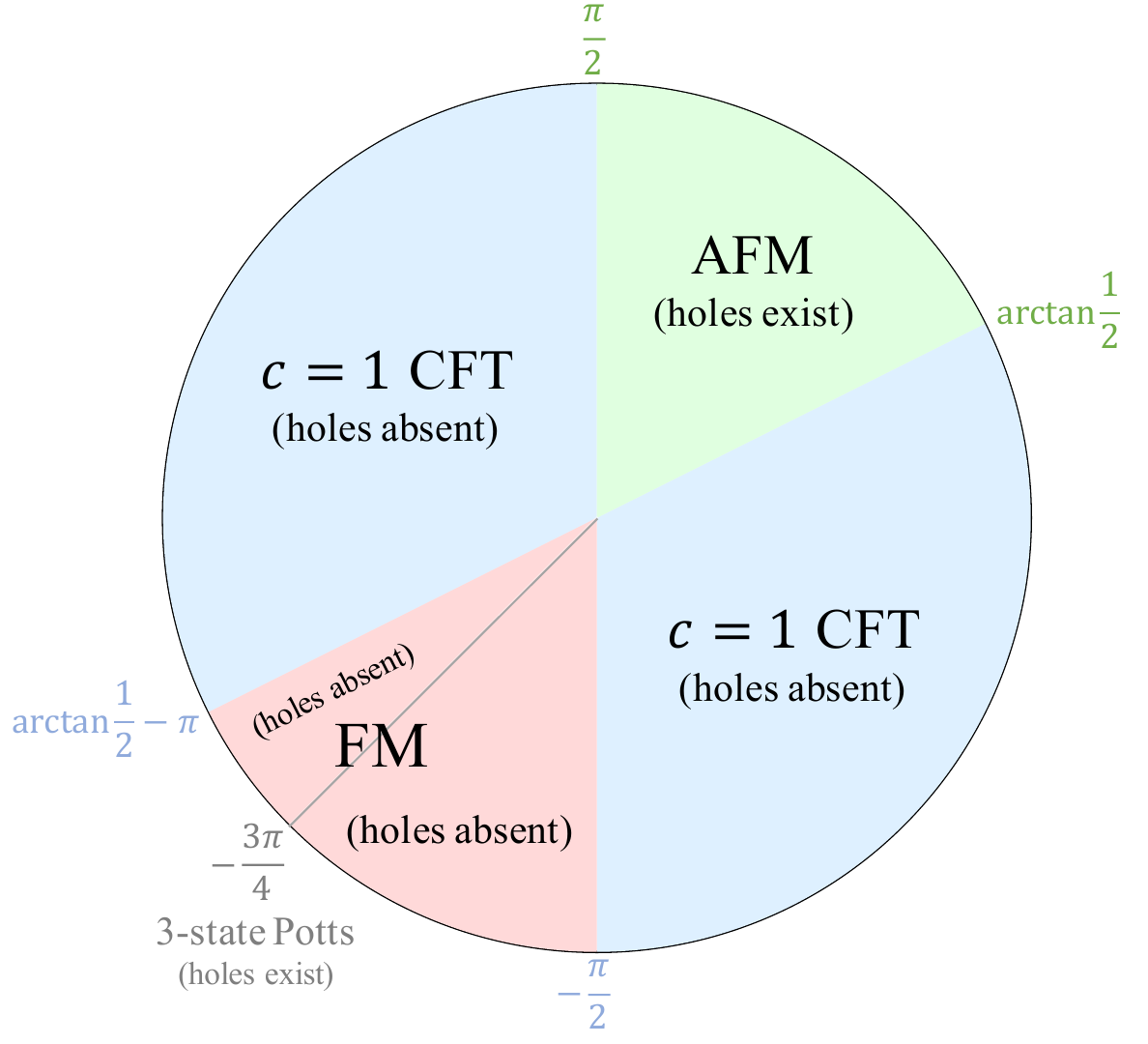}
\caption{\label{SD_PD} Phase diagram for the GS of $\HSD(\theta)$ in \Eq{HSD}. Holes are absent from the low-energy eigenstates when 
$\theta\in (\pi/2, 2\pi+\arctan\frac{1}{2})\backslash\{5\pi/4\}$,
which means that the low-energy physics of $\HSD$ 
in this region is exactly the same as the spin-1/2 XXZ model.
Note that 
%$\HSD$ and $H'_{\text{SD}}$ are related to each other
%by $V=\prod_{k=\text{odd}} \sigy_k$, so that 
an FM (AFM) GS of $\HSD$
corresponds to an AFM (FM) GS of $H'_{\text{SD}}$.}
\end{figure}

A proof of the {Proposition}
is presented here.
Although our proof might not be entirely rigorous 
from a mathematical point of view,
it nevertheless makes sense for physicists.
%Recall that $V=\prod_{k=\text{odd}} \sigy_k$;
%According to \Eq{HSD}, 
%$\mathcal{H}_{\text{SD}}(\theta) =  V\HSD(\theta)V$
Let us begin by first noting that
by properly rotating the spins, 
$\HSD$ in \Eq{HSD} is always unitarily equivalent to
\begin{widetext}
	\begin{equation}
	H'_{\text{SD}}(\theta) = 2\sqrt{2} \left| \sin(\frac{\pi}{4}-\theta) \right| \times H_{\text{XXZ}} (\theta) 
	+ \sin\theta\sum_{j=1}^{L-1} \left( 
	 2 h_j h_{j+1} + n_j n_{j+1} +2 \right),
	 \label{mathcal_HSD}
\end{equation}
\end{widetext}
%\begin{equation}
%\begin{split}
%	H'_{\text{SD}}(\theta) = &2\sqrt{2} \left| \sin(\frac{\pi}{4}-\theta) \right| \times H_{\text{XXZ}} (\theta) \\
%	&+ \sin\theta\sum_{j=1}^{L-1} \left( 
%	 2 h_j h_{j+1} + n_j n_{j+1} +2 \right),
%	 \label{mathcal_HSD}
%\end{split}
%\end{equation}
where 
\begin{equation}
	H_{\text{XXZ}} (\theta) =  \frac{1}{2}
	\sum_{j=1}^{L-1} \Big[\sigx_j \sigx_{j+1} + \sigy_{j}\sigy_{j+1} - \Delta(\theta) \sigz_{j}\sigz_{j+1}  \Big]. \label{XXZ_supp}
\end{equation}
Note that the definition of $\XXZ$ 
here is different from the main text, and
within this appendix we will adapt the definition in \Eq{XXZ_supp}.
The parameter
\begin{equation}
	\Delta(\theta) 
	= \frac{\sin\theta}{|\cos\theta - \sin\theta|} 
\end{equation}
determines the phase of $H_{\text{XXZ}}(\theta)$ 
and hence $H_{\text{SD}}(\theta)$.
The results are summarized in \Fig{SD_PD}.

\subsection{Absence of holes}

Let
\begin{align}
	\mathcal{A} &= 
	[-\frac{\pi}{2} , \arctan\frac{1}{2} ) \cup 
	(\frac{\pi}{2} , \arctan\frac{1}{2} + \pi ], \label{S_theta1}\\
	\mathcal{B} &= 
	(\arctan\frac{1}{2}-\pi  , -\frac{\pi}{2} )\backslash\{-\frac{3\pi}{4}\} .
	\label{S_theta2}
\end{align}
When $\theta\in \mathcal{A}$, we have
$-1 \leqslant \Delta(\theta)<1$, and thus $H_{\text{XXZ}}$
is gapless and the low-energy physics
is described by a $c=1$ CFT.
When $\theta\in \mathcal{B}$, $\Delta(\theta)<-1$ and $H_{\text{XXZ}}$
has two degenerate and AFM ground states
in the thermodynamic limit $L\to\infty$.
Let $\gamma = \arccos[-\Delta(\theta)]$ when $\theta\in \mathcal{A}$
and $\xi = \mathrm{arccosh}[-\Delta(\theta)]$ when $\theta\in \mathcal{B}$.
The ground-state (GS) energy density of $H_{\text{XXZ}} $ 
in the thermodynamic limit,
denoted as $e_{\infty}$,
was exactly obtained by 
Yang and Yang back in 1966~\cite{PhysRev.150.321, PhysRev.150.327}:
\begin{widetext}
	\begin{equation}
e_{\infty}(\theta) =
\begin{dcases}
	&\dfrac{1}{2}\cos\gamma - (\sin\gamma)^2
	\int_{-\infty}^\infty \dfrac{\mathrm{d}x}{\cosh(\pi x)[ \cosh(2\gamma x) -\cos\gamma ]}, \quad -1\leqslant -\cos\gamma=\Delta(\theta)<1, \\
	&\dfrac{1}{2} \cosh\xi - 
	\bigg[ 1 + 4 \sum_{n=1}^{\infty} \dfrac{1}{ 1 + e^{2\xi n} }  \bigg] \sinh\xi,
	\quad -\cosh\xi = \Delta(\theta) <-1.
\end{dcases}
\end{equation}
\end{widetext}
When $1\ll L<\infty$, we need to consider finite-size corrections.
For OBC, the finite-size GS energy density 
$e^{\text{OBC}}_{L}$ takes the form~\cite{hamer1987conformal,Kapustin_1996}:
\begin{equation}
	e^{\text{OBC}}_{L} = e_{\infty} + \frac{f}{L} + o\left(\frac{1}{L} \right),
\end{equation}
where $f$ is called the ``surface energy" and is given by
\index{surface energy}
%\begin{equation}
%	f = \frac{\pi \sin \gamma}{2\gamma} 
%	- \frac{\cos\gamma}{2}
%	- \frac{\sin\gamma}{4}\int_{-\infty}^\infty 
%	\mathrm{d}x\bigg[ 1- \tanh\left(\frac{\pi x}{4}\right) \tanh\left(\frac{\gamma x}{2}\right) \bigg].
%	\label{surface_1}
%\end{equation}
\begin{widetext}
	\begin{equation}
f(\theta)=
	\begin{dcases}
		\frac{\pi \sin \gamma}{2\gamma} 
	- \frac{\cos\gamma}{2}
	- \frac{\sin\gamma}{4}\int_{-\infty}^\infty 
	\mathrm{d}x\bigg[ 1- \tanh\left(\frac{\pi x}{4}\right) \tanh\left(\frac{\gamma x}{2}\right) \bigg], \quad -1\leqslant -\cos\gamma=\Delta(\theta)<1, \\
	-\frac{1}{2} \cosh\xi + 
	4\bigg[ \frac{1}{4} + \sum_{n=1}^{\infty}  \frac{e^{2n\xi}-1}{1+e^{4n\xi}} 
	+ \sum_{n=1}^{\infty} \frac{(-1)^n}{1+e^{2n\xi}}  \bigg] \  \sinh\xi,
	\quad -\cosh\xi = \Delta(\theta) <-1
	\end{dcases}.
	\label{surface_1}
\end{equation}
\end{widetext}

\begin{figure}
\centering
\includegraphics[width=0.5\textwidth]{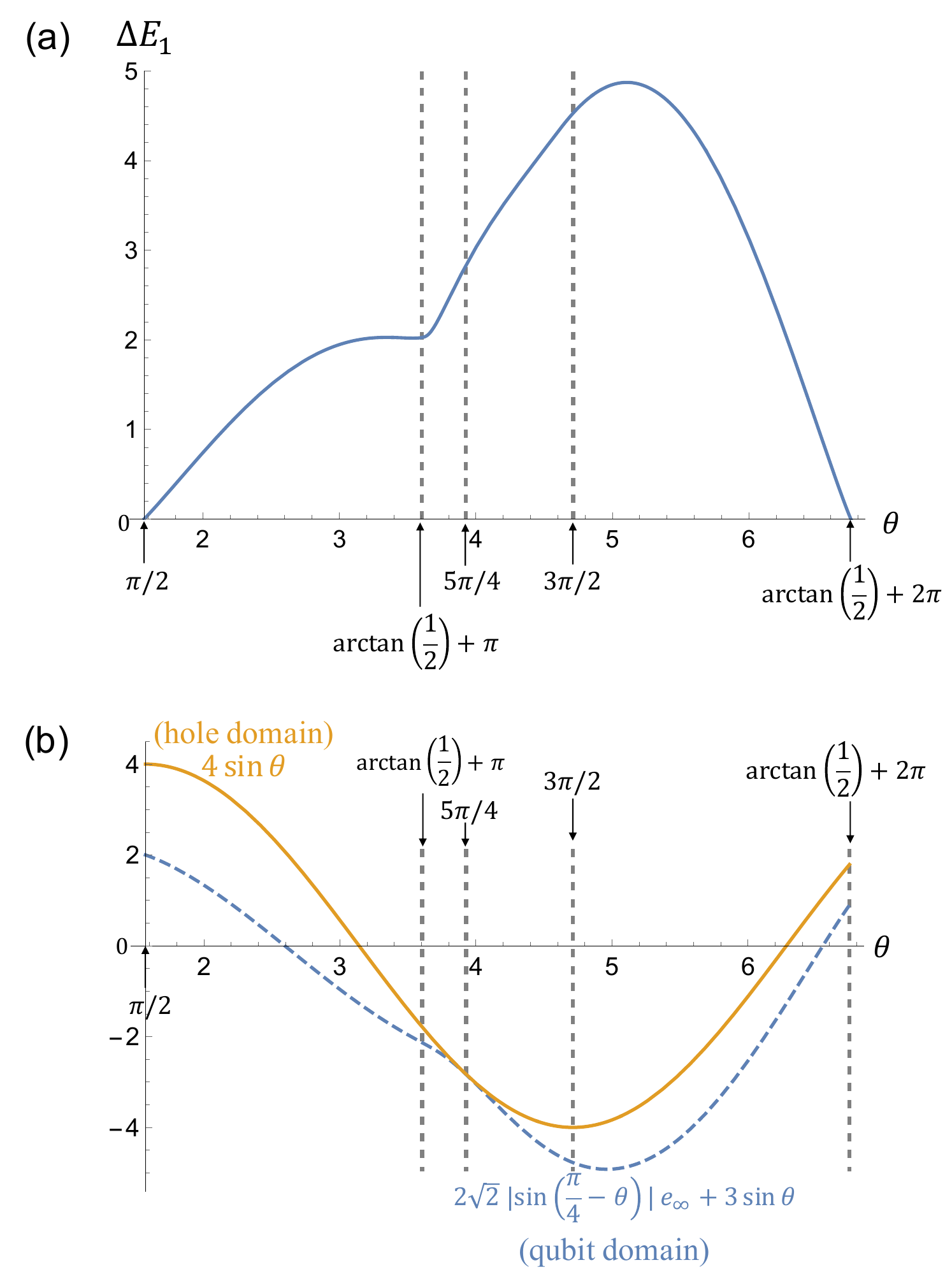}
\caption{\label{fig_E} (a) $\Delta E_1>0$ for all 
$\theta\in(\pi/2, \arctan\frac{1}{2}+2\pi)$.
(b) Energy density of the hole domain is higher than that of the qubit domain
when $\theta\in A\cup B = (\pi/2, \arctan\frac{1}{2}+2\pi) \backslash \{5\pi/4\}$, implying the absence of the phase separation.}
\end{figure}

Note that \Eq{mathcal_HSD} is defined on a chain with OBC.
For a sufficiently long chain,
in $\Hqbit$ (the subspace without holes),
the GS energy of $H_{\text{SD}}$ 
(up to the order of $L^0$) is given by
\begin{equation}
	E_{0} = 2\sqrt{2}\ \left|\sin(\frac{\pi}{4}-\theta) \right|\ \left( L e_{\infty} + f \right) + 3(L-1)\sin\theta. 
\end{equation}
Now consider the subspace of $m$ holes.
When the $m$ holes are disjoint, 
sufficiently far away from each other, and sufficiently far away from the boundary,
the GS energy of $H_{\text{SD}}$ in this subspace is
(up to the order of $L^0$)
\begin{equation}
\begin{split}
	E_m = &2\sqrt{2}\ \left|\sin(\frac{\pi}{4}-\theta) \right| \ \big[ (L-m) e_{\infty} + (m+1)f \big]\\
	&+ (L-1-2m)\sin\theta + 2(L-1) \sin\theta.
\end{split}
\end{equation}
The energy difference is given by
\begin{equation}
\begin{split}
	\Delta E_m &= 
	E_m - E_0 \\ &= m \bigg[ 2\sqrt{2}\ \left|\sin(\frac{\pi}{4}-\theta) \right| \ (f-e_{\infty}) - 2\sin\theta \bigg] \\
	 &= m \Delta E_1.
\end{split}
\end{equation}
As long as $\Delta E_1 = 2\sqrt{2} |\sin(\pi/4-\theta)| (f-e_{\infty}) - 2\sin\theta >0$,
eigenstates with disjoint holes are gapped from the ground state.
The value of $\Delta E_1$ can be easily obtained numerically, 
see \Fig{fig_E}(a). We can see that $\Delta E_1$ is indeed
positive when $\theta\in \mathcal{A} \cup \mathcal{B}$.

We have not yet ruled out the possibility of
``phase separation", meaning that holes 
form a domain, like 
$\ket{...\uparrow \uparrow \text{hhh...hh} \downarrow\uparrow ...}$.
In fact, the energy density of the hole domain 
is given by $4\sin\theta$,
while the energy density of
the spin-1/2 domain is $2\sqrt{2} |\sin(\pi/4-\theta)| e_{\infty} + 3\sin\theta$
(up to the order of $L^0$).
We can numerically show that the former energy density is always higher than
the latter when $\theta\in \mathcal{A} \cup \mathcal{B}$ [see \Fig{fig_E}(b)], 
which means that the phase separation does not occur.

To sum up, holes $\{\ket{\text{h}}_j\}$ are gapped from the ground state of 
$\HSD(\theta)$ when $\theta$ is in the region described by 
\Eq{S_theta1} and \Eq{S_theta2},
which is in accordance with our DMRG results.

\subsection{Existence of holes}

At two special points $\theta=\pi/4$ and $-3\pi/4$,
$\HSD$ (but not $H'_{\text{SD}}$) reduces to 
the classical \textit{three-state Potts model}.
\index{Potts model}
This can be seen by noting that
\begin{equation}
\begin{split}
	&\sigz_j \sigz_{j+1} + 2h_j h_{j+1} + n_j n_{j+1} \\
	=& 2\Big( \ket{\uparrow\uparrow}\bra{\uparrow\uparrow}
	+ \ket{\downarrow\downarrow}\bra{\downarrow\downarrow} + \ket{\text{hh}}\bra{\text{hh}} \Big),
\end{split} \label{eq_3statepotts}
\end{equation}
where $\theta=\pi/4$ is AFM and $\theta=-3\pi/4$ is FM.
The ground states at these two points are thus degenerate,
and holes $\{\ket{\text{h}}_j\}$ appear in the GS eigenspace.

In the region
\begin{equation}
	\theta \in \bigg[ \arctan\frac{1}{2},  \frac{\pi}{2} \bigg],
\end{equation}
$\Delta(\theta) \geqslant 1$, and the ground state of $H_{\text{XXZ}}$
is FM. The GS energy of $H_{\text{XXZ}}$ exactly equals
$-(L-1)\Delta(\theta)/2$.
In the subspace without holes,
the GS energy of $H_{\text{SD}}$ 
is then given by
\begin{equation}
\begin{split}
	E_0 &= 2\sqrt{2} | \sin(\frac{\pi}{4}-\theta) | \times  [-(L-1)\Delta(\theta)/2]
	+ 3 (L-1) \sin\theta\\
	&= 2(L-1)\sin\theta.
\end{split}
\end{equation}
In the subspace of $m$ holes, when the holes are all disjoint, the GS energy becomes
\begin{equation}
\begin{split}
	E_{m} &= 2\sqrt{2} | \sin(\frac{\pi}{4}-\theta) | \times  [-(L-1-2m)\Delta(\theta)/2]\\
	&\quad +  (L-1-2m) \sin\theta + 2(L-1)\sin\theta \\
	&= 2(L-1)\sin\theta = E_0,
\end{split}
\end{equation}
which means that adding disjoint holes to the ground state 
does not cost energy. 
On the other hand, if the $m$ holes form a domain,
\begin{equation}
\begin{split}
	E_{m} &= 2\sqrt{2} | \sin(\frac{\pi}{4}-\theta) | \times  [-(L-m-2)\Delta(\theta)/2]\\
	&\quad +2(m-1)\sin\theta + (L-m-2) \sin\theta + 2(L-1)\sin\theta \\
	&= 2(L+m-2)\sin\theta. \label{E_m_holes}
\end{split}
\end{equation}
We see that $E_m > E_0$ as long as $m \geqslant 2$.
In other words, phase separation does not occur in the ground state.

\section{KT duality in low-energy theory} \label{app_KT_lowE}

As shown in the previous sections,
when 
\begin{equation}
|\lambda| \ll 1, \quad
	\theta \in \mathcal{R}=\left(-\frac{\pi}{4}, \arctan\frac{1}{2} \right), 
\end{equation}
the low-energy eigenspace of the 
model 
$ \big( H(\lambda,\theta ), \Hone \big)$ completely lies in $\Hqbit$.
The projection onto $\Hqbit$ gives
\begin{equation}
	P \ \KT  P = 
	\prod_{1 \leqslant u<v \leqslant L} 
	P \exp \big[\frac{i\pi}{4} ( 1+ \sigz_u )  ( 1- \sigz_v )  \big] P.
% e^{\frac{i\pi}{4} \big[ 1+(-1)^u \sigz_u \big] \big[ 1-(-1)^v \sigz_v \big] }
\label{UKT_qubit}
\end{equation}
From \Eq{UKT_qubit}, it can be shown that within $\Hqbit$,
the following duality holds:
\begin{equation}
\begin{split}
	-{\sigy_j} \sigy_{j+1} \ &\xleftrightarrow[\ 1\leqslant j< L-1\ ]{\ \ \KT  \ \  } \ \sigx_j \sigx_{j+1}, \\
	(-1)^{L-1} {\sigy_L} \sigy_{1} \ &\xleftrightarrow[\quad \quad \quad \ \ ]{\ \ \KT  \ \  } \ \sigx_L \sigx_{1}.
\end{split} \label{PBC_qubit}
\end{equation}
Interestingly, although we have been dealing with the case of OBC,
\Eq{PBC_qubit} shows that even if we impose
PBC on \Eq{XYZ_eff}, 
the effective theory around the self-dual point
still respects the KT duality
as long as $L$ is even.

Within $\mathcal{R}$, it follows from
$(H_{\text{XXZ}},\mathscr{H}_{1/2})$ 
that the low-energy theory of 
$(\mathcal{H}_{\text{SD}} = V\HSD V,\mathscr{H}_{1})$ 
can be exactly mapped
to a U(1) symmetric spinless fermion chain 
by the Jordan-Wigner transformation
\begin{equation}
	c_j^\dagger = \sigma^+_j \prod_{k<j}(-\sigz_k),
\end{equation}
where $V$ is defined in \Eq{V_odd} and
$c_j^\dagger$ is a fermion creation operator.
Let $\mathcal{U}_{\text{KT}} = V \KT V$ satisfying
\mbox{$[\mathcal{H}_{\text{SD}}, \mathcal{U}_{\text{KT}}] = 0$.}
It then follows from 
\Eq{UKT_qubit} that within $\Hqbit$,
the following duality holds for fermions:
\begin{equation}
	  \ c_j^\dagger \ 
	\xleftrightarrow{\ \ \mathcal{U}_{\text{KT}}  \ \  }
	 (-1)^{L(L-1)/2} \  c_j^\dagger \ (-1)^F,
\end{equation}
where $F=\sum_j c^\dagger_j c_j$.
Note that $(-1)^F$ cannot be simply regarded as a phase,
because it anti-commutes with $c_j^\dagger$.

\FloatBarrier 
\section{Numerical results}

\subsection{Gaussian criticality}

\begin{figure}[hbt]
  \includegraphics[width=0.5\textwidth]{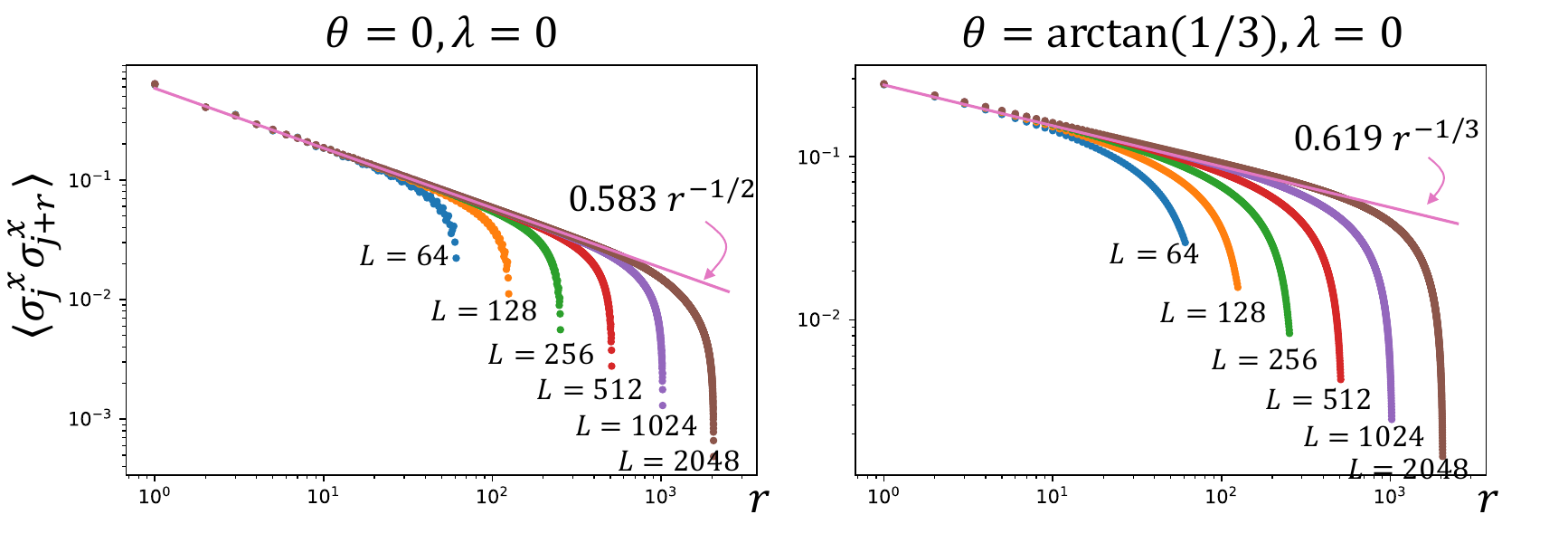}
  \caption{The correlation function $\langle \sigx_j \sigx_{j+r} \rangle$ on an open chain with length $L$ is calculated with various $L$ at $\theta=0$ (Left) and $\arctan(1/3)$ (Right). From the log-log plots, it is clear that the data are well fitted by $r^{-\eta(\theta)}$ when $1 \ll r \ll L$.} \label{gaussian}
\end{figure}

The fact that the self-dual point in the region $\theta \in \mathcal{R}$
stands for a Gaussian criticality ($c=1$ CFT) is also supported by 
our density matrix renormalization group (DMRG) calculations of the critical exponent $\eta$. 
\index{CFT!Gaussian ($c=1$)}
According to $(\XXZ, \Hqbit)$
in~\Eq{XXZXXZXXZ}
[see also the effective Hamiltonian in~\Eq{XYZ_eff}
with $\lambda =0$)],
the spin correlation function behaves as
\begin{equation}
	\langle \sigma^x_j \sigma^x_{j+r} \rangle \sim r^{-\eta(\theta)},
\end{equation}
where~\cite{PhysRevB.12.3908}
\begin{equation}
	\eta(\theta) = \frac{1}{2} - \frac{1}{\pi} \arcsin[ \Delta(\theta)].
\end{equation}
In particular, $\eta(0)=1/2$ and $\eta(\arctan\frac{1}{3})=1/3$,
which are consistent with the numerical results presented in
\Fig{gaussian}.

\subsection{Ising criticality} \label{app_num_Ising}

\begin{figure}[hbt]
  \includegraphics[width=0.5\textwidth]{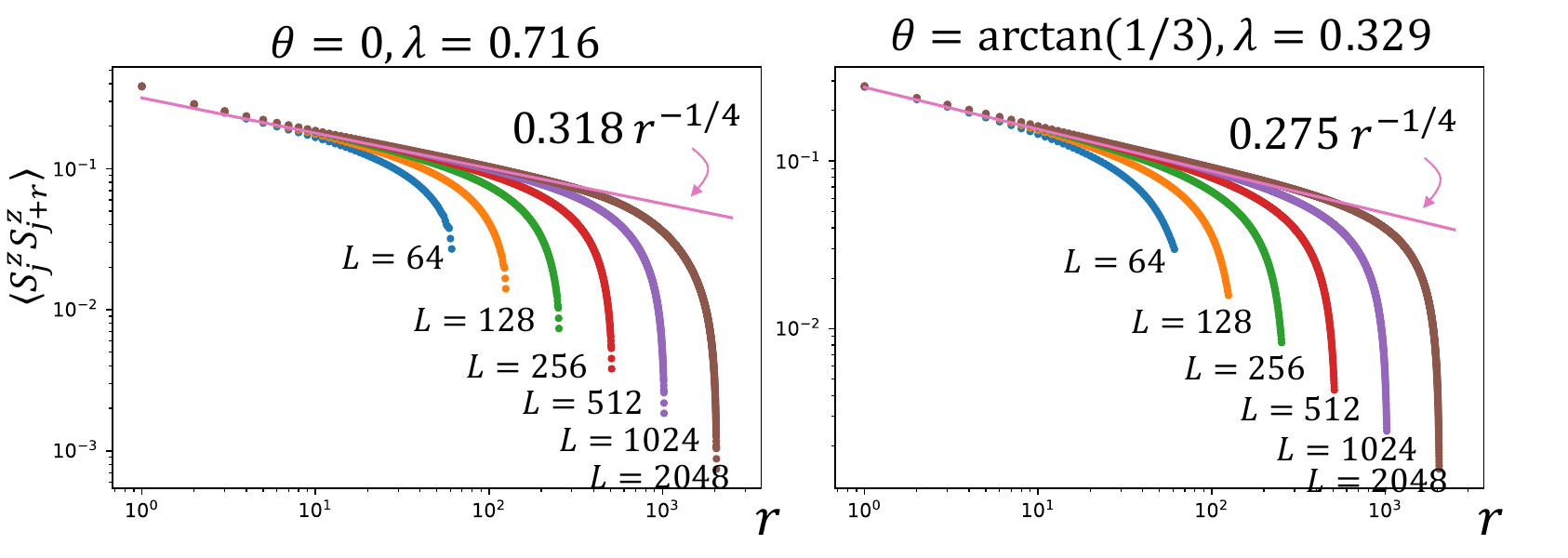}
  \caption{The correlation function $\langle \sz_j \sz_{j+r} \rangle$ on an open chain with length $L$ is calculated with various $L$ at $\theta=0$ and $\arctan(1/3)$. From the log-log plots, it is clear that the data are well fitted by $r^{-1/4}$ when $1 \ll r \ll L$.} \label{fig_lambda_c}
\end{figure}

Our DMRG results in
\Fig{fig_lambda_c} show that
$\langle \sz_j \sz_{j+r} \rangle = \langle \sx_j \sx_{j+r} \rangle \sim r^{-1/4} $
at the critical point $\lambda_c>0$,
which indeed suggests the Ising universality class~\cite{francesco2012conformal}.
The fact that $\mathcal{O}_{\text{FM}}^{x'}>0$ at $-\lambda_c$
and $\mathcal{O}_{\text{str}}^{y}>0$ at $\lambda_c$ is also 
supported by DMRG calculations;
see~\Fig{phase_diagram}(b).

\subsection{Direct transition between the SPT phase and the
$\ZZ$ SSB phase}

\begin{figure}[hbt]
\centering
  \includegraphics[width=0.5\textwidth]{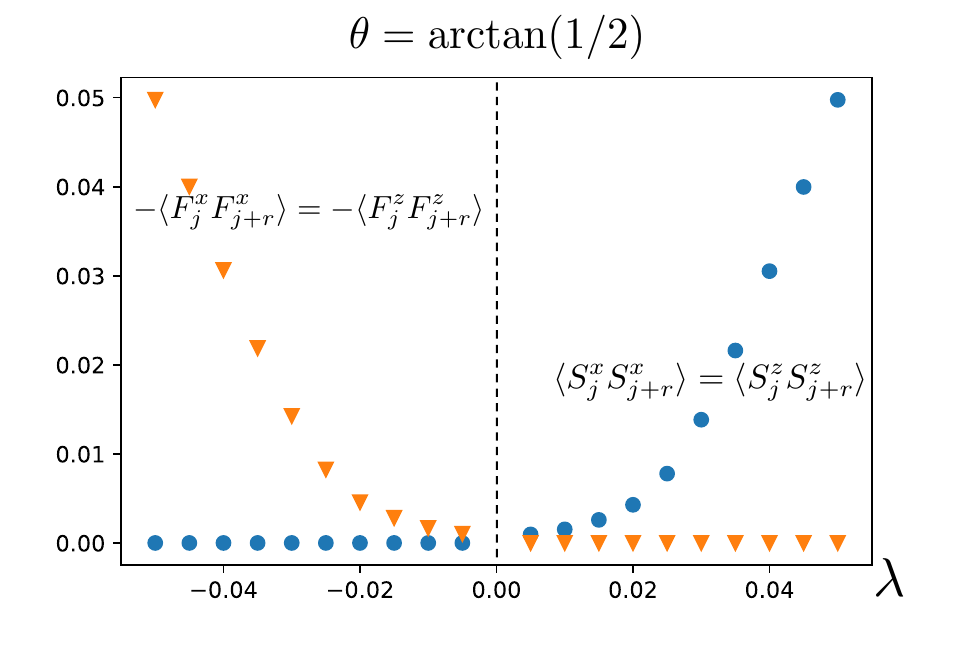}
  \caption{DMRG calculations at $\theta=\arctan(1/2)$ and $-0.05 \leqslant \lambda \leqslant 0.05$. The order parameters $\mathcal{O}^{x,z}_{\text{str}}$
and $\mathcal{O}^{x,z}_{\text{FM}}$ are estimated by taking $L=1024$ and $r=512$.} \label{fig_direct_tran}
\end{figure}

To numerically show that there is indeed a direct transition between the Haldane phase and the
$\ZZ$ SSB phase at
$(\lambda,\theta)=(0,\arctan\frac{1}{2})$,
we estimate $\mathcal{O}^{x,z}_{\text{str}}$
and $\mathcal{O}^{x,z}_{\text{FM}}$
around that point; see~\Fig{fig_direct_tran}.
Compared to \Fig{phase_diagram}(b),
the result in~\Fig{fig_direct_tran} suggests that the two $\mathbb{Z}_2$
SSB phases vanish at $\theta=\arctan(1/2)$.

\section{Hidden symmetry breaking}
\label{hidd_symm_brk}

The KT transformation $\KT$ provides a 
hidden symmetry breaking interpretation of gapped SPT phases
not only because it defines an SPT-SSB duality,
but also because the KT dual of a trivially gapped phase is again trivial,
meaning that a trivially gapped phase has no hidden symmetry breaking
[see~\Eq{hsb_gapped}],
\begin{equation}
\begin{split}
	\z \times \z \text{ gapped SPT} \ &\xleftrightarrow{\ \ U_{\text{KT}}  \ \  } \ \z \times \z \text{ SSB},\\
	\z \times \z \text{ gapped trivial} \ &\xleftrightarrow{\ \ U_{\text{KT}}  \ \  } \ \z \times \z \text{ gapped trivial}.
\end{split} \label{hsb_gapped}
\end{equation}

\begin{figure}[hbt]
\centering
  \includegraphics[width=0.45\textwidth]{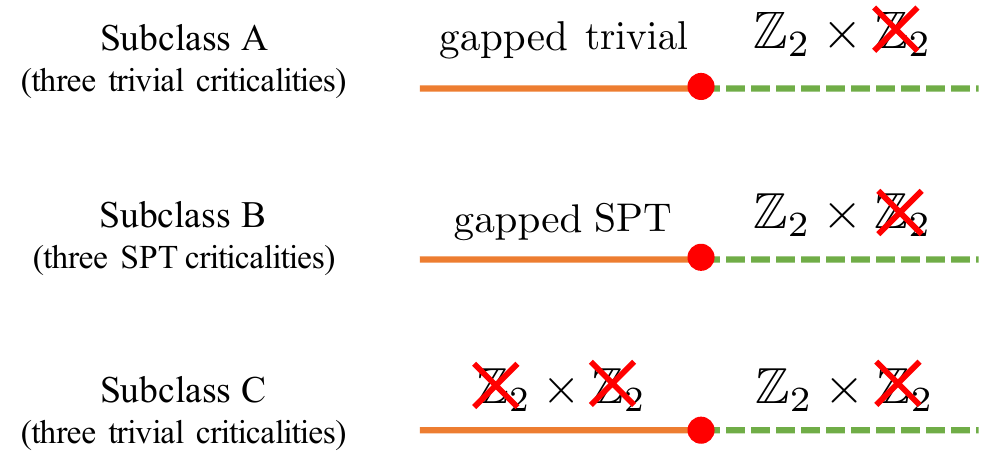}
  \caption{With $\z \times \z$ symmetry, the Ising universality class splits into three subclasses, and each subclass contains three symmetry-enriched criticalities~\cite{PhysRevX.11.041059}. Subclasses A and C are trivial, while subclass B is SPT.} \label{IsingCFTclass}
\end{figure}

Can $\KT$ also provide a 
hidden symmetry breaking interpretation of SPT Ising critical phases?
To answer this question, let us first briefly review
the classification of Ising criticalities with $\z \times \z$ symmetry.
There are nine different $\z \times \z$ Ising criticalities, which can be
divided into three subclasses A, B, and C~\cite{PhysRevX.11.041059}; see~\Fig{IsingCFTclass}.
Subclass A contains the criticalities between the trivially gapped phase and
a $\z$ SSB phase.
There are three $\z$ SSB phases represented by
\begin{equation}
H_x= \sum_j \sx_j \sx_{j+1}, \ H_y=\sum_j \sy_j \sy_{j+1},  \
H_z= \sum_j \sz_j \sz_{j+1}.
\end{equation}
Subclass B contains the criticalities between the gapped SPT phase and
a $\z$ SSB phase. 
Subclass C contains the criticalities between the fully $\z \times \z$ symmetry breaking
phase and
a $\z$ SSB phase.
The critical line $-\lambda_c(\theta)<0$ 
in~\Fig{phase_diagram}(a) belongs to subclass B,
while the critical line $\lambda_c(\theta)>0$ 
belongs to subclass C.

An example of subclass A is given by~\cite{PhysRevB.67.172402, PhysRevLett.100.067203}
\begin{equation}
	H_A = \sum_j \bigg[ (\sz_j)^2 + a \ \sx_j \sx_{j+1} \bigg]
\end{equation}
with $a=\pm 2$. 
According to~\Eq{KT_transf_integer_spin},
we have
\begin{equation}
	H_A \ \xleftrightarrow{\ \ U_{\text{KT}}  \ \  } \ 
	\widetilde{H}_A = 
	\sum_j \bigg[ (\sz_j)^2 - a \ \sx_j \sx_{j+1} \bigg],
\end{equation}
where $\widetilde{H}_A$ still belongs to subclass A.
It is thus clear that,
as summarized in~\Eq{hsb_gapless}, the
$\z \times \z$ SPT Ising criticalities (subclass B)
have hidden symmetry breaking, 
while subclass A has no hidden symmetry breaking.
\begin{equation}
\begin{split}
	\text{Subclass B} \ &\xleftrightarrow{\ \ U_{\text{KT}}  \ \  } \  \text{Subclass C},\\
	\text{Subclass A} \ &\xleftrightarrow{\ \ U_{\text{KT}}  \ \  } \  \text{Subclass A}.
\end{split} \label{hsb_gapless}
\end{equation}

\section{KT transformation, Kramers-Wannier duality, and domain wall decoration}
\label{KT_KW_DW}

\begin{figure}[hbt]
  \includegraphics[width=0.4\textwidth]{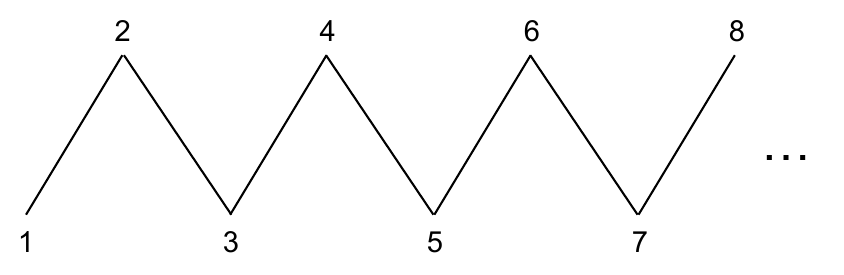}
  \caption{Zigzag chain with a spin-1/2 on each vertex.} \label{zigzag}
\end{figure}

In this section, we show that the KT transformation,
the Kramers-Wannier (KW) duality, and the domain wall (DW) decoration
are closely related to each other.

\begin{figure}[hbt]
  \includegraphics[width=0.5\textwidth]{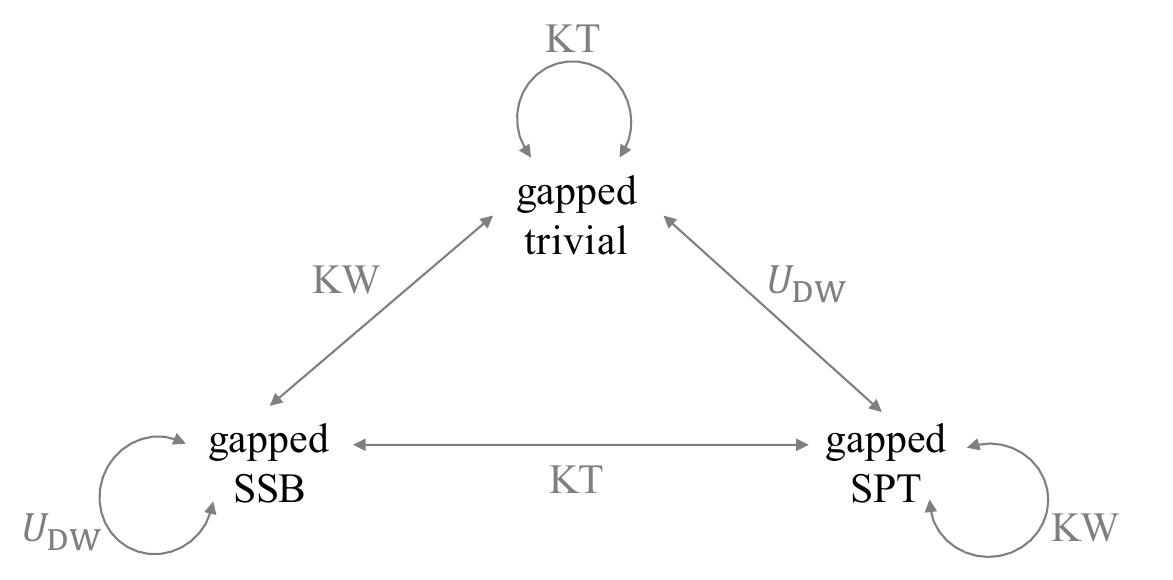}
  \caption{Three $\mathbb{Z}_2 \times \mathbb{Z}_2$ gapped phases and the dualities between them.
  KT transformation can be defined as a combination of KW duality and $U_{\text{DW}}$.
  Note that the KW duality in the figure applies to both $\mathbb{Z}_2$ symmetries.} \label{KWDWKT}
\end{figure}

Let us consider spin-1/2 models 
defined on a zigzag chain, see~\Fig{zigzag}. 
We require the models to have $\mathbb{Z}_2 \times \mathbb{Z}_2$ symmetry,
where
\begin{equation}
\begin{split}
	\mathbb{Z}_2 \times \mathbb{Z}_2 
	= \{ 1, A, B, AB \},\\
	A = \prod_{k=1}^\infty \sigx_{2k-1},\ 
	B = \prod_{k=1}^\infty \sigx_{2k}.
\end{split}
\end{equation}
On the zigzag chain, the trivially gapped phase is represented by the Hamiltonian
\begin{equation}
	H_{\text{triv}} = - \sum_{j} \sigma^x_j.
\end{equation}
The gapped SPT phase is represented by the cluster model~\cite{zeng2019quantum,Son_2011, PhysRevB.96.165124}
\begin{equation}
	H_{\text{SPT}} = - \sum_{j} \sigz_{j-1} \sigx_{j} \sigz_{j+1}.
\end{equation}
The fully $\z \times \z $ symmetry breaking phase is represented by
\begin{equation}
	H_{\text{SSB}} = - \sum_{k} \left( \sigz_{2k-1}\sigz_{2k+1} + \sigz_{2k}\sigz_{2k+2} \right).
\end{equation}
It is well known that $H_{\text{triv}}$ and $H_{\text{SSB}}$
can be transformed into each other by applying the KW duality
to both $\z$ symmetries~\cite{PhysRev.60.252, Aasen_2016}.
In fact, $H_{\text{triv}}$ can also be transformed into
$H_{\text{SPT}}$
via the so-called DW decoration~\cite{tantivasadakarn2021pivot, chen2014symmetry,wang2021domain}
\begin{equation}
	U_{\text{DW}}^\dagger H_{\text{triv}} U_{\text{DW}}
	= H_{\text{SPT}},
\end{equation}
where
\begin{equation}
	U_{\text{DW}} = \exp \bigg[  \frac{i\pi}{4} \sum_{j=2}^\infty (-1)^j \sigz_{j-1} \sigz_{j} \bigg].
\end{equation}
KT transformation is a duality between the SPT and SSB phases.
One may realize that by properly combining 
KW duality and DW decoration,
the KT transformation can be defined.
Indeed, if we define~\cite{LLi1, LLi2}
\begin{equation}
	\text{KT} = \text{KW} \times U_{\text{DW}} \times \text{KW},
	\label{eq_KT_spin-1/2}
\end{equation}
then KT in~\Eq{eq_KT_spin-1/2} gives the correct 
SPT-SSB and trivial-trivial mappings.
The relations between KT, KW, and DW are summarized in~\Fig{KWDWKT}.
For more details about~\Eq{eq_KT_spin-1/2}, see
Refs.~\cite{LLi1, LLi2}.

\bibliography{Ref_selfdual}

\begin{thebibliography}{88}%
\makeatletter
\providecommand \@ifxundefined [1]{%
 \@ifx{#1\undefined}
}%
\providecommand \@ifnum [1]{%
 \ifnum #1\expandafter \@firstoftwo
 \else \expandafter \@secondoftwo
 \fi
}%
\providecommand \@ifx [1]{%
 \ifx #1\expandafter \@firstoftwo
 \else \expandafter \@secondoftwo
 \fi
}%
\providecommand \natexlab [1]{#1}%
\providecommand \enquote  [1]{``#1''}%
\providecommand \bibnamefont  [1]{#1}%
\providecommand \bibfnamefont [1]{#1}%
\providecommand \citenamefont [1]{#1}%
\providecommand \href@noop [0]{\@secondoftwo}%
\providecommand \href [0]{\begingroup \@sanitize@url \@href}%
\providecommand \@href[1]{\@@startlink{#1}\@@href}%
\providecommand \@@href[1]{\endgroup#1\@@endlink}%
\providecommand \@sanitize@url [0]{\catcode `\\12\catcode `\$12\catcode
  `\&12\catcode `\#12\catcode `\^12\catcode `\_12\catcode `\%12\relax}%
\providecommand \@@startlink[1]{}%
\providecommand \@@endlink[0]{}%
\providecommand \url  [0]{\begingroup\@sanitize@url \@url }%
\providecommand \@url [1]{\endgroup\@href {#1}{\urlprefix }}%
\providecommand \urlprefix  [0]{URL }%
\providecommand \Eprint [0]{\href }%
\providecommand \doibase [0]{https://doi.org/}%
\providecommand \selectlanguage [0]{\@gobble}%
\providecommand \bibinfo  [0]{\@secondoftwo}%
\providecommand \bibfield  [0]{\@secondoftwo}%
\providecommand \translation [1]{[#1]}%
\providecommand \BibitemOpen [0]{}%
\providecommand \bibitemStop [0]{}%
\providecommand \bibitemNoStop [0]{.\EOS\space}%
\providecommand \EOS [0]{\spacefactor3000\relax}%
\providecommand \BibitemShut  [1]{\csname bibitem#1\endcsname}%
\let\auto@bib@innerbib\@empty
%</preamble>
\bibitem [{\citenamefont {Pollmann}\ \emph {et~al.}(2010)\citenamefont
  {Pollmann}, \citenamefont {Turner}, \citenamefont {Berg},\ and\ \citenamefont
  {Oshikawa}}]{PTBO_2010}%
  \BibitemOpen
  \bibfield  {author} {\bibinfo {author} {\bibfnamefont {F.}~\bibnamefont
  {Pollmann}}, \bibinfo {author} {\bibfnamefont {A.~M.}\ \bibnamefont
  {Turner}}, \bibinfo {author} {\bibfnamefont {E.}~\bibnamefont {Berg}},\ and\
  \bibinfo {author} {\bibfnamefont {M.}~\bibnamefont {Oshikawa}},\ }\bibfield
  {title} {\bibinfo {title} {Entanglement spectrum of a topological phase in
  one dimension},\ }\href {https://doi.org/10.1103/PhysRevB.81.064439}
  {\bibfield  {journal} {\bibinfo  {journal} {Phys. Rev. B}\ }\textbf {\bibinfo
  {volume} {81}},\ \bibinfo {pages} {064439} (\bibinfo {year}
  {2010})}\BibitemShut {NoStop}%
\bibitem [{\citenamefont {Pollmann}\ \emph {et~al.}(2012)\citenamefont
  {Pollmann}, \citenamefont {Berg}, \citenamefont {Turner},\ and\ \citenamefont
  {Oshikawa}}]{PTBO_2012}%
  \BibitemOpen
  \bibfield  {author} {\bibinfo {author} {\bibfnamefont {F.}~\bibnamefont
  {Pollmann}}, \bibinfo {author} {\bibfnamefont {E.}~\bibnamefont {Berg}},
  \bibinfo {author} {\bibfnamefont {A.~M.}\ \bibnamefont {Turner}},\ and\
  \bibinfo {author} {\bibfnamefont {M.}~\bibnamefont {Oshikawa}},\ }\bibfield
  {title} {\bibinfo {title} {Symmetry protection of topological phases in
  one-dimensional quantum spin systems},\ }\href
  {https://doi.org/10.1103/PhysRevB.85.075125} {\bibfield  {journal} {\bibinfo
  {journal} {Phys. Rev. B}\ }\textbf {\bibinfo {volume} {85}},\ \bibinfo
  {pages} {075125} (\bibinfo {year} {2012})}\BibitemShut {NoStop}%
\bibitem [{\citenamefont {Gu}\ and\ \citenamefont
  {Wen}(2009)}]{PhysRevB.80.155131}%
  \BibitemOpen
  \bibfield  {author} {\bibinfo {author} {\bibfnamefont {Z.-C.}\ \bibnamefont
  {Gu}}\ and\ \bibinfo {author} {\bibfnamefont {X.-G.}\ \bibnamefont {Wen}},\
  }\bibfield  {title} {\bibinfo {title} {Tensor-entanglement-filtering
  renormalization approach and symmetry-protected topological order},\ }\href
  {https://doi.org/10.1103/PhysRevB.80.155131} {\bibfield  {journal} {\bibinfo
  {journal} {Phys. Rev. B}\ }\textbf {\bibinfo {volume} {80}},\ \bibinfo
  {pages} {155131} (\bibinfo {year} {2009})}\BibitemShut {NoStop}%
\bibitem [{\citenamefont {Chen}\ \emph {et~al.}(2012)\citenamefont {Chen},
  \citenamefont {Gu}, \citenamefont {Liu},\ and\ \citenamefont
  {Wen}}]{Chen1604}%
  \BibitemOpen
  \bibfield  {author} {\bibinfo {author} {\bibfnamefont {X.}~\bibnamefont
  {Chen}}, \bibinfo {author} {\bibfnamefont {Z.-C.}\ \bibnamefont {Gu}},
  \bibinfo {author} {\bibfnamefont {Z.-X.}\ \bibnamefont {Liu}},\ and\ \bibinfo
  {author} {\bibfnamefont {X.-G.}\ \bibnamefont {Wen}},\ }\bibfield  {title}
  {\bibinfo {title} {Symmetry-protected topological orders in interacting
  bosonic systems},\ }\href {https://doi.org/10.1126/science.1227224}
  {\bibfield  {journal} {\bibinfo  {journal} {Science}\ }\textbf {\bibinfo
  {volume} {338}},\ \bibinfo {pages} {1604} (\bibinfo {year}
  {2012})}\BibitemShut {NoStop}%
\bibitem [{\citenamefont {Chen}\ \emph {et~al.}(2011)\citenamefont {Chen},
  \citenamefont {Gu},\ and\ \citenamefont {Wen}}]{PhysRevB.83.035107}%
  \BibitemOpen
  \bibfield  {author} {\bibinfo {author} {\bibfnamefont {X.}~\bibnamefont
  {Chen}}, \bibinfo {author} {\bibfnamefont {Z.-C.}\ \bibnamefont {Gu}},\ and\
  \bibinfo {author} {\bibfnamefont {X.-G.}\ \bibnamefont {Wen}},\ }\bibfield
  {title} {\bibinfo {title} {Classification of gapped symmetric phases in
  one-dimensional spin systems},\ }\href
  {https://doi.org/10.1103/PhysRevB.83.035107} {\bibfield  {journal} {\bibinfo
  {journal} {Phys. Rev. B}\ }\textbf {\bibinfo {volume} {83}},\ \bibinfo
  {pages} {035107} (\bibinfo {year} {2011})}\BibitemShut {NoStop}%
\bibitem [{\citenamefont {Chen}\ \emph {et~al.}(2013)\citenamefont {Chen},
  \citenamefont {Gu}, \citenamefont {Liu},\ and\ \citenamefont
  {Wen}}]{PhysRevB.87.155114}%
  \BibitemOpen
  \bibfield  {author} {\bibinfo {author} {\bibfnamefont {X.}~\bibnamefont
  {Chen}}, \bibinfo {author} {\bibfnamefont {Z.-C.}\ \bibnamefont {Gu}},
  \bibinfo {author} {\bibfnamefont {Z.-X.}\ \bibnamefont {Liu}},\ and\ \bibinfo
  {author} {\bibfnamefont {X.-G.}\ \bibnamefont {Wen}},\ }\bibfield  {title}
  {\bibinfo {title} {Symmetry protected topological orders and the group
  cohomology of their symmetry group},\ }\href
  {https://doi.org/10.1103/PhysRevB.87.155114} {\bibfield  {journal} {\bibinfo
  {journal} {Phys. Rev. B}\ }\textbf {\bibinfo {volume} {87}},\ \bibinfo
  {pages} {155114} (\bibinfo {year} {2013})}\BibitemShut {NoStop}%
\bibitem [{\citenamefont {Zeng}\ \emph {et~al.}(2019)\citenamefont {Zeng},
  \citenamefont {Chen}, \citenamefont {Zhou},\ and\ \citenamefont
  {Wen}}]{zeng2019quantum}%
  \BibitemOpen
  \bibfield  {author} {\bibinfo {author} {\bibfnamefont {B.}~\bibnamefont
  {Zeng}}, \bibinfo {author} {\bibfnamefont {X.}~\bibnamefont {Chen}}, \bibinfo
  {author} {\bibfnamefont {D.-L.}\ \bibnamefont {Zhou}},\ and\ \bibinfo
  {author} {\bibfnamefont {X.-G.}\ \bibnamefont {Wen}},\ }\href@noop {} {\emph
  {\bibinfo {title} {{Quantum Information Meets Quantum Matter}}}}\ (\bibinfo
  {publisher} {Springer, New York},\ \bibinfo {year} {2019})\BibitemShut
  {NoStop}%
\bibitem [{\citenamefont {Else}\ \emph {et~al.}(2013)\citenamefont {Else},
  \citenamefont {Bartlett},\ and\ \citenamefont
  {Doherty}}]{PhysRevB.88.085114}%
  \BibitemOpen
  \bibfield  {author} {\bibinfo {author} {\bibfnamefont {D.~V.}\ \bibnamefont
  {Else}}, \bibinfo {author} {\bibfnamefont {S.~D.}\ \bibnamefont {Bartlett}},\
  and\ \bibinfo {author} {\bibfnamefont {A.~C.}\ \bibnamefont {Doherty}},\
  }\bibfield  {title} {\bibinfo {title} {Hidden symmetry-breaking picture of
  symmetry-protected topological order},\ }\href
  {https://doi.org/10.1103/PhysRevB.88.085114} {\bibfield  {journal} {\bibinfo
  {journal} {Phys. Rev. B}\ }\textbf {\bibinfo {volume} {88}},\ \bibinfo
  {pages} {085114} (\bibinfo {year} {2013})}\BibitemShut {NoStop}%
\bibitem [{\citenamefont {Kennedy}\ and\ \citenamefont
  {Tasaki}(1992{\natexlab{a}})}]{Kennedy_Tasaki}%
  \BibitemOpen
  \bibfield  {author} {\bibinfo {author} {\bibfnamefont {T.}~\bibnamefont
  {Kennedy}}\ and\ \bibinfo {author} {\bibfnamefont {H.}~\bibnamefont
  {Tasaki}},\ }\bibfield  {title} {\bibinfo {title} {Hidden
  ${{Z}}_{2}$\ifmmode\times\else\texttimes\fi{}${{Z}}_{2}$ symmetry breaking in
  {Haldane}-gap antiferromagnets},\ }\href
  {https://doi.org/10.1103/PhysRevB.45.304} {\bibfield  {journal} {\bibinfo
  {journal} {Phys. Rev. B}\ }\textbf {\bibinfo {volume} {45}},\ \bibinfo
  {pages} {304} (\bibinfo {year} {1992}{\natexlab{a}})}\BibitemShut {NoStop}%
\bibitem [{\citenamefont {Kennedy}\ and\ \citenamefont
  {Tasaki}(1992{\natexlab{b}})}]{kennedy1992hidden}%
  \BibitemOpen
  \bibfield  {author} {\bibinfo {author} {\bibfnamefont {T.}~\bibnamefont
  {Kennedy}}\ and\ \bibinfo {author} {\bibfnamefont {H.}~\bibnamefont
  {Tasaki}},\ }\bibfield  {title} {\bibinfo {title} {Hidden symmetry breaking
  and the {Haldane} phase in {$S=1$} quantum spin chains},\ }\href
  {https://doi.org/https://doi.org/10.1007/BF02097239} {\bibfield  {journal}
  {\bibinfo  {journal} {Comm. Math. Phys.}\ }\textbf {\bibinfo {volume}
  {147}},\ \bibinfo {pages} {431} (\bibinfo {year}
  {1992}{\natexlab{b}})}\BibitemShut {NoStop}%
\bibitem [{\citenamefont {Oshikawa}(1992)}]{Oshikawa_1992}%
  \BibitemOpen
  \bibfield  {author} {\bibinfo {author} {\bibfnamefont {M.}~\bibnamefont
  {Oshikawa}},\ }\bibfield  {title} {\bibinfo {title} {Hidden {$Z_2 \times
  Z_2$} symmetry in quantum spin chains with arbitrary integer spin},\ }\href
  {https://doi.org/10.1088/0953-8984/4/36/019} {\bibfield  {journal} {\bibinfo
  {journal} {J. Phys. Condens. Matter}\ }\textbf {\bibinfo {volume} {4}},\
  \bibinfo {pages} {7469} (\bibinfo {year} {1992})}\BibitemShut {NoStop}%
\bibitem [{\citenamefont {Verresen}\ \emph {et~al.}(2021)\citenamefont
  {Verresen}, \citenamefont {Thorngren}, \citenamefont {Jones},\ and\
  \citenamefont {Pollmann}}]{PhysRevX.11.041059}%
  \BibitemOpen
  \bibfield  {author} {\bibinfo {author} {\bibfnamefont {R.}~\bibnamefont
  {Verresen}}, \bibinfo {author} {\bibfnamefont {R.}~\bibnamefont {Thorngren}},
  \bibinfo {author} {\bibfnamefont {N.~G.}\ \bibnamefont {Jones}},\ and\
  \bibinfo {author} {\bibfnamefont {F.}~\bibnamefont {Pollmann}},\ }\bibfield
  {title} {\bibinfo {title} {Gapless topological phases and symmetry-enriched
  quantum criticality},\ }\href {https://doi.org/10.1103/PhysRevX.11.041059}
  {\bibfield  {journal} {\bibinfo  {journal} {Phys. Rev. X}\ }\textbf {\bibinfo
  {volume} {11}},\ \bibinfo {pages} {041059} (\bibinfo {year}
  {2021})}\BibitemShut {NoStop}%
\bibitem [{\citenamefont {Scaffidi}\ \emph {et~al.}(2017)\citenamefont
  {Scaffidi}, \citenamefont {Parker},\ and\ \citenamefont
  {Vasseur}}]{PhysRevX.7.041048}%
  \BibitemOpen
  \bibfield  {author} {\bibinfo {author} {\bibfnamefont {T.}~\bibnamefont
  {Scaffidi}}, \bibinfo {author} {\bibfnamefont {D.~E.}\ \bibnamefont
  {Parker}},\ and\ \bibinfo {author} {\bibfnamefont {R.}~\bibnamefont
  {Vasseur}},\ }\bibfield  {title} {\bibinfo {title} {Gapless
  symmetry-protected topological order},\ }\href
  {https://doi.org/10.1103/PhysRevX.7.041048} {\bibfield  {journal} {\bibinfo
  {journal} {Phys. Rev. X}\ }\textbf {\bibinfo {volume} {7}},\ \bibinfo {pages}
  {041048} (\bibinfo {year} {2017})}\BibitemShut {NoStop}%
\bibitem [{\citenamefont {Verresen}\ \emph {et~al.}(2018)\citenamefont
  {Verresen}, \citenamefont {Jones},\ and\ \citenamefont
  {Pollmann}}]{PhysRevLett.120.057001}%
  \BibitemOpen
  \bibfield  {author} {\bibinfo {author} {\bibfnamefont {R.}~\bibnamefont
  {Verresen}}, \bibinfo {author} {\bibfnamefont {N.~G.}\ \bibnamefont
  {Jones}},\ and\ \bibinfo {author} {\bibfnamefont {F.}~\bibnamefont
  {Pollmann}},\ }\bibfield  {title} {\bibinfo {title} {Topology and edge modes
  in quantum critical chains},\ }\href
  {https://doi.org/10.1103/PhysRevLett.120.057001} {\bibfield  {journal}
  {\bibinfo  {journal} {Phys. Rev. Lett.}\ }\textbf {\bibinfo {volume} {120}},\
  \bibinfo {pages} {057001} (\bibinfo {year} {2018})}\BibitemShut {NoStop}%
\bibitem [{\citenamefont {Parker}\ \emph {et~al.}(2018)\citenamefont {Parker},
  \citenamefont {Scaffidi},\ and\ \citenamefont
  {Vasseur}}]{PhysRevB.97.165114}%
  \BibitemOpen
  \bibfield  {author} {\bibinfo {author} {\bibfnamefont {D.~E.}\ \bibnamefont
  {Parker}}, \bibinfo {author} {\bibfnamefont {T.}~\bibnamefont {Scaffidi}},\
  and\ \bibinfo {author} {\bibfnamefont {R.}~\bibnamefont {Vasseur}},\
  }\bibfield  {title} {\bibinfo {title} {Topological {Luttinger} liquids from
  decorated domain walls},\ }\href {https://doi.org/10.1103/PhysRevB.97.165114}
  {\bibfield  {journal} {\bibinfo  {journal} {Phys. Rev. B}\ }\textbf {\bibinfo
  {volume} {97}},\ \bibinfo {pages} {165114} (\bibinfo {year}
  {2018})}\BibitemShut {NoStop}%
\bibitem [{\citenamefont {Jones}\ and\ \citenamefont
  {Verresen}(2019)}]{jones2019asymptotic}%
  \BibitemOpen
  \bibfield  {author} {\bibinfo {author} {\bibfnamefont {N.~G.}\ \bibnamefont
  {Jones}}\ and\ \bibinfo {author} {\bibfnamefont {R.}~\bibnamefont
  {Verresen}},\ }\bibfield  {title} {\bibinfo {title} {Asymptotic correlations
  in gapped and critical topological phases of {1D} quantum systems},\ }\href
  {https://doi.org/https://doi.org/10.1007/s10955-019-02257-9} {\bibfield
  {journal} {\bibinfo  {journal} {J. Stat. Phys.}\ }\textbf {\bibinfo {volume}
  {175}},\ \bibinfo {pages} {1164} (\bibinfo {year} {2019})}\BibitemShut
  {NoStop}%
\bibitem [{\citenamefont {Duque}\ \emph {et~al.}(2021)\citenamefont {Duque},
  \citenamefont {Hu}, \citenamefont {You}, \citenamefont {Khemani},
  \citenamefont {Verresen},\ and\ \citenamefont
  {Vasseur}}]{PhysRevB.103.L100207}%
  \BibitemOpen
  \bibfield  {author} {\bibinfo {author} {\bibfnamefont {C.~M.}\ \bibnamefont
  {Duque}}, \bibinfo {author} {\bibfnamefont {H.-Y.}\ \bibnamefont {Hu}},
  \bibinfo {author} {\bibfnamefont {Y.-Z.}\ \bibnamefont {You}}, \bibinfo
  {author} {\bibfnamefont {V.}~\bibnamefont {Khemani}}, \bibinfo {author}
  {\bibfnamefont {R.}~\bibnamefont {Verresen}},\ and\ \bibinfo {author}
  {\bibfnamefont {R.}~\bibnamefont {Vasseur}},\ }\bibfield  {title} {\bibinfo
  {title} {Topological and symmetry-enriched random quantum critical points},\
  }\href {https://doi.org/10.1103/PhysRevB.103.L100207} {\bibfield  {journal}
  {\bibinfo  {journal} {Phys. Rev. B}\ }\textbf {\bibinfo {volume} {103}},\
  \bibinfo {pages} {L100207} (\bibinfo {year} {2021})}\BibitemShut {NoStop}%
\bibitem [{\citenamefont {Thorngren}\ \emph {et~al.}(2021)\citenamefont
  {Thorngren}, \citenamefont {Vishwanath},\ and\ \citenamefont
  {Verresen}}]{PhysRevB.104.075132}%
  \BibitemOpen
  \bibfield  {author} {\bibinfo {author} {\bibfnamefont {R.}~\bibnamefont
  {Thorngren}}, \bibinfo {author} {\bibfnamefont {A.}~\bibnamefont
  {Vishwanath}},\ and\ \bibinfo {author} {\bibfnamefont {R.}~\bibnamefont
  {Verresen}},\ }\bibfield  {title} {\bibinfo {title} {Intrinsically gapless
  topological phases},\ }\href {https://doi.org/10.1103/PhysRevB.104.075132}
  {\bibfield  {journal} {\bibinfo  {journal} {Phys. Rev. B}\ }\textbf {\bibinfo
  {volume} {104}},\ \bibinfo {pages} {075132} (\bibinfo {year}
  {2021})}\BibitemShut {NoStop}%
\bibitem [{\citenamefont {Lu}\ \emph {et~al.}(2021)\citenamefont {Lu},
  \citenamefont {Xu},\ and\ \citenamefont {You}}]{PhysRevB.104.205142}%
  \BibitemOpen
  \bibfield  {author} {\bibinfo {author} {\bibfnamefont {D.-C.}\ \bibnamefont
  {Lu}}, \bibinfo {author} {\bibfnamefont {C.}~\bibnamefont {Xu}},\ and\
  \bibinfo {author} {\bibfnamefont {Y.-Z.}\ \bibnamefont {You}},\ }\bibfield
  {title} {\bibinfo {title} {Self-duality protected multicriticality in
  deconfined quantum phase transitions},\ }\href
  {https://doi.org/10.1103/PhysRevB.104.205142} {\bibfield  {journal} {\bibinfo
   {journal} {Phys. Rev. B}\ }\textbf {\bibinfo {volume} {104}},\ \bibinfo
  {pages} {205142} (\bibinfo {year} {2021})}\BibitemShut {NoStop}%
\bibitem [{\citenamefont {O'Brien}\ and\ \citenamefont
  {Fendley}(2020)}]{10.21468/SciPostPhys.9.6.088}%
  \BibitemOpen
  \bibfield  {author} {\bibinfo {author} {\bibfnamefont {E.}~\bibnamefont
  {O'Brien}}\ and\ \bibinfo {author} {\bibfnamefont {P.}~\bibnamefont
  {Fendley}},\ }\bibfield  {title} {\bibinfo {title} {{Self-dual
  $S_3$-invariant quantum chains}},\ }\href
  {https://doi.org/10.21468/SciPostPhys.9.6.088} {\bibfield  {journal}
  {\bibinfo  {journal} {SciPost Phys.}\ }\textbf {\bibinfo {volume} {9}},\
  \bibinfo {pages} {88} (\bibinfo {year} {2020})}\BibitemShut {NoStop}%
\bibitem [{\citenamefont {Tantivasadakarn}\ \emph {et~al.}(2021)\citenamefont
  {Tantivasadakarn}, \citenamefont {Thorngren}, \citenamefont {Vishwanath},\
  and\ \citenamefont {Verresen}}]{tantivasadakarn2021pivot}%
  \BibitemOpen
  \bibfield  {author} {\bibinfo {author} {\bibfnamefont {N.}~\bibnamefont
  {Tantivasadakarn}}, \bibinfo {author} {\bibfnamefont {R.}~\bibnamefont
  {Thorngren}}, \bibinfo {author} {\bibfnamefont {A.}~\bibnamefont
  {Vishwanath}},\ and\ \bibinfo {author} {\bibfnamefont {R.}~\bibnamefont
  {Verresen}},\ }\bibfield  {title} {\bibinfo {title} {Pivot hamiltonians as
  generators of symmetry and entanglement},\ }\href
  {https://doi.org/10.48550/arXiv.2110.07599} {\bibfield  {journal} {\bibinfo
  {journal} {arXiv:2110.07599}\ } (\bibinfo {year} {2021})}\BibitemShut
  {NoStop}%
\bibitem [{\citenamefont {Aasen}\ \emph {et~al.}(2020)\citenamefont {Aasen},
  \citenamefont {Fendley},\ and\ \citenamefont {Mong}}]{aasen2020topological}%
  \BibitemOpen
  \bibfield  {author} {\bibinfo {author} {\bibfnamefont {D.}~\bibnamefont
  {Aasen}}, \bibinfo {author} {\bibfnamefont {P.}~\bibnamefont {Fendley}},\
  and\ \bibinfo {author} {\bibfnamefont {R.~S.}\ \bibnamefont {Mong}},\
  }\bibfield  {title} {\bibinfo {title} {Topological defects on the lattice:
  dualities and degeneracies},\ }\href
  {https://doi.org/10.48550/arXiv.2008.08598} {\bibfield  {journal} {\bibinfo
  {journal} {arXiv:2008.08598}\ } (\bibinfo {year} {2020})}\BibitemShut
  {NoStop}%
\bibitem [{\citenamefont {Bridgeman}\ and\ \citenamefont
  {Williamson}(2017)}]{PhysRevB.96.125104}%
  \BibitemOpen
  \bibfield  {author} {\bibinfo {author} {\bibfnamefont {J.~C.}\ \bibnamefont
  {Bridgeman}}\ and\ \bibinfo {author} {\bibfnamefont {D.~J.}\ \bibnamefont
  {Williamson}},\ }\bibfield  {title} {\bibinfo {title} {Anomalies and
  entanglement renormalization},\ }\href
  {https://doi.org/10.1103/PhysRevB.96.125104} {\bibfield  {journal} {\bibinfo
  {journal} {Phys. Rev. B}\ }\textbf {\bibinfo {volume} {96}},\ \bibinfo
  {pages} {125104} (\bibinfo {year} {2017})}\BibitemShut {NoStop}%
\bibitem [{\citenamefont {Kramers}\ and\ \citenamefont
  {Wannier}(1941)}]{PhysRev.60.252}%
  \BibitemOpen
  \bibfield  {author} {\bibinfo {author} {\bibfnamefont {H.~A.}\ \bibnamefont
  {Kramers}}\ and\ \bibinfo {author} {\bibfnamefont {G.~H.}\ \bibnamefont
  {Wannier}},\ }\bibfield  {title} {\bibinfo {title} {{Statistics of the
  Two-Dimensional Ferromagnet. Part I}},\ }\href
  {https://doi.org/10.1103/PhysRev.60.252} {\bibfield  {journal} {\bibinfo
  {journal} {Phys. Rev.}\ }\textbf {\bibinfo {volume} {60}},\ \bibinfo {pages}
  {252} (\bibinfo {year} {1941})}\BibitemShut {NoStop}%
\bibitem [{\citenamefont {Aasen}\ \emph {et~al.}(2016)\citenamefont {Aasen},
  \citenamefont {Mong},\ and\ \citenamefont {Fendley}}]{Aasen_2016}%
  \BibitemOpen
  \bibfield  {author} {\bibinfo {author} {\bibfnamefont {D.}~\bibnamefont
  {Aasen}}, \bibinfo {author} {\bibfnamefont {R.~S.~K.}\ \bibnamefont {Mong}},\
  and\ \bibinfo {author} {\bibfnamefont {P.}~\bibnamefont {Fendley}},\
  }\bibfield  {title} {\bibinfo {title} {Topological defects on the lattice:
  {I. The Ising model}},\ }\href
  {https://doi.org/10.1088/1751-8113/49/35/354001} {\bibfield  {journal}
  {\bibinfo  {journal} {J. Phys. A: Math. Theor.}\ }\textbf {\bibinfo {volume}
  {49}},\ \bibinfo {pages} {354001} (\bibinfo {year} {2016})}\BibitemShut
  {NoStop}%
\bibitem [{\citenamefont {Cho}\ \emph {et~al.}(2017)\citenamefont {Cho},
  \citenamefont {Hsieh},\ and\ \citenamefont {Ryu}}]{PhysRevB.96.195105}%
  \BibitemOpen
  \bibfield  {author} {\bibinfo {author} {\bibfnamefont {G.~Y.}\ \bibnamefont
  {Cho}}, \bibinfo {author} {\bibfnamefont {C.-T.}\ \bibnamefont {Hsieh}},\
  and\ \bibinfo {author} {\bibfnamefont {S.}~\bibnamefont {Ryu}},\ }\bibfield
  {title} {\bibinfo {title} {Anomaly manifestation of {Lieb-Schultz-Mattis}
  theorem and topological phases},\ }\href
  {https://doi.org/10.1103/PhysRevB.96.195105} {\bibfield  {journal} {\bibinfo
  {journal} {Phys. Rev. B}\ }\textbf {\bibinfo {volume} {96}},\ \bibinfo
  {pages} {195105} (\bibinfo {year} {2017})}\BibitemShut {NoStop}%
\bibitem [{\citenamefont {Metlitski}\ and\ \citenamefont
  {Thorngren}(2018)}]{PhysRevB.98.085140}%
  \BibitemOpen
  \bibfield  {author} {\bibinfo {author} {\bibfnamefont {M.~A.}\ \bibnamefont
  {Metlitski}}\ and\ \bibinfo {author} {\bibfnamefont {R.}~\bibnamefont
  {Thorngren}},\ }\bibfield  {title} {\bibinfo {title} {Intrinsic and emergent
  anomalies at deconfined critical points},\ }\href
  {https://doi.org/10.1103/PhysRevB.98.085140} {\bibfield  {journal} {\bibinfo
  {journal} {Phys. Rev. B}\ }\textbf {\bibinfo {volume} {98}},\ \bibinfo
  {pages} {085140} (\bibinfo {year} {2018})}\BibitemShut {NoStop}%
\bibitem [{\citenamefont {Wang}\ \emph {et~al.}(2018)\citenamefont {Wang},
  \citenamefont {Wen},\ and\ \citenamefont {Witten}}]{PhysRevX.8.031048}%
  \BibitemOpen
  \bibfield  {author} {\bibinfo {author} {\bibfnamefont {J.}~\bibnamefont
  {Wang}}, \bibinfo {author} {\bibfnamefont {X.-G.}\ \bibnamefont {Wen}},\ and\
  \bibinfo {author} {\bibfnamefont {E.}~\bibnamefont {Witten}},\ }\bibfield
  {title} {\bibinfo {title} {Symmetric gapped interfaces of {SPT} and {SET}
  states: Systematic constructions},\ }\href
  {https://doi.org/10.1103/PhysRevX.8.031048} {\bibfield  {journal} {\bibinfo
  {journal} {Phys. Rev. X}\ }\textbf {\bibinfo {volume} {8}},\ \bibinfo {pages}
  {031048} (\bibinfo {year} {2018})}\BibitemShut {NoStop}%
\bibitem [{\citenamefont {Prakash}\ \emph {et~al.}(2018)\citenamefont
  {Prakash}, \citenamefont {Wang},\ and\ \citenamefont
  {Wei}}]{PhysRevB.98.125108}%
  \BibitemOpen
  \bibfield  {author} {\bibinfo {author} {\bibfnamefont {A.}~\bibnamefont
  {Prakash}}, \bibinfo {author} {\bibfnamefont {J.}~\bibnamefont {Wang}},\ and\
  \bibinfo {author} {\bibfnamefont {T.-C.}\ \bibnamefont {Wei}},\ }\bibfield
  {title} {\bibinfo {title} {Unwinding short-range entanglement},\ }\href
  {https://doi.org/10.1103/PhysRevB.98.125108} {\bibfield  {journal} {\bibinfo
  {journal} {Phys. Rev. B}\ }\textbf {\bibinfo {volume} {98}},\ \bibinfo
  {pages} {125108} (\bibinfo {year} {2018})}\BibitemShut {NoStop}%
\bibitem [{Note1()}]{Note1}%
  \BibitemOpen
  \bibinfo {note} {The direct product group $\protect \mathbb {Z}_2^y \times
  \protect \mathbb {Z}_2^z$ is isomorphic to the dihedral group $D_2$ (the
  dihedral group of order $4$), while the semidirect product group $\protect
  \mathbb {Z}_4^y \rtimes \protect \mathbb {Z}_2^z$ is isomorphic to
  $D_4$.}\BibitemShut {Stop}%
\bibitem [{\citenamefont {den Nijs}\ and\ \citenamefont
  {Rommelse}(1989)}]{StringOrder}%
  \BibitemOpen
  \bibfield  {author} {\bibinfo {author} {\bibfnamefont {M.}~\bibnamefont {den
  Nijs}}\ and\ \bibinfo {author} {\bibfnamefont {K.}~\bibnamefont {Rommelse}},\
  }\bibfield  {title} {\bibinfo {title} {Preroughening transitions in crystal
  surfaces and valence-bond phases in quantum spin chains},\ }\href
  {https://doi.org/10.1103/PhysRevB.40.4709} {\bibfield  {journal} {\bibinfo
  {journal} {Phys. Rev. B}\ }\textbf {\bibinfo {volume} {40}},\ \bibinfo
  {pages} {4709} (\bibinfo {year} {1989})}\BibitemShut {NoStop}%
\bibitem [{\citenamefont {Pollmann}\ and\ \citenamefont
  {Turner}(2012)}]{TopoIndex_PhysRevB.86.125441}%
  \BibitemOpen
  \bibfield  {author} {\bibinfo {author} {\bibfnamefont {F.}~\bibnamefont
  {Pollmann}}\ and\ \bibinfo {author} {\bibfnamefont {A.~M.}\ \bibnamefont
  {Turner}},\ }\bibfield  {title} {\bibinfo {title} {Detection of
  symmetry-protected topological phases in one dimension},\ }\href
  {https://doi.org/10.1103/PhysRevB.86.125441} {\bibfield  {journal} {\bibinfo
  {journal} {Phys. Rev. B}\ }\textbf {\bibinfo {volume} {86}},\ \bibinfo
  {pages} {125441} (\bibinfo {year} {2012})}\BibitemShut {NoStop}%
\bibitem [{\citenamefont {P\'erez-Garc\'{\i}a}\ \emph
  {et~al.}(2008)\citenamefont {P\'erez-Garc\'{\i}a}, \citenamefont {Wolf},
  \citenamefont {Sanz}, \citenamefont {Verstraete},\ and\ \citenamefont
  {Cirac}}]{PhysRevLett.100.167202}%
  \BibitemOpen
  \bibfield  {author} {\bibinfo {author} {\bibfnamefont {D.}~\bibnamefont
  {P\'erez-Garc\'{\i}a}}, \bibinfo {author} {\bibfnamefont {M.~M.}\
  \bibnamefont {Wolf}}, \bibinfo {author} {\bibfnamefont {M.}~\bibnamefont
  {Sanz}}, \bibinfo {author} {\bibfnamefont {F.}~\bibnamefont {Verstraete}},\
  and\ \bibinfo {author} {\bibfnamefont {J.~I.}\ \bibnamefont {Cirac}},\
  }\bibfield  {title} {\bibinfo {title} {String order and symmetries in quantum
  spin lattices},\ }\href {https://doi.org/10.1103/PhysRevLett.100.167202}
  {\bibfield  {journal} {\bibinfo  {journal} {Phys. Rev. Lett.}\ }\textbf
  {\bibinfo {volume} {100}},\ \bibinfo {pages} {167202} (\bibinfo {year}
  {2008})}\BibitemShut {NoStop}%
\bibitem [{\citenamefont {Okunishi}(2011)}]{PhysRevB.83.104411}%
  \BibitemOpen
  \bibfield  {author} {\bibinfo {author} {\bibfnamefont {K.}~\bibnamefont
  {Okunishi}},\ }\bibfield  {title} {\bibinfo {title} {Topological disentangler
  for the valence-bond-solid chain},\ }\href
  {https://doi.org/10.1103/PhysRevB.83.104411} {\bibfield  {journal} {\bibinfo
  {journal} {Phys. Rev. B}\ }\textbf {\bibinfo {volume} {83}},\ \bibinfo
  {pages} {104411} (\bibinfo {year} {2011})}\BibitemShut {NoStop}%
\bibitem [{\citenamefont {Tasaki}(2020)}]{Tasaki2020}%
  \BibitemOpen
  \bibfield  {author} {\bibinfo {author} {\bibfnamefont {H.}~\bibnamefont
  {Tasaki}},\ }\href@noop {} {\emph {\bibinfo {title} {{Physics and Mathematics
  of Quantum Many-Body Systems}}}}\ (\bibinfo  {publisher} {Springer},\
  \bibinfo {address} {New York},\ \bibinfo {year} {2020})\BibitemShut {NoStop}%
\bibitem [{Note2()}]{Note2}%
  \BibitemOpen
  \bibinfo {note} {\label {footnote0}When $S=1$, we have $\protect \qopname
  \relax o{exp}(i\pi {S}^x_{j+1}) {S}^x_{j+1} = - {S}^x_{j+1}$ and $\protect
  \qopname \relax o{exp}(i\pi {S}^z_{j}) {S}^z_j = -{S}^z_j$.}\BibitemShut
  {Stop}%
\bibitem [{\citenamefont {Imambekov}\ \emph {et~al.}(2003)\citenamefont
  {Imambekov}, \citenamefont {Lukin},\ and\ \citenamefont
  {Demler}}]{PhysRevA.68.063602}%
  \BibitemOpen
  \bibfield  {author} {\bibinfo {author} {\bibfnamefont {A.}~\bibnamefont
  {Imambekov}}, \bibinfo {author} {\bibfnamefont {M.}~\bibnamefont {Lukin}},\
  and\ \bibinfo {author} {\bibfnamefont {E.}~\bibnamefont {Demler}},\
  }\bibfield  {title} {\bibinfo {title} {Spin-exchange interactions of spin-one
  bosons in optical lattices: Singlet, nematic, and dimerized phases},\ }\href
  {https://doi.org/10.1103/PhysRevA.68.063602} {\bibfield  {journal} {\bibinfo
  {journal} {Phys. Rev. A}\ }\textbf {\bibinfo {volume} {68}},\ \bibinfo
  {pages} {063602} (\bibinfo {year} {2003})}\BibitemShut {NoStop}%
\bibitem [{\citenamefont {L\"auchli}\ \emph {et~al.}(2006)\citenamefont
  {L\"auchli}, \citenamefont {Schmid},\ and\ \citenamefont
  {Trebst}}]{BLBQ_Lauchli}%
  \BibitemOpen
  \bibfield  {author} {\bibinfo {author} {\bibfnamefont {A.}~\bibnamefont
  {L\"auchli}}, \bibinfo {author} {\bibfnamefont {G.}~\bibnamefont {Schmid}},\
  and\ \bibinfo {author} {\bibfnamefont {S.}~\bibnamefont {Trebst}},\
  }\bibfield  {title} {\bibinfo {title} {Spin nematics correlations in
  bilinear-biquadratic {$S=1$} spin chains},\ }\href
  {https://doi.org/10.1103/PhysRevB.74.144426} {\bibfield  {journal} {\bibinfo
  {journal} {Phys. Rev. B}\ }\textbf {\bibinfo {volume} {74}},\ \bibinfo
  {pages} {144426} (\bibinfo {year} {2006})}\BibitemShut {NoStop}%
\bibitem [{\citenamefont {Yang}\ \emph {et~al.}(2021)\citenamefont {Yang},
  \citenamefont {Nakano},\ and\ \citenamefont
  {Katsura}}]{PhysRevResearch.3.023210}%
  \BibitemOpen
  \bibfield  {author} {\bibinfo {author} {\bibfnamefont {H.}~\bibnamefont
  {Yang}}, \bibinfo {author} {\bibfnamefont {H.}~\bibnamefont {Nakano}},\ and\
  \bibinfo {author} {\bibfnamefont {H.}~\bibnamefont {Katsura}},\ }\bibfield
  {title} {\bibinfo {title} {Symmetry-protected topological phases in spinful
  bosons with a flat band},\ }\href
  {https://doi.org/10.1103/PhysRevResearch.3.023210} {\bibfield  {journal}
  {\bibinfo  {journal} {Phys. Rev. Research}\ }\textbf {\bibinfo {volume}
  {3}},\ \bibinfo {pages} {023210} (\bibinfo {year} {2021})}\BibitemShut
  {NoStop}%
\bibitem [{\citenamefont {Affleck}\ \emph {et~al.}(1987)\citenamefont
  {Affleck}, \citenamefont {Kennedy}, \citenamefont {Lieb},\ and\ \citenamefont
  {Tasaki}}]{AKLT1987}%
  \BibitemOpen
  \bibfield  {author} {\bibinfo {author} {\bibfnamefont {I.}~\bibnamefont
  {Affleck}}, \bibinfo {author} {\bibfnamefont {T.}~\bibnamefont {Kennedy}},
  \bibinfo {author} {\bibfnamefont {E.~H.}\ \bibnamefont {Lieb}},\ and\
  \bibinfo {author} {\bibfnamefont {H.}~\bibnamefont {Tasaki}},\ }\bibfield
  {title} {\bibinfo {title} {Rigorous results on valence-bond ground states in
  antiferromagnets},\ }\href {https://doi.org/10.1103/PhysRevLett.59.799}
  {\bibfield  {journal} {\bibinfo  {journal} {Phys. Rev. Lett.}\ }\textbf
  {\bibinfo {volume} {59}},\ \bibinfo {pages} {799} (\bibinfo {year}
  {1987})}\BibitemShut {NoStop}%
\bibitem [{\citenamefont {Affleck}\ \emph {et~al.}(1988)\citenamefont
  {Affleck}, \citenamefont {Kennedy}, \citenamefont {Lieb},\ and\ \citenamefont
  {Tasaki}}]{AKLT1988}%
  \BibitemOpen
  \bibfield  {author} {\bibinfo {author} {\bibfnamefont {I.}~\bibnamefont
  {Affleck}}, \bibinfo {author} {\bibfnamefont {T.}~\bibnamefont {Kennedy}},
  \bibinfo {author} {\bibfnamefont {E.~H.}\ \bibnamefont {Lieb}},\ and\
  \bibinfo {author} {\bibfnamefont {H.}~\bibnamefont {Tasaki}},\ }\bibfield
  {title} {\bibinfo {title} {Valence bond ground states in isotropic quantum
  antiferromagnets},\ }\href {https://doi.org/10.1007/BF01218021} {\bibfield
  {journal} {\bibinfo  {journal} {Commun. Math. Phys.}\ }\textbf {\bibinfo
  {volume} {115}},\ \bibinfo {pages} {477} (\bibinfo {year}
  {1988})}\BibitemShut {NoStop}%
\bibitem [{Note3()}]{Note3}%
  \BibitemOpen
  \bibinfo {note} {\label {footnote2} The Hamiltonian $\DOTSB \sum@ \slimits@
  _j \protect \bm {S}_j \cdot \protect \bm {S}_{j+1}$ has an on-site $\protect
  \text {SO(3)}$ symmetry whose group element $g$ looks like $g = \DOTSB \prod@
  \slimits@ _j \protect \qopname \relax o{exp}(-i \theta _x {S}^x_j - i \theta
  _y {S}^y_j - i \theta _z {S}^z_j)$. The dual Hamiltonian $ U_{\protect
  \textnormal {KT}}(\DOTSB \sum@ \slimits@ _j \protect \bm {S}_j \cdot \protect
  \bm {S}_{j+1}) U_{\protect \textnormal {KT}}$ also has an $\protect \text
  {SO(3)}$ symmetry, but elements in this $\protect \text {SO(3)}$ group take
  the form $\protect \tilde {g}= U_{\protect \textnormal {KT}}\ g \ U_{\protect
  \textnormal {KT}}$ which are \protect \textit {not} on-site in general. We
  are only interested in on-site symmetries. The on-site $\protect \mathbb
  {Z}_2^y \times \protect \mathbb {Z}_2^z$ symmetry of $U_{\protect \textnormal
  {KT}}$ guarantees the on-site $\protect \mathbb {Z}_2^y \times \protect
  \mathbb {Z}_2^z$ symmetry of $ U_{\protect \textnormal {KT}}\ (\DOTSB \sum@
  \slimits@ _j \protect \bm {S}_j \cdot \protect \bm {S}_{j+1}) \ U_{\protect
  \textnormal {KT}}$. However, the on-site $\protect \mathbb {Z}_4^y \rtimes
  \protect \mathbb {Z}_2^z$ symmetry of $ U_{\protect \textnormal {KT}}(\DOTSB
  \sum@ \slimits@ _j \protect \bm {S}_j \cdot \protect \bm {S}_{j+1})
  U_{\protect \textnormal {KT}}$ is rather a coincidence for spin-1 chains due
  to the identities in footnote~[36]. For an integer spin $S>1$, $ U_{\protect
  \textnormal {KT}}\ (\DOTSB \sum@ \slimits@ _j \protect \bm {S}_j \cdot
  \protect \bm {S}_{j+1}) \ U_{\protect \textnormal {KT}}$ in general has no
  on-site $\protect \mathbb {Z}_4^y \rtimes \protect \mathbb {Z}_2^z$
  symmetry.}\BibitemShut {Stop}%
\bibitem [{\citenamefont {Kennedy}(1994)}]{Kennedy_1994}%
  \BibitemOpen
  \bibfield  {author} {\bibinfo {author} {\bibfnamefont {T.}~\bibnamefont
  {Kennedy}},\ }\bibfield  {title} {\bibinfo {title} {Non-positive matrix
  elements for {Hamiltonians} of spin-1 chains},\ }\href
  {https://doi.org/10.1088/0953-8984/6/39/020} {\bibfield  {journal} {\bibinfo
  {journal} {J. Phys. Condens. Matter}\ }\textbf {\bibinfo {volume} {6}},\
  \bibinfo {pages} {8015} (\bibinfo {year} {1994})}\BibitemShut {NoStop}%
\bibitem [{\citenamefont {Okunishi}\ and\ \citenamefont
  {Harada}(2014)}]{PhysRevB.89.134422}%
  \BibitemOpen
  \bibfield  {author} {\bibinfo {author} {\bibfnamefont {K.}~\bibnamefont
  {Okunishi}}\ and\ \bibinfo {author} {\bibfnamefont {K.}~\bibnamefont
  {Harada}},\ }\bibfield  {title} {\bibinfo {title} {Symmetry-protected
  topological order and negative-sign problem for {$\mathrm{SO(}N)$}
  bilinear-biquadratic chains},\ }\href
  {https://doi.org/10.1103/PhysRevB.89.134422} {\bibfield  {journal} {\bibinfo
  {journal} {Phys. Rev. B}\ }\textbf {\bibinfo {volume} {89}},\ \bibinfo
  {pages} {134422} (\bibinfo {year} {2014})}\BibitemShut {NoStop}%
\bibitem [{\citenamefont {Yang}\ and\ \citenamefont
  {Yang}(1966{\natexlab{a}})}]{PhysRev.150.321}%
  \BibitemOpen
  \bibfield  {author} {\bibinfo {author} {\bibfnamefont {C.~N.}\ \bibnamefont
  {Yang}}\ and\ \bibinfo {author} {\bibfnamefont {C.~P.}\ \bibnamefont
  {Yang}},\ }\bibfield  {title} {\bibinfo {title} {{One-Dimensional Chain of
  Anisotropic Spin-Spin Interactions. I. Proof of Bethe's Hypothesis for Ground
  State in a Finite System}},\ }\href {https://doi.org/10.1103/PhysRev.150.321}
  {\bibfield  {journal} {\bibinfo  {journal} {Phys. Rev.}\ }\textbf {\bibinfo
  {volume} {150}},\ \bibinfo {pages} {321} (\bibinfo {year}
  {1966}{\natexlab{a}})}\BibitemShut {NoStop}%
\bibitem [{\citenamefont {Yang}\ and\ \citenamefont
  {Yang}(1966{\natexlab{b}})}]{PhysRev.150.327}%
  \BibitemOpen
  \bibfield  {author} {\bibinfo {author} {\bibfnamefont {C.~N.}\ \bibnamefont
  {Yang}}\ and\ \bibinfo {author} {\bibfnamefont {C.~P.}\ \bibnamefont
  {Yang}},\ }\bibfield  {title} {\bibinfo {title} {{One-Dimensional Chain of
  Anisotropic Spin-Spin Interactions. II. Properties of the Ground-State Energy
  Per Lattice Site for an Infinite System}},\ }\href
  {https://doi.org/10.1103/PhysRev.150.327} {\bibfield  {journal} {\bibinfo
  {journal} {Phys. Rev.}\ }\textbf {\bibinfo {volume} {150}},\ \bibinfo {pages}
  {327} (\bibinfo {year} {1966}{\natexlab{b}})}\BibitemShut {NoStop}%
\bibitem [{\citenamefont {Hamer}\ \emph {et~al.}(1987)\citenamefont {Hamer},
  \citenamefont {Quispel},\ and\ \citenamefont
  {Batchelor}}]{hamer1987conformal}%
  \BibitemOpen
  \bibfield  {author} {\bibinfo {author} {\bibfnamefont {C.}~\bibnamefont
  {Hamer}}, \bibinfo {author} {\bibfnamefont {G.}~\bibnamefont {Quispel}},\
  and\ \bibinfo {author} {\bibfnamefont {M.}~\bibnamefont {Batchelor}},\
  }\bibfield  {title} {\bibinfo {title} {Conformal anomaly and surface energy
  for {Potts} and {Ashkin-Teller} quantum chains},\ }\href
  {https://doi.org/https://doi.org/10.1088/0305-4470/20/16/040} {\bibfield
  {journal} {\bibinfo  {journal} {J. Phys. A: Math. Gen.}\ }\textbf {\bibinfo
  {volume} {20}},\ \bibinfo {pages} {5677} (\bibinfo {year}
  {1987})}\BibitemShut {NoStop}%
\bibitem [{\citenamefont {Lieb}\ \emph {et~al.}(1961)\citenamefont {Lieb},
  \citenamefont {Schultz},\ and\ \citenamefont {Mattis}}]{LSM1961}%
  \BibitemOpen
  \bibfield  {author} {\bibinfo {author} {\bibfnamefont {E.}~\bibnamefont
  {Lieb}}, \bibinfo {author} {\bibfnamefont {T.}~\bibnamefont {Schultz}},\ and\
  \bibinfo {author} {\bibfnamefont {D.}~\bibnamefont {Mattis}},\ }\bibfield
  {title} {\bibinfo {title} {Two soluble models of an antiferromagnetic
  chain},\ }\href
  {https://doi.org/https://doi.org/10.1016/0003-4916(61)90115-4} {\bibfield
  {journal} {\bibinfo  {journal} {Ann. Phys.}\ }\textbf {\bibinfo {volume}
  {16}},\ \bibinfo {pages} {407 } (\bibinfo {year} {1961})}\BibitemShut
  {NoStop}%
\bibitem [{\citenamefont {Affleck}\ and\ \citenamefont {Lieb}(1986)}]{LSM1986}%
  \BibitemOpen
  \bibfield  {author} {\bibinfo {author} {\bibfnamefont {I.}~\bibnamefont
  {Affleck}}\ and\ \bibinfo {author} {\bibfnamefont {E.~H.}\ \bibnamefont
  {Lieb}},\ }\bibfield  {title} {\bibinfo {title} {A proof of part of
  {Haldane's} conjecture on spin chains},\ }\href
  {https://doi.org/https://doi.org/10.1007/BF00400304} {\bibfield  {journal}
  {\bibinfo  {journal} {Lett. Math. Phys.}\ }\textbf {\bibinfo {volume} {12}},\
  \bibinfo {pages} {57} (\bibinfo {year} {1986})}\BibitemShut {NoStop}%
\bibitem [{\citenamefont {Oshikawa}(2000)}]{PhysRevLett.84.1535}%
  \BibitemOpen
  \bibfield  {author} {\bibinfo {author} {\bibfnamefont {M.}~\bibnamefont
  {Oshikawa}},\ }\bibfield  {title} {\bibinfo {title} {Commensurability,
  excitation gap, and topology in quantum many-particle systems on a periodic
  lattice},\ }\href {https://doi.org/10.1103/PhysRevLett.84.1535} {\bibfield
  {journal} {\bibinfo  {journal} {Phys. Rev. Lett.}\ }\textbf {\bibinfo
  {volume} {84}},\ \bibinfo {pages} {1535} (\bibinfo {year}
  {2000})}\BibitemShut {NoStop}%
\bibitem [{\citenamefont {Fuji}(2016)}]{PhysRevB.93.104425}%
  \BibitemOpen
  \bibfield  {author} {\bibinfo {author} {\bibfnamefont {Y.}~\bibnamefont
  {Fuji}},\ }\bibfield  {title} {\bibinfo {title} {Effective field theory for
  one-dimensional valence-bond-solid phases and their symmetry protection},\
  }\href {https://doi.org/10.1103/PhysRevB.93.104425} {\bibfield  {journal}
  {\bibinfo  {journal} {Phys. Rev. B}\ }\textbf {\bibinfo {volume} {93}},\
  \bibinfo {pages} {104425} (\bibinfo {year} {2016})}\BibitemShut {NoStop}%
\bibitem [{\citenamefont {Ogata}\ and\ \citenamefont
  {Tasaki}(2019)}]{ogata2019lieb}%
  \BibitemOpen
  \bibfield  {author} {\bibinfo {author} {\bibfnamefont {Y.}~\bibnamefont
  {Ogata}}\ and\ \bibinfo {author} {\bibfnamefont {H.}~\bibnamefont {Tasaki}},\
  }\bibfield  {title} {\bibinfo {title} {{Lieb--Schultz--Mattis} type theorems
  for quantum spin chains without continuous symmetry},\ }\href
  {https://link.springer.com/article/10.1007/s00220-019-03343-5#citeas}
  {\bibfield  {journal} {\bibinfo  {journal} {Commun. Math. Phys.}\ }\textbf
  {\bibinfo {volume} {372}},\ \bibinfo {pages} {951} (\bibinfo {year}
  {2019})}\BibitemShut {NoStop}%
\bibitem [{\citenamefont {Ogata}\ \emph {et~al.}(2021)\citenamefont {Ogata},
  \citenamefont {Tachikawa},\ and\ \citenamefont {Tasaki}}]{ogata2021general}%
  \BibitemOpen
  \bibfield  {author} {\bibinfo {author} {\bibfnamefont {Y.}~\bibnamefont
  {Ogata}}, \bibinfo {author} {\bibfnamefont {Y.}~\bibnamefont {Tachikawa}},\
  and\ \bibinfo {author} {\bibfnamefont {H.}~\bibnamefont {Tasaki}},\
  }\bibfield  {title} {\bibinfo {title} {General {Lieb--Schultz--Mattis} type
  theorems for quantum spin chains},\ }\href
  {https://doi.org/https://doi.org/10.1007/s00220-021-04116-9} {\bibfield
  {journal} {\bibinfo  {journal} {Commun. Math. Phys.}\ }\textbf {\bibinfo
  {volume} {385}},\ \bibinfo {pages} {79} (\bibinfo {year} {2021})}\BibitemShut
  {NoStop}%
\bibitem [{\citenamefont {Yao}\ and\ \citenamefont
  {Oshikawa}(2021)}]{yao2020twisted}%
  \BibitemOpen
  \bibfield  {author} {\bibinfo {author} {\bibfnamefont {Y.}~\bibnamefont
  {Yao}}\ and\ \bibinfo {author} {\bibfnamefont {M.}~\bibnamefont {Oshikawa}},\
  }\bibfield  {title} {\bibinfo {title} {Twisted boundary condition and
  {Lieb--Schultz--Mattis} ingappability for discrete symmetries},\ }\href
  {https://doi.org/10.1103/PhysRevLett.126.217201} {\bibfield  {journal}
  {\bibinfo  {journal} {Phys. Rev. Lett.}\ }\textbf {\bibinfo {volume} {126}},\
  \bibinfo {pages} {217201} (\bibinfo {year} {2021})}\BibitemShut {NoStop}%
\bibitem [{\citenamefont {Lukyanov}\ and\ \citenamefont
  {Terras}(2003)}]{LUKYANOV2003323}%
  \BibitemOpen
  \bibfield  {author} {\bibinfo {author} {\bibfnamefont {S.}~\bibnamefont
  {Lukyanov}}\ and\ \bibinfo {author} {\bibfnamefont {V.}~\bibnamefont
  {Terras}},\ }\bibfield  {title} {\bibinfo {title} {Long-distance asymptotics
  of spin–spin correlation functions for the {XXZ} spin chain},\ }\href
  {https://doi.org/https://doi.org/10.1016/S0550-3213(02)01141-0} {\bibfield
  {journal} {\bibinfo  {journal} {Nucl. Phys. B}\ }\textbf {\bibinfo {volume}
  {654}},\ \bibinfo {pages} {323} (\bibinfo {year} {2003})}\BibitemShut
  {NoStop}%
\bibitem [{\citenamefont {Baxter}\ and\ \citenamefont
  {Kelland}(1974)}]{Baxter_1974}%
  \BibitemOpen
  \bibfield  {author} {\bibinfo {author} {\bibfnamefont {R.~J.}\ \bibnamefont
  {Baxter}}\ and\ \bibinfo {author} {\bibfnamefont {S.~B.}\ \bibnamefont
  {Kelland}},\ }\bibfield  {title} {\bibinfo {title} {Spontaneous polarization
  of the eight-vertex model},\ }\href
  {https://doi.org/10.1088/0022-3719/7/22/003} {\bibfield  {journal} {\bibinfo
  {journal} {J. Phys. C: Solid State Phys.}\ }\textbf {\bibinfo {volume} {7}},\
  \bibinfo {pages} {L403} (\bibinfo {year} {1974})}\BibitemShut {NoStop}%
\bibitem [{\citenamefont {Baxter}(1971{\natexlab{a}})}]{PhysRevLett.26.832}%
  \BibitemOpen
  \bibfield  {author} {\bibinfo {author} {\bibfnamefont {R.~J.}\ \bibnamefont
  {Baxter}},\ }\bibfield  {title} {\bibinfo {title} {Eight-vertex model in
  lattice statistics},\ }\href {https://doi.org/10.1103/PhysRevLett.26.832}
  {\bibfield  {journal} {\bibinfo  {journal} {Phys. Rev. Lett.}\ }\textbf
  {\bibinfo {volume} {26}},\ \bibinfo {pages} {832} (\bibinfo {year}
  {1971}{\natexlab{a}})}\BibitemShut {NoStop}%
\bibitem [{\citenamefont {Baxter}(1971{\natexlab{b}})}]{PhysRevLett.26.834}%
  \BibitemOpen
  \bibfield  {author} {\bibinfo {author} {\bibfnamefont {R.~J.}\ \bibnamefont
  {Baxter}},\ }\bibfield  {title} {\bibinfo {title} {One-dimensional
  anisotropic {Heisenberg} chain},\ }\href
  {https://doi.org/10.1103/PhysRevLett.26.834} {\bibfield  {journal} {\bibinfo
  {journal} {Phys. Rev. Lett.}\ }\textbf {\bibinfo {volume} {26}},\ \bibinfo
  {pages} {834} (\bibinfo {year} {1971}{\natexlab{b}})}\BibitemShut {NoStop}%
\bibitem [{\citenamefont {Cheng}\ \emph {et~al.}(2016)\citenamefont {Cheng},
  \citenamefont {Zaletel}, \citenamefont {Barkeshli}, \citenamefont
  {Vishwanath},\ and\ \citenamefont {Bonderson}}]{PhysRevX.6.041068}%
  \BibitemOpen
  \bibfield  {author} {\bibinfo {author} {\bibfnamefont {M.}~\bibnamefont
  {Cheng}}, \bibinfo {author} {\bibfnamefont {M.}~\bibnamefont {Zaletel}},
  \bibinfo {author} {\bibfnamefont {M.}~\bibnamefont {Barkeshli}}, \bibinfo
  {author} {\bibfnamefont {A.}~\bibnamefont {Vishwanath}},\ and\ \bibinfo
  {author} {\bibfnamefont {P.}~\bibnamefont {Bonderson}},\ }\bibfield  {title}
  {\bibinfo {title} {Translational symmetry and microscopic constraints on
  symmetry-enriched topological phases: A view from the surface},\ }\href
  {https://doi.org/10.1103/PhysRevX.6.041068} {\bibfield  {journal} {\bibinfo
  {journal} {Phys. Rev. X}\ }\textbf {\bibinfo {volume} {6}},\ \bibinfo {pages}
  {041068} (\bibinfo {year} {2016})}\BibitemShut {NoStop}%
\bibitem [{\citenamefont {Yao}\ \emph {et~al.}(2019)\citenamefont {Yao},
  \citenamefont {Hsieh},\ and\ \citenamefont
  {Oshikawa}}]{PhysRevLett.123.180201}%
  \BibitemOpen
  \bibfield  {author} {\bibinfo {author} {\bibfnamefont {Y.}~\bibnamefont
  {Yao}}, \bibinfo {author} {\bibfnamefont {C.-T.}\ \bibnamefont {Hsieh}},\
  and\ \bibinfo {author} {\bibfnamefont {M.}~\bibnamefont {Oshikawa}},\
  }\bibfield  {title} {\bibinfo {title} {Anomaly matching and
  symmetry-protected critical phases in {$SU(N)$} spin systems in $1+1$
  dimensions},\ }\href {https://doi.org/10.1103/PhysRevLett.123.180201}
  {\bibfield  {journal} {\bibinfo  {journal} {Phys. Rev. Lett.}\ }\textbf
  {\bibinfo {volume} {123}},\ \bibinfo {pages} {180201} (\bibinfo {year}
  {2019})}\BibitemShut {NoStop}%
\bibitem [{\citenamefont {Thorngren}\ and\ \citenamefont
  {Else}(2018)}]{PhysRevX.8.011040}%
  \BibitemOpen
  \bibfield  {author} {\bibinfo {author} {\bibfnamefont {R.}~\bibnamefont
  {Thorngren}}\ and\ \bibinfo {author} {\bibfnamefont {D.~V.}\ \bibnamefont
  {Else}},\ }\bibfield  {title} {\bibinfo {title} {Gauging spatial symmetries
  and the classification of topological crystalline phases},\ }\href
  {https://doi.org/10.1103/PhysRevX.8.011040} {\bibfield  {journal} {\bibinfo
  {journal} {Phys. Rev. X}\ }\textbf {\bibinfo {volume} {8}},\ \bibinfo {pages}
  {011040} (\bibinfo {year} {2018})}\BibitemShut {NoStop}%
\bibitem [{\citenamefont {Zou}\ \emph {et~al.}(2021)\citenamefont {Zou},
  \citenamefont {He},\ and\ \citenamefont {Wang}}]{PhysRevX.11.031043}%
  \BibitemOpen
  \bibfield  {author} {\bibinfo {author} {\bibfnamefont {L.}~\bibnamefont
  {Zou}}, \bibinfo {author} {\bibfnamefont {Y.-C.}\ \bibnamefont {He}},\ and\
  \bibinfo {author} {\bibfnamefont {C.}~\bibnamefont {Wang}},\ }\bibfield
  {title} {\bibinfo {title} {Stiefel liquids: Possible non-{Lagrangian} quantum
  criticality from intertwined orders},\ }\href
  {https://doi.org/10.1103/PhysRevX.11.031043} {\bibfield  {journal} {\bibinfo
  {journal} {Phys. Rev. X}\ }\textbf {\bibinfo {volume} {11}},\ \bibinfo
  {pages} {031043} (\bibinfo {year} {2021})}\BibitemShut {NoStop}%
\bibitem [{Note4()}]{Note4}%
  \BibitemOpen
  \bibinfo {note} {In general, the partition function of an SPT phase on $M_{d}
  \times S^1$ and the partition function of its corresponding anomalous theory
  on $\partial M_{d} \times S^1$ differ by a gauge invariant term, which is not
  important.}\BibitemShut {Stop}%
\bibitem [{Note5()}]{Note5}%
  \BibitemOpen
  \bibinfo {note} {\label {footnote_cochain}Strictly speaking, the gauge field
  of a discrete group is a cochain rather than a differential form. Therefore,
  the product of such gauge fields should be the cup product $\smile $ instead
  of the wedge product $\wedge $. See Appendix~B of Ref.~\cite
  {kapustin2014coupling}, Appendix~J.4.e of Ref.~\cite {PhysRevB.87.155114},
  and Refs.~\cite {steenrod1947products, PhysRevB.95.205142} for details.
  Nevertheless, the calculus on a differential manifold and the calculus on a
  simplicial complex are almost parallel. In this article, we sacrifice the
  mathematical rigor and use differential forms and wedge products for
  simplicity.}\BibitemShut {Stop}%
\bibitem [{\citenamefont {Wang}\ \emph {et~al.}(2014)\citenamefont {Wang},
  \citenamefont {Gu},\ and\ \citenamefont {Wen}}]{PhysRevLett.114.031601}%
  \BibitemOpen
  \bibfield  {author} {\bibinfo {author} {\bibfnamefont {J.~C.}\ \bibnamefont
  {Wang}}, \bibinfo {author} {\bibfnamefont {Z.-C.}\ \bibnamefont {Gu}},\ and\
  \bibinfo {author} {\bibfnamefont {X.-G.}\ \bibnamefont {Wen}},\ }\bibfield
  {title} {\bibinfo {title} {Field-theory representation of gauge-gravity
  symmetry-protected topological invariants, group cohomology, and beyond},\
  }\href {https://arxiv.org/abs/1405.7689} {\bibfield  {journal} {\bibinfo
  {journal} {arXiv:1405.7689}\ } (\bibinfo {year} {2014})}\BibitemShut
  {NoStop}%
\bibitem [{Note6()}]{Note6}%
  \BibitemOpen
  \bibinfo {note} {See footnote [64].}\BibitemShut {Stop}%
\bibitem [{\citenamefont {Tachikawa}(2020)}]{10.21468/SciPostPhys.8.1.015}%
  \BibitemOpen
  \bibfield  {author} {\bibinfo {author} {\bibfnamefont {Y.}~\bibnamefont
  {Tachikawa}},\ }\bibfield  {title} {\bibinfo {title} {{On gauging finite
  subgroups}},\ }\href {https://doi.org/10.21468/SciPostPhys.8.1.015}
  {\bibfield  {journal} {\bibinfo  {journal} {SciPost Phys.}\ }\textbf
  {\bibinfo {volume} {8}},\ \bibinfo {pages} {15} (\bibinfo {year}
  {2020})}\BibitemShut {NoStop}%
\bibitem [{\citenamefont {Di~Francesco}\ \emph {et~al.}(2012)\citenamefont
  {Di~Francesco}, \citenamefont {Mathieu},\ and\ \citenamefont
  {S{\'e}n{\'e}chal}}]{francesco2012conformal}%
  \BibitemOpen
  \bibfield  {author} {\bibinfo {author} {\bibfnamefont {P.}~\bibnamefont
  {Di~Francesco}}, \bibinfo {author} {\bibfnamefont {P.}~\bibnamefont
  {Mathieu}},\ and\ \bibinfo {author} {\bibfnamefont {D.}~\bibnamefont
  {S{\'e}n{\'e}chal}},\ }\href@noop {} {\emph {\bibinfo {title} {{Conformal
  Field Theory}}}}\ (\bibinfo  {publisher} {Springer Science \& Business
  Media},\ \bibinfo {address} {New York},\ \bibinfo {year} {2012})\BibitemShut
  {NoStop}%
\bibitem [{Note7()}]{Note7}%
  \BibitemOpen
  \bibinfo {note} {\label {footnote2} Instead of $\protect \mathbb {Z}_4^y
  \rtimes \protect \mathbb {Z}_2^z$, it is sufficient to consider its subgroup
  $\protect \mathbb {Z}_2^y \times \protect \mathbb {Z}_2^z$, because their
  second cohomology groups are the same: $H^2[\protect \mathbb {Z}_4^y \rtimes
  \protect \mathbb {Z}_2^z,U(1)] = H^2[\protect \mathbb {Z}_2^y \times \protect
  \mathbb {Z}_2^z,U(1)] = \protect \mathbb {Z}_2$~\cite {PhysRevB.87.155114,
  PhysRevX.11.041059}.}\BibitemShut {Stop}%
\bibitem [{\citenamefont {Takhtajan}(1982)}]{TAKHTAJAN1982479}%
  \BibitemOpen
  \bibfield  {author} {\bibinfo {author} {\bibfnamefont {L.}~\bibnamefont
  {Takhtajan}},\ }\bibfield  {title} {\bibinfo {title} {The picture of
  low-lying excitations in the isotropic {Heisenberg} chain of arbitrary
  spins},\ }\href
  {https://doi.org/https://doi.org/10.1016/0375-9601(82)90764-2} {\bibfield
  {journal} {\bibinfo  {journal} {Phys. Lett. A}\ }\textbf {\bibinfo {volume}
  {87}},\ \bibinfo {pages} {479} (\bibinfo {year} {1982})}\BibitemShut
  {NoStop}%
\bibitem [{\citenamefont {Babujian}(1982)}]{BABUJIAN1982479}%
  \BibitemOpen
  \bibfield  {author} {\bibinfo {author} {\bibfnamefont {H.}~\bibnamefont
  {Babujian}},\ }\bibfield  {title} {\bibinfo {title} {Exact solution of the
  one-dimensional isotropic {Heisenberg} chain with arbitrary spins ${S}$},\
  }\href {https://doi.org/https://doi.org/10.1016/0375-9601(82)90403-0}
  {\bibfield  {journal} {\bibinfo  {journal} {Phys. Lett. A}\ }\textbf
  {\bibinfo {volume} {90}},\ \bibinfo {pages} {479} (\bibinfo {year}
  {1982})}\BibitemShut {NoStop}%
\bibitem [{\citenamefont {Kitazawa}\ and\ \citenamefont
  {Nomura}(1999)}]{PhysRevB.59.11358}%
  \BibitemOpen
  \bibfield  {author} {\bibinfo {author} {\bibfnamefont {A.}~\bibnamefont
  {Kitazawa}}\ and\ \bibinfo {author} {\bibfnamefont {K.}~\bibnamefont
  {Nomura}},\ }\bibfield  {title} {\bibinfo {title} {Bifurcation at the $c=3/2$
  {Takhtajan-Babujian} point to the $c=1$ critical line},\ }\href
  {https://doi.org/10.1103/PhysRevB.59.11358} {\bibfield  {journal} {\bibinfo
  {journal} {Phys. Rev. B}\ }\textbf {\bibinfo {volume} {59}},\ \bibinfo
  {pages} {11358} (\bibinfo {year} {1999})}\BibitemShut {NoStop}%
\bibitem [{\citenamefont {Tu}\ \emph {et~al.}(2008)\citenamefont {Tu},
  \citenamefont {Zhang},\ and\ \citenamefont {Xiang}}]{PhysRevB.78.094404}%
  \BibitemOpen
  \bibfield  {author} {\bibinfo {author} {\bibfnamefont {H.-H.}\ \bibnamefont
  {Tu}}, \bibinfo {author} {\bibfnamefont {G.-M.}\ \bibnamefont {Zhang}},\ and\
  \bibinfo {author} {\bibfnamefont {T.}~\bibnamefont {Xiang}},\ }\bibfield
  {title} {\bibinfo {title} {Class of exactly solvable {SO($n$)} symmetric spin
  chains with matrix product ground states},\ }\href
  {https://doi.org/10.1103/PhysRevB.78.094404} {\bibfield  {journal} {\bibinfo
  {journal} {Phys. Rev. B}\ }\textbf {\bibinfo {volume} {78}},\ \bibinfo
  {pages} {094404} (\bibinfo {year} {2008})}\BibitemShut {NoStop}%
\bibitem [{\citenamefont {Duivenvoorden}\ and\ \citenamefont
  {Quella}(2013)}]{PhysRevB.88.125115}%
  \BibitemOpen
  \bibfield  {author} {\bibinfo {author} {\bibfnamefont {K.}~\bibnamefont
  {Duivenvoorden}}\ and\ \bibinfo {author} {\bibfnamefont {T.}~\bibnamefont
  {Quella}},\ }\bibfield  {title} {\bibinfo {title} {From symmetry-protected
  topological order to {Landau} order},\ }\href
  {https://doi.org/10.1103/PhysRevB.88.125115} {\bibfield  {journal} {\bibinfo
  {journal} {Phys. Rev. B}\ }\textbf {\bibinfo {volume} {88}},\ \bibinfo
  {pages} {125115} (\bibinfo {year} {2013})}\BibitemShut {NoStop}%
\bibitem [{\citenamefont {Rao}\ \emph {et~al.}(2014)\citenamefont {Rao},
  \citenamefont {Zhang},\ and\ \citenamefont {Yang}}]{PhysRevB.89.125112}%
  \BibitemOpen
  \bibfield  {author} {\bibinfo {author} {\bibfnamefont {W.-J.}\ \bibnamefont
  {Rao}}, \bibinfo {author} {\bibfnamefont {G.-M.}\ \bibnamefont {Zhang}},\
  and\ \bibinfo {author} {\bibfnamefont {K.}~\bibnamefont {Yang}},\ }\bibfield
  {title} {\bibinfo {title} {Disentangling topological degeneracy in the
  entanglement spectrum of one-dimensional symmetry-protected topological
  phases},\ }\href {https://doi.org/10.1103/PhysRevB.89.125112} {\bibfield
  {journal} {\bibinfo  {journal} {Phys. Rev. B}\ }\textbf {\bibinfo {volume}
  {89}},\ \bibinfo {pages} {125112} (\bibinfo {year} {2014})}\BibitemShut
  {NoStop}%
\bibitem [{\citenamefont {Kapustin}\ and\ \citenamefont
  {Skorik}(1996)}]{Kapustin_1996}%
  \BibitemOpen
  \bibfield  {author} {\bibinfo {author} {\bibfnamefont {A.}~\bibnamefont
  {Kapustin}}\ and\ \bibinfo {author} {\bibfnamefont {S.}~\bibnamefont
  {Skorik}},\ }\bibfield  {title} {\bibinfo {title} {Surface excitations and
  surface energy of the antiferromagnetic {XXZ} chain by the {Bethe} ansatz
  approach},\ }\href {https://doi.org/10.1088/0305-4470/29/8/011} {\bibfield
  {journal} {\bibinfo  {journal} {J. Phys. A: Math. Gen.}\ }\textbf {\bibinfo
  {volume} {29}},\ \bibinfo {pages} {1629} (\bibinfo {year}
  {1996})}\BibitemShut {NoStop}%
\bibitem [{\citenamefont {Luther}\ and\ \citenamefont
  {Peschel}(1975)}]{PhysRevB.12.3908}%
  \BibitemOpen
  \bibfield  {author} {\bibinfo {author} {\bibfnamefont {A.}~\bibnamefont
  {Luther}}\ and\ \bibinfo {author} {\bibfnamefont {I.}~\bibnamefont
  {Peschel}},\ }\bibfield  {title} {\bibinfo {title} {Calculation of critical
  exponents in two dimensions from quantum field theory in one dimension},\
  }\href {https://doi.org/10.1103/PhysRevB.12.3908} {\bibfield  {journal}
  {\bibinfo  {journal} {Phys. Rev. B}\ }\textbf {\bibinfo {volume} {12}},\
  \bibinfo {pages} {3908} (\bibinfo {year} {1975})}\BibitemShut {NoStop}%
\bibitem [{\citenamefont {Oitmaa}\ and\ \citenamefont {von
  Brasch}(2003)}]{PhysRevB.67.172402}%
  \BibitemOpen
  \bibfield  {author} {\bibinfo {author} {\bibfnamefont {J.}~\bibnamefont
  {Oitmaa}}\ and\ \bibinfo {author} {\bibfnamefont {A.~M.~A.}\ \bibnamefont
  {von Brasch}},\ }\bibfield  {title} {\bibinfo {title} {Spin-1 {Ising} model
  in a transverse crystal field},\ }\href
  {https://doi.org/10.1103/PhysRevB.67.172402} {\bibfield  {journal} {\bibinfo
  {journal} {Phys. Rev. B}\ }\textbf {\bibinfo {volume} {67}},\ \bibinfo
  {pages} {172402} (\bibinfo {year} {2003})}\BibitemShut {NoStop}%
\bibitem [{\citenamefont {Yang}\ \emph {et~al.}(2008)\citenamefont {Yang},
  \citenamefont {Yang}, \citenamefont {Dai},\ and\ \citenamefont
  {Xiang}}]{PhysRevLett.100.067203}%
  \BibitemOpen
  \bibfield  {author} {\bibinfo {author} {\bibfnamefont {Z.}~\bibnamefont
  {Yang}}, \bibinfo {author} {\bibfnamefont {L.}~\bibnamefont {Yang}}, \bibinfo
  {author} {\bibfnamefont {J.}~\bibnamefont {Dai}},\ and\ \bibinfo {author}
  {\bibfnamefont {T.}~\bibnamefont {Xiang}},\ }\bibfield  {title} {\bibinfo
  {title} {Rigorous solution of the spin-1 quantum ising model with single-ion
  anisotropy},\ }\href {https://doi.org/10.1103/PhysRevLett.100.067203}
  {\bibfield  {journal} {\bibinfo  {journal} {Phys. Rev. Lett.}\ }\textbf
  {\bibinfo {volume} {100}},\ \bibinfo {pages} {067203} (\bibinfo {year}
  {2008})}\BibitemShut {NoStop}%
\bibitem [{\citenamefont {Son}\ \emph {et~al.}(2011)\citenamefont {Son},
  \citenamefont {Amico}, \citenamefont {Fazio}, \citenamefont {Hamma},
  \citenamefont {Pascazio},\ and\ \citenamefont {Vedral}}]{Son_2011}%
  \BibitemOpen
  \bibfield  {author} {\bibinfo {author} {\bibfnamefont {W.}~\bibnamefont
  {Son}}, \bibinfo {author} {\bibfnamefont {L.}~\bibnamefont {Amico}}, \bibinfo
  {author} {\bibfnamefont {R.}~\bibnamefont {Fazio}}, \bibinfo {author}
  {\bibfnamefont {A.}~\bibnamefont {Hamma}}, \bibinfo {author} {\bibfnamefont
  {S.}~\bibnamefont {Pascazio}},\ and\ \bibinfo {author} {\bibfnamefont
  {V.}~\bibnamefont {Vedral}},\ }\bibfield  {title} {\bibinfo {title} {Quantum
  phase transition between cluster and antiferromagnetic states},\ }\href
  {https://doi.org/10.1209/0295-5075/95/50001} {\bibfield  {journal} {\bibinfo
  {journal} {{EPL}}\ }\textbf {\bibinfo {volume} {95}},\ \bibinfo {pages}
  {50001} (\bibinfo {year} {2011})}\BibitemShut {NoStop}%
\bibitem [{\citenamefont {Verresen}\ \emph {et~al.}(2017)\citenamefont
  {Verresen}, \citenamefont {Moessner},\ and\ \citenamefont
  {Pollmann}}]{PhysRevB.96.165124}%
  \BibitemOpen
  \bibfield  {author} {\bibinfo {author} {\bibfnamefont {R.}~\bibnamefont
  {Verresen}}, \bibinfo {author} {\bibfnamefont {R.}~\bibnamefont {Moessner}},\
  and\ \bibinfo {author} {\bibfnamefont {F.}~\bibnamefont {Pollmann}},\
  }\bibfield  {title} {\bibinfo {title} {One-dimensional symmetry protected
  topological phases and their transitions},\ }\href
  {https://doi.org/10.1103/PhysRevB.96.165124} {\bibfield  {journal} {\bibinfo
  {journal} {Phys. Rev. B}\ }\textbf {\bibinfo {volume} {96}},\ \bibinfo
  {pages} {165124} (\bibinfo {year} {2017})}\BibitemShut {NoStop}%
\bibitem [{\citenamefont {Chen}\ \emph {et~al.}(2014)\citenamefont {Chen},
  \citenamefont {Lu},\ and\ \citenamefont {Vishwanath}}]{chen2014symmetry}%
  \BibitemOpen
  \bibfield  {author} {\bibinfo {author} {\bibfnamefont {X.}~\bibnamefont
  {Chen}}, \bibinfo {author} {\bibfnamefont {Y.-M.}\ \bibnamefont {Lu}},\ and\
  \bibinfo {author} {\bibfnamefont {A.}~\bibnamefont {Vishwanath}},\ }\bibfield
   {title} {\bibinfo {title} {Symmetry-protected topological phases from
  decorated domain walls},\ }\href
  {https://doi.org/https://doi.org/10.1038/ncomms4507} {\bibfield  {journal}
  {\bibinfo  {journal} {Nat. Commun.}\ }\textbf {\bibinfo {volume} {5}},\
  \bibinfo {pages} {1} (\bibinfo {year} {2014})}\BibitemShut {NoStop}%
\bibitem [{\citenamefont {Wang}\ \emph {et~al.}(2021)\citenamefont {Wang},
  \citenamefont {Ning},\ and\ \citenamefont {Cheng}}]{wang2021domain}%
  \BibitemOpen
  \bibfield  {author} {\bibinfo {author} {\bibfnamefont {Q.-R.}\ \bibnamefont
  {Wang}}, \bibinfo {author} {\bibfnamefont {S.-Q.}\ \bibnamefont {Ning}},\
  and\ \bibinfo {author} {\bibfnamefont {M.}~\bibnamefont {Cheng}},\ }\bibfield
   {title} {\bibinfo {title} {Domain wall decorations, anomalies and spectral
  sequences in bosonic topological phases},\ }\href
  {https://arxiv.org/abs/2104.13233} {\bibfield  {journal} {\bibinfo  {journal}
  {arXiv:2104.13233}\ } (\bibinfo {year} {2021})}\BibitemShut {NoStop}%
\bibitem [{\citenamefont {Li}\ \emph {et~al.}(2023)\citenamefont {Li},
  \citenamefont {Oshikawa},\ and\ \citenamefont {Zheng}}]{LLi1}%
  \BibitemOpen
  \bibfield  {author} {\bibinfo {author} {\bibfnamefont {L.}~\bibnamefont
  {Li}}, \bibinfo {author} {\bibfnamefont {M.}~\bibnamefont {Oshikawa}},\ and\
  \bibinfo {author} {\bibfnamefont {Y.}~\bibnamefont {Zheng}},\ }\bibfield
  {title} {\bibinfo {title} {{Non-Invertible Duality Transformation Between SPT
  and SSB Phases}},\ }\href {https://arxiv.org/abs/2301.07899} {\bibfield
  {journal} {\bibinfo  {journal} {arXiv:2301.07899}\ } (\bibinfo {year}
  {2023})}\BibitemShut {NoStop}%
\bibitem [{\citenamefont {Li}\ \emph {et~al.}()\citenamefont {Li},
  \citenamefont {Oshikawa},\ and\ \citenamefont {Zheng}}]{LLi2}%
  \BibitemOpen
  \bibfield  {author} {\bibinfo {author} {\bibfnamefont {L.}~\bibnamefont
  {Li}}, \bibinfo {author} {\bibfnamefont {M.}~\bibnamefont {Oshikawa}},\ and\
  \bibinfo {author} {\bibfnamefont {Y.}~\bibnamefont {Zheng}},\ }\bibfield
  {title} {\bibinfo {title} {{Non-Invertible Kennedy-Tasaki Transformation and
  Applications to Gapless-SPT}},\ }\href@noop {} {\ }\bibinfo {note} {In
  preparation}\BibitemShut {NoStop}%
\bibitem [{\citenamefont {Kapustin}\ and\ \citenamefont
  {Seiberg}(2014)}]{kapustin2014coupling}%
  \BibitemOpen
  \bibfield  {author} {\bibinfo {author} {\bibfnamefont {A.}~\bibnamefont
  {Kapustin}}\ and\ \bibinfo {author} {\bibfnamefont {N.}~\bibnamefont
  {Seiberg}},\ }\bibfield  {title} {\bibinfo {title} {Coupling a {QFT} to a
  {TQFT} and duality},\ }\href
  {https://doi.org/https://doi.org/10.1007/JHEP04(2014)001} {\bibfield
  {journal} {\bibinfo  {journal} {J. High Energy Phys.}\ }\textbf {\bibinfo
  {volume} {2014}}\bibinfo  {number} { (4)},\ \bibinfo {pages} {1}}\BibitemShut
  {NoStop}%
\bibitem [{\citenamefont {Steenrod}(1947)}]{steenrod1947products}%
  \BibitemOpen
\bibfield  {number} {  }\bibfield  {author} {\bibinfo {author} {\bibfnamefont
  {N.~E.}\ \bibnamefont {Steenrod}},\ }\bibfield  {title} {\bibinfo {title}
  {Products of cocycles and extensions of mappings},\ }\href
  {https://doi.org/https://doi.org/10.2307/1969172} {\bibfield  {journal}
  {\bibinfo  {journal} {Ann. Math.}\ }\textbf {\bibinfo {volume} {48}},\
  \bibinfo {pages} {290} (\bibinfo {year} {1947})}\BibitemShut {NoStop}%
\bibitem [{\citenamefont {Wen}(2017)}]{PhysRevB.95.205142}%
  \BibitemOpen
  \bibfield  {author} {\bibinfo {author} {\bibfnamefont {X.-G.}\ \bibnamefont
  {Wen}},\ }\bibfield  {title} {\bibinfo {title} {Exactly soluble local bosonic
  cocycle models, statistical transmutation, and simplest time-reversal
  symmetric topological orders in 3+1 dimensions},\ }\href
  {https://doi.org/10.1103/PhysRevB.95.205142} {\bibfield  {journal} {\bibinfo
  {journal} {Phys. Rev. B}\ }\textbf {\bibinfo {volume} {95}},\ \bibinfo
  {pages} {205142} (\bibinfo {year} {2017})}\BibitemShut {NoStop}%
\end{thebibliography}%

\end{document}